\documentclass[12pt]{article}
\usepackage{graphicx}
\usepackage{morefloats}
\usepackage{amssymb}

\title{Auger neutralization and ionization processes for charge exchange between slow noble gas atoms and solid surfaces}
\author{R. Carmina Monreal \\
Departamento de F\'{\i}sica Te\'orica de la Materia Condensada C-5,  and  \\
Condensed Matter Physics Center (IFIMAC), \\
Universidad Aut\'onoma de Madrid. E-28049 Madrid. Spain.\\
}

\begin{document} 

\maketitle



\begin{abstract}

Electron and energy transfer processes between an atom or molecule and a surface are extremely important for many applications in physics and chemistry.  Therefore a profound understanding of these processes is essential in order to analyze a large variety of physical systems. The microscopic description of the two-electron Auger processes, leading to neutralization/ionization of an ion/neutral atom in front of a solid surface, has been a long-standing problem.  It can be dated back to the 1950s when H. D. Hagstrum proposed to use the information contained in the spectrum of the electrons emitted during the neutralization of slow noble gas ions as a surface analytical tool complementing photoelectron spectroscopy.  However, only recently a comprehensive description of the Auger neutralization mechanism has been achieved by the combined efforts of theoretical and experimental methods. In this article we review the theoretical models for this problem, stressing how their outcome compare with experimental results. We also analyze the inverse problem of Auger ionization.
 We emphasize the understanding of the key quantities governing the processes and outline the challenges remaining. This opens new perspectives for future developments of theoretical and experimental work in this field.

\end{abstract}



\newpage

\tableofcontents


\section{Introduction}

Electron and energy transfer processes between an atom or molecule and
a surface are extremely important for many applications
in physics and chemistry. In nanoscience, ion implantation in microelectronics has become an often used technique 
for fabrication of nanostructures and focused ion beams are used today for nano-patterning and nano-lithography. 
In plasma physics, the knowledge of plasma wall interactions
is of paramount importance to achieve high temperatures and confined plasmas and also, the sputtering and
ionization properties of the vessel walls and the plasma have a strong influence on the plasma temperature. 
In space science and astrophysics, solar wind exchanges charge and energy with satellites. In surface science,
Low Energy Ion Scattering (LEIS) has become a powerful tool for the analysis and characterization of surfaces and chemical composition. Likewise,   
Ion Neutralization Spectroscopy (INS) is considered to be a technique complementary to photoemission electron spectroscopy (PES).
 In chemistry, 
catalytic reactions are extremely sensitive to charge exchange. All these applications make the field of particle-surface interactions a very
interesting field. However, from the theoretical point of view, this is in general a problem of great  complexity. 
On the one hand, target and projectile have a complicated internal structure 
which is revealed during the collision. On the other hand, new structure appears in the interaction of the particle with the surface which
in turn is a dynamical (time-dependent) situation. Therefore, a substantial body of work is
devoted to the understanding of the relevant microscopic mechanisms leading to charge transfer 
\cite{EFR,losGeerlings,niehusSSRep,winterPhysRep2002,monrealFlores,gauyacqPSS2007,brongersmaSSRep2007}. 
\newline

The two basic charge transfer mechanisms between an atom or molecule and a solid surface are known as resonant and Auger processes.
Resonant processes are single electron  mechanisms
in which an electron tunnels from/to the atom to/from the solid when
the energy level of the atom is in resonance with the continuum
of states of the solid (see Fig. 1). Resonant processes, being one-electron ones,
have been described abundantly in the literature, practically for any atom/solid combination, using different techniques
\cite{gadzukSS1967,gadzuk1SS1967,janevJOPB1974,blandin1976,tullyPRB1977,janev1PLA1978,brakoNewnsSS1981,geelingsLosSS1986,
nordlanderTullyPRL1988,sulstonDavidsonPRB1988,sulstonDavidsonSS1989,brakoNewnsRPP1989,
nordlanderTullyPRB1990,teilletGauyacqSS1990,
langrethNordlanderPRB1991,borisovTeilletPRL1992,marstonPRB1993,shaoPRB1994,shao1PRB1994,bolcattoPRA1994,
authBorisovPRL1995,winterMertensPRA1996,merinoPRB1996,deutcherPRA1997,kurpickPRA1997,merinoMarstonPRB1998,
borisovKazanskyPRB1999,borisovKazanskySS1999,wangGarciaPRA2001,carterJCP2004,goldFloresMonrealPRB2005}.
\newline

Besides resonant tunneling, the two-electron Auger processes comprise the other fundamental electron transfer processes for ion-surface
interactions. 
In the Auger Neutralization (AN) process depicted in Fig. 2a, one electron from the surface is
transferred to a bound state (often the ground state) of the atom
while, by virtue of electron-electron interaction, energy
and momentum are transferred to the solid creating
surface excitations (electron-hole pairs and plasmons).
In the inverse process of Auger Ionization (AI) (Fig. 2b), an electron bound to the atom is transferred
to a state above the Fermi energy with the creation of surface excitations.
Energy conservation requires kinetic energy from the atom and
therefore AI is only possible above a threshold kinetic energy.
Auger deexcitation (AD) is another two-electron mechanism of charge and energy transfer. 
Here an atom initially in an excited state decays to its ground state via electron emission.
This process can proceed in a direct or indirect way. In the direct AD (Fig. 2c) the excited atom 
decays to the atomic ground state with energy and momentum taken by an electron from the solid, while
in the indirect AD (Fig. 2d) an electron from the solid is transferred to the ground state and the atomic excited electron is emitted.
In contrast to resonant, the Auger two-electron processes  are much more difficult to describe theoretically  for 
two reasons:
i) they involve four different electronic states and ii)
the long range of the Coulomb electron-electron interaction has to be appropriately screened which is particularly important
at metal surfaces. Then, Auger processes actually do not involve only two but many electrons in the system. 
These problems, added to the breaking of spatial symmetry brought about by the surface and the atom, have made the  path to    
a thorough theory for these processes to be long and full of different approximations, leading in some cases
to apparent discrepancies with experiments. In this
respect, it is desirable to realize experimentally systems where single
processes can be studied under well defined and controlled conditions. This is the case for singly charged noble gas ions in the ground state and metal surfaces of high work function.  For these systems the atomic ground state is non-degenerate with the occupied 
electronic states of the surface and the atomic excited states
are resonant with the empty states of the metal. Fig. \ref{figHeMetal} illustrates schematically the relative positions of the different energy levels for the case of He$^+$. 
 This is a case study system since Auger
neutralization  is the only possible mechanism of charge transfer. However, in spite of its apparent simplicity,
 a complete microscopic understanding of this system has been achieved only recently.
\newline

This article is devoted to review the progress made during the last years in the theoretical description of Auger charge transfer processes between
slow singly charged noble gas ions and metal surfaces of high work function. The manuscript is organized in the following way. Section 2 gives a historical 
overview of the problem, describing briefly both the experimental methods and the theoretical approaches to Auger neutralization employed
until the 1990s. Section 3 reviews in detail the general formalism of a multielectron theory of Auger neutralization, 
presenting first the theory and calculations for the Auger  
neutralization rate of He$^+$ in front of a jellium surface and then on a corrugated surface.
 After addressing the problem of energy level variation in Section 4, Section 5 is devoted to show the comparison between theory and experiment for
ion fractions and energy gains of He$^+$ interacting with Al and also with noble metal surfaces, both under grazing scattering and in the LEIS regime.
In Section 6 we discuss the inverse problem of Auger ionization, providing comparison with measurements of ion fractions. Finally the conclusions and
outlook are presented in Section 7. 
Atomic units ($e=\hbar=m_e=1$) are used throughout this article unless otherwise stated.

\section{Historical Overview}
 
\subsection{Experimental Methods} 
\label{secExperimentalMethods}

Since Shekhter \cite{shekhter1937} proposed the two-electron Auger mechanism, different theoretical work was performed until the 1950s 
\cite{shekhter1937,cobasLamb}. 
In that decade the pioneering work by Hagstrum \cite{hagstrumPR1953,hagstrum1PR1953,hagstrumPR1954,hagstrum1PR1954,hagstrumPR1956,
hagstrumPR1961,hagstrumPR1966,hagstrumBeckerPR1967}
 developed the experimental technique for INS and also established the 
 ingredients necessary to describe ion neutralization in the same way we use today.
 It took him several years of research to design and construct a
new apparatus that incorporated a low-energy
electron-diffraction insert for being able to investigate
the surface symmetry and reconstruction.  He introduced the concept of a turret within which
the sample could be manipulated to allow a number of different
probes or coatings to be applied to the same surface.
 In the experiments, Hagstrum directed slow noble gas ions (usually in the ground state) 
at a metal surface at normal incidence, and
 the emitted electrons were collected using a hemi-spherical cup and analyzed in energy. 

In addition to his experimental work,
 Hagstrum interpreted his results using a minimum number of assumptions \cite{hagstrum1PR1954}. By comparing the energy spectra of the electrons emitted
 under impact of different noble gas ions, 
 he noticed that the high energy end of the electron distribution was
 directly related to the ionization potential of the noble gas atom and  assumed these electrons were emitted in the Auger neutralization of the 
 impinging ion. Then, the maximum energy of an emitted electron, $E_{max}$, should correspond 
 to a transition in which the two electrons involved in the process are at the Fermi level (see Fig. 2a), 
  $E_{max}=-E_a-2W$, where $-E_a$ is the ionization potential of the atom and $W$ is the work function 
 of the metal surface. However the experimental value of $E_{max}$ was lower by about 2 eV with respect to the value one would obtain using the ionization potentials
 of free atoms. This fact was interpreted in terms of a change in the energy level of
 the active atomic electron due to the its interaction with the surface. 
In the simplest approximation, the electron interacts with its own image charge and with that of the nucleus, resulting in a distance-dependent energy level
of the form  
 
 \begin{equation}
E_a(z)=E_a(\infty)+e^2/4z, 
\label{Ea-image}
\end{equation}
 where it has been assumed that the interaction energy 
of the neutral atom and the surface is essentially of the van der Waals type and thus negligibly small at the distances were neutralization occurs. 
In Eq. (\ref{Ea-image}), 
$e$ is the electron charge  
and $z$  is the perpendicular distance between the ion and the surface.
Then, the observed change of 2 eV in the ionization potential has to be approximately the value of the image potential energy
at the distance $z_{m}$ where the ions were most probable neutralized, obtaining  in this way $z_{m}\simeq 3.5$ a.u. 

 Hagstrum also introduced the Auger transition rate which is  the 
probability per unit time that an electron at a given band energy in the
solid will be involved in the neutralization process. It depends upon the initial
and final densities of states and upon the transition matrix
elements and final state interactions as in PES. It also
depends on the overlap between the  wave functions of the solid electron and those of the ion outside the surface and is
thus more surface sensitive than other spectroscopies. Actually the Auger transition rate $\Gamma$ has to decay exponentially away of
 the surface as the overlap does
 
\begin{equation}
\Gamma(z)=\Gamma_0 e^{-\frac{z}{d}},
\label{Gamma-general}
\end{equation}
 with $\Gamma_0$ and $d$ being the parameters describing the 
 Auger interaction for a given ion-solid combination. 
 The most probable distance of neutralization is related to these parameters in a simple way.
 The fraction of the incoming ions that have survived Auger neutralization at a distance $z$, $n_{+}(z)$,  has to fulfill the rate equation, 
 \begin{equation}
\frac{dn_{+}(z)}{dz}=-\frac{\Gamma(z)}{v} n_{+}(z) ,
\label{eq-rate}
\end{equation}
where $v$ is the ion velocity perpendicular to the surface.  With  $\Gamma(z)$ given by Eq. (\ref{Gamma-general}), the integration of the rate equation
 gives
 
 \begin{equation}
n_{+}(z)=exp[-\frac{\Gamma_0 d}{v} exp(-\frac{z}{d})]
\label{n+}
\end{equation}
  
 Note in Eq. (\ref{n+}) that $v_c= \Gamma_0 d$ plays the role of a characteristic velocity for the AN process.  Finally,  the most
 probable distance for neutralization $z_{m}$ is the maximum of the function $n_{+}(z) \Gamma(z)$, 
since this function is directly proportional to the probability
 that an ion be neutralized at a distance $z$ from the surface.  Eqs. (\ref{Gamma-general}) 
and (\ref{n+}) give  $z_{m}=d\; ln(\frac{v_c}{v})$.
Then for a 10 eV He$^+$ ion and taking $z_{m}=3.5 a.u.$ and $d\simeq 1 a.u.$, the experimental estimate of the characteristic velocity  is 
$v_c \simeq 10^9 cm/s$. When comparing to theoretical calculations by Shekhter \cite{shekhter1937} and
Cobas and Lamb \cite{cobasLamb}, Hagstrum pointed out that in general the experimental estimates of $v_c$ were considerably larger
than the theoretical and that the much shorter theoretical estimates of $z_{m}$ appeared unrealistic. 
It was also noticed that with these experimental values of the parameters virtually no ions would survive Auger neutralization already
in the incoming part of the trajectory.
 The concepts of distance dependent energy level (or ionization potential), which is today known as "levelshift", 
 and Auger transition rates introduced by Hagstrum have been almost invariable used in all
 the subsequent work. Moreover, the belief that slow ions are Auger-neutralized far from the surface,
 at typical distances of 3-4 a.u. with respect to the image plane
(6-7 a.u. with respect to the first atomic layer), has been at the core of the interpretation
 of many experiments concerning ion neutralization of slow ions at surfaces until recently.  Subsequent experiments by different groups measuring 
 electrons emitted in the interaction of incident ions or excited atoms of noble gases and different metal surfaces  
\cite{hagstrumBeckerPR1967,sesselmannPRL1983,sesselmannPRB1987,hagstrumPRB1988} did not substantially alter Hagstrum's earlier findings. 

Other experiments measured the spin-polarization of the electrons emitted when a beam of metastable 
He (2 $^3$S) atoms are scattered off metal surfaces \cite{onellionPRL1984,hartPRB1989,lancasterPRB2003}. The basic idea in these experiments is that He$^*$ 
loses resonantly the excited electron at a large distance from the surface, leaving a He$^+$ ion with a well defined electron spin. 
When this ion is Auger- neutralized the emitted Auger electron has information on the spin dependence of the surface density of states. This kind
of experiments were generally interpreted in terms of Hagstrum's model with the same assumptions.
\newline

A different  kind of experimental work has been conducted to deduce charge transfer rates and interaction potentials 
from the analysis of the angular distributions of scattered particles in grazing collisions 
\cite{winterPRA1992,winterJPCM1993,winterJPCM1996}. 
The basic idea is to bombard the metal surface with ions and neutrals at a 
very grazing angle of incidence $\Phi_{in}$ of typically  $1^{\circ}$ 
with respect to the surface and along high-index ("random") direction.  Under these conditions
scattering proceeds in the surface channeling regime \cite{winterPhysRep2002,gemmell} where the projectiles spend a long
time traveling  well above  the topmost layer of atoms and  parallel and perpendicular motions are widely decoupled. 
The parallel motion takes place
with constant velocity while the perpendicular velocity of the ions changes due to the electric force exerted by the metal surface before
neutralization. 
Since incident neutral atoms do not  experience such Coulomb forces, the trajectories of ions and neutrals with identical energies
 and angles of incidence will differ after scattering with the surface, see Fig. \ref{figWinterJP}.
Thus, by measuring polar angular distributions of the neutral atoms collected from the scattering of incident ions and neutrals,
 it is possible to directly obtain the amount of kinetic energy
gained by the ions prior to their neutralization in the following way.
The polar angular distribution for incident neutrals has a  nearly gaussian shape, peaking at a scattering angle $\Phi^{0} $ which is close to the elastic value 
$2 \Phi_{in}$, and having a width controlled essentially by lattice vibrations.
The peak of the distribution for incident ions  is at a different scattering angle $\Phi^{+}$, usually larger than $\Phi^{0} $ for 
the smaller incident normal energies, and the width of the distribution is larger,  
reflecting the width in the the distribution of neutralizing distances (Fig. \ref{figWinterJP}). From the peak positions, the amount of
energy gained by the ions prior to neutralization is obtained via

\begin{equation}
E_{gain}=E_{\perp}^{out}-E_{\perp}^{in}=E_{0}[sin^2(\Phi^ {+}-\Phi^{0}/2)-sin^2(\Phi^ {0}/2)]
\label{Egain} 
\end{equation}
where $E_{\perp}^{out}$ is the normal energy of the outgoing neutralized ions and $E_{\perp}^{in}$ is their incident normal energy.
Experiments in which  a beam of $^4$He$^+$ projectiles at grazing incidence and with typical 
perpendicular incident energies of a few eV were scattered off Al(111) surfaces \cite{winterJPCM1993,hechtSS1998}, 
gave values of the energy gain close to 2eV, in agreement with Hagstrum findings.  
In fact, since the ion and surface interact via the electric charges they constitute a conservative system
in which the increase in the ion kinetic energy due
to its attraction by the surface has to be compensated by a decrease in its potential energy. 
The increase in kinetic energy was
again interpreted in terms of a pure ion- image charge Coulomb interaction similar to Eq. (\ref{Ea-image}). With this assumption,   
and using an Auger transition rate of the exponential form Eq. (\ref{Gamma-general}), 
molecular dynamics simulations of trajectories and polar angular distributions of scattered particles were performed with $\Gamma_0$ and $d$ 
as adjustable parameters. In this way, the Auger neutralization rate   
$\Gamma(z)$ for He$^+$ on Al was retrieved from the experiments in \cite{hechtSS1998} and compared to the most sophisticated  calculations existing
to that date \cite{lorenteMonrealSS1997} with identical conclusion as by Hagstrum: theory gave too small values of the AN rates to account for the experiments.
\newline

A widely used technique for surface analysis is Low Energy Ions Scattering (LEIS), also known as Ion Scattering Spectroscopy (ISS). 
This technique operates in a very different regime of incident normal energies than INS.  In LEIS the  sample is bombarded with noble
gas ions at nearly normal incidence and the particles which are back-scattered at a certain solid angle are analyzed. 
Typical incident energies range from 0.5 keV to 10 keV and the
angle of incidence, measured with respect to the surface normal, is smaller than 60$^{\circ}$. In a common experimental setup the ions are
incident normal to the surface and the backscattered particles are collected at a large scattering angle, 
of the order of 140$^{\circ}$ (40$^{\circ}$ with respect to the surface normal).  
Under these conditions, the 
scattering of the incident projectiles takes place at the outermost atomic layers and this is the basic reason why LEIS is a non-destructive
technique with pronounced surface sensitivity. Its  field of application is mostly analysis of surface structure or surface composition
\cite{niehusSSRep,brongersmaSSRep2007}. 
There are two standard ways to analyze the energy of the back-scattered particles. One is to measure their Time of Flight (TOF) and to calculate from it 
the corresponding energies of all scattered particles (ions and neutrals) and the other is to determine directly the energy using an
Electrostatic Energy Analyzer (ESA) in which case only ions are selected. The ESA setup is mostly used for analysis of the surface composition 
while TOF is used for analysis of surface structure. 
From the energy spectra of scattered particles, one can obtain the 
charge fraction $P^{+}$ which is the fraction of the particles that reach the detector in an ionic state and therefore depends on the neutralization probability. In order to obtain quantitative results with LEIS, detailed knowledge of charge exchange
processes at surfaces is of crucial importance \cite{oetelaar1998,cortenraad1999}.
 In the AN regime of LEIS it is common practice to use Hagstrum's model and  Eq. (\ref{eq-rate}), yielding the ion fraction as

\begin{equation}
P^{+}=exp(-\int \Gamma(z)\frac{dz}{v_{\perp}})\simeq exp(-2 \frac{\Gamma_0 d}{v_{\perp}})
\label{eq-P+}
\end{equation}
where the factor of 2 in the second exponential on the right hand side accounts for survival in the incoming and outgoing trajectories,
assuming for simplicity that the perpendicular velocity $ v_{\perp}$ is the same. 
Then, a plot of the experimental results for $P^{+}$ as a function
of $\frac{1}{v_{\perp}}$ fitted to the exponential form provided by Eq. (\ref{eq-P+}) yields the value
of the characteristic velocity which can be compared to theory \cite{draxlerPRL2002}. 
It has been commonly assumed in LEIS that the neutralization probability only depends upon the chemical species of the incident ion and target atom
and  is independent of the environment  where the target atom is placed. This particular behavior is often termed as "absence of matrix-effects" in LEIS 
\cite{brongersmaSSRep2007}.
One might ask, however, why band structure and  density of states of the solid play no role in LEIS while this is not the case in INS.  

\subsection{Theoretical Models}
\label{secTheoreticalModels}

As we have already stressed, the understanding of the Auger neutralization processes at surfaces has been a long standing problem. 
The first works by Shekhter and Cobas and Lamb
\cite{shekhter1937,cobasLamb} were later improved by Propst \cite{propst1963}. He calculated the Auger matrix elements using a WKB approximation
for the tunneling through the ion-surface barrier, considered as one-dimensional problem. Later on, Heine \cite{heine1966} established some basic properties
of Auger neutralization. He pointed out the importance of the screening of the Coulomb potential at the surface and argued that the emitted 
electron came from the first atomic layer. 
\newline

Many works concentrated on the calculations of the spectrum of the electrons emitted in the neutralization
process in order to ascertain to which extent they carry information on the surface electronic structure. 
 Following Appelbaum and Hamman \cite{appelbaum1975}, Modinos and Easa \cite{modinos1986} calculated electron spectra 
assuming that one of the two-electrons was localized near the ion while the second one was emitted from the surface.
However, other works \cite{hood1985} parametrized the density of states and assumed an Auger interaction to be fully local in the sense
that the wave functions of both electrons are sensitive to the spatial region around the ion. 
In other cases, both transition rates and transition energies were taken as adjustable parameters to fit the whole electron emission spectra, particularly  when the projectiles are metastable atoms or doubly charged ions, due to the fact that different resonant and Auger 
processes can happen in the most common atom/surface combinations
\cite{niehausSS1988,brentenPRL1993,kempterSS1996}.

In the case of experiments measuring the spin polarization  of the emitted electrons, 
the experimental spin-asymmetry found for magnetic Ni surfaces \cite{onellionPRL1984}, was analyzed by Penn and Apell 
\cite{pennApellPRB1990} on the basis of Hagstrum's model of image-potential energy level shift, using two adjustable parameters.
The authors conclude that the spin-asymmetry measures the Ni magnetization outside the surface at an ion position of about 9 a.u., a value
of the most probable distance of neutralization which is even larger than  Hagstrum's estimates.   
For the case of non-magnetic surfaces, the experiments of \cite{hartPRB1989} were analyzed in terms of the exchange interaction at the surface
\cite{salmiSSC1991,salmiPRB1992}. Other experiments \cite{lancasterPRB2003} have been analyzed in terms of 
the exchange interaction not at the surface but in the volume of a free-electron gas, including the production  
of secondary (cascade) electrons and the transport to the surface \cite{roslerAlducinNIMB2007}.  
\newline

In this article we will be mainly concerned with calculations of the Auger neutralization rate of slow ions at metal surfaces. 
This means that the ion velocity is much smaller than the Fermi velocity of the electrons in the metal and thus
the ion is considered to be "at rest" at a position $\vec{R_a}$ with respect to the surface. 
To first order in
perturbation theory the rate is given by Fermi's golden rule as

\begin{equation}
\Gamma(\vec{R_a})=  2\pi \sum_{\vec{k}_1} \sum_{\vec{k}_2} \sum_{\vec{k}_A} \sum_{\sigma} f_{\vec{k}_1} f_{\vec{k}_2}
(1-f_{\vec{k}_A})
|M_{\vec{k}_A,a,\vec{k}_1,\vec{k}_2,\sigma}(\vec{R}_a)|^2
 \delta(E_{\vec{k}_A,\sigma}+E_a(\vec{R}_a)-E_{\vec{k}_2,\sigma}-E_{\vec{k}_1}),
\label{rate-general}
\end{equation}
with the matrix elements given by

\begin{equation}
M_{\vec{k}_A,a,\vec{k}_1,\vec{k}_2,\sigma}(\vec{R}_a)=
\int d \vec{r}_1 \int d \vec{r}_2 \Psi^{*}_{\vec{k}_A,\sigma}(\vec{r}_2) \Psi^{*}_{a}(\vec{R}_a,\vec{r}_1) V_{SC}(\vec{r}_2,\vec{r}_1)
\Psi_{\vec{k}_1}(\vec{r}_1) \Psi_{\vec{k}_2,\sigma}(\vec{r}_2).
\label{matrix-general}
\end{equation}

 Eq. (\ref{matrix-general}) is the matrix element representing an Auger neutralization event in which 
an electron of the solid in a state $\vec{k}_1$ of energy $E_{\vec{k}_1}$ below the Fermi level (wave vector
$k_F$ and energy $E_F$) ,
described by the wave function $\Psi_{\vec{k}_1}$, is transferred to an atomic state of energy $E_a(\vec{R}_a)$
 described by the wave function $\Psi_{a}$ which is localized around the atomic position $\vec{R}_a$, while another electron 
in a state of wave vector $\vec{k}_2$ and spin $\sigma$ of energy $E_{\vec{k}_2,\sigma}$ below the Fermi level,
described by the wave function $\Psi_{\vec{k}_2,\sigma}$ is excited above the Fermi level to a state of the same spin,
momentum $\vec{k}_A$, energy $E_{\vec{k}_A,\sigma}$ and
 wave function $\Psi_{\vec{k}_A,\sigma}$. 
The two-electron scattering potential $V_{SC}(\vec{r}_2, \vec{r}_1)$
is a Coulombic one with appropriate screening.  In Eq. (\ref{rate-general}), $f_{\vec{k}}$ is the Fermi-Dirac distribution function and
the $\delta$-function  expresses energy conservation in the scattering process.
One can appreciate the difficulty of the problem by noting that, first, 
the calculation of the matrix elements involve a 6-dimensional integral in coordinate space and 
then the calculation of $\Gamma$ a 9-dimensional integral in momentum space.

Snowdon et al. \cite{hentschke1986,snowdon1986} performed one of the first "modern" calculations of the Auger neutralization rate. 
They used a free-electron model for the metal in which the electron wave functions were that of a simple step-potential surface.
Screening was introduced by using a parametrized Thomas-Fermi type of potential which allowed the authors to calculate matrix elements analytically.
Furthermore some simplifications in the summations over electronic states had to be done, as to consider only states having momentum
perpendicular to the surface. Janev and Nedeljkovic \cite{janevNedel1985} 
reduced drastically the phase space of integration by considering only the dipolar term in an expansion of the Coulomb electron-electron
interaction, thus obtaining an analytical expression for the Auger rate. 

Other authors were interested in the effects of
the ion velocity under grazing incidence conditions.  Miskovic and Janev \cite{miskovicJanevSS1986,miskovicJanevSS1989}
introduced the effect of the ion motion in the Auger rates, and Zimny et al. \cite{zimnySS1991} 
derived an "universal" function of the ion velocity to be included in the calculation of the rate.
Effects of the corrugation of the solid surface, have been studied by Kaji et al. \cite{kajiSS1990,kajiSS1992} using simplified wave functions.

A major step forward was taken in the work by Fond\'en and Zwartkruis \cite{fondenZwartkruisSS1992,fondenZwartkruisSS1993,fondenZwartkruisPRB1993}
who used realistic wave functions for the electrons, obtained within the jellium model in the Local Density Approximation (LDA)
and  computed the multidimensional integrals using Monte Carlo techniques. 
However, like many of the approaches above, these authors used a bare Coulomb potential or described the screening of the electron-electron interaction 
in the static Thomas-Fermi approximation. They showed that the Auger rate is strongly sensitive to the value of the Thomas-Fermi screening length and
thus could lead to unphysical results \cite{fondenZwartkruisPRB1993}. This made evident that a proper treatment of screening, essential to get reliable results, was still lacking.

\section{Multielectron Theory of Auger Neutralization}
\label{secMultielectron}

The most striking effect of the screening properties of a many-electron system to a dynamical perturbation is the building up of collective
excitations known as plasmons. These are charge oscillations of the electrons with frequency $\omega_p=\sqrt{\frac{4 \pi e^2 n}{m}}$ where $n$ is
the electronic density. 
The quantum nature of plasmons was discovered in 1941 as multiple discrete energy losses
of electrons traversing thin films \cite{ruthemann1941} and first explained theoretically using quantum theory by Pines and Bohm in 1952 
\cite{pinesBohm1952}. 
These excitations are evidenced not only as electron energy losses but also in the "shake-up" excitation produced by sudden changes in the occupation
of localized electron states \cite{langreth,cini}, eg., in the photoionization and decay of inner shells \cite{almbladh, almbladh1}.

The presence of a boundary between the metal and its surroundings introduces charge oscillations at the surface, known as surface plasmons. 
 They were first predicted by Ritchie in 1957 \cite{ritchiePR1957} and have been the object of an 
intense research ever since \cite{raether1980,feibelmanPSS1982,raether1988,rocca1995,liebsch1997}, 
giving rise in our days to the emergent field of plasmonics \cite{maier2007,fjgdeabajo2007,fjgarciavidal2010}.
The frequency of a surface plasmon depends, of course, on the 
geometry of the boundary. For the case of a planar surface we are interested in, and assuming the external perturbation 
to be of infinite wavelength or, equivalently very short wave vector, 
it is given by  $\omega_{sp}=\omega_p/\sqrt 2$.  It can be shown from  general grounds \cite{harrisGriffin} that 
the frequency of the surface plasmon depends on the component of the wave vector parallel to the surface $q$ as

\begin{equation}
\omega_{sp}(q)=\frac{\omega_p}{\sqrt 2}(1+\frac{1}{2}q Re[d_{\perp}]),
\label{surface-plasmon}
\end{equation}
where the length $d_{\perp}$ is in general a complex quantity describing the screening properties of the metal surface.
Its real part, $Re [d_{\perp}]$, which is directly responsible for the surface plasmon relation of dispersion 
expressed by Eq. (\ref{surface-plasmon}), 
is related to the center of gravity of the charge density induced at the surface,  
and its imaginary part describes the absorption of energy by the surface \cite{feibelmanPSS1982}. 
The fact that the surface plasmon energy acquires an imaginary part indicates that the collective mode decays into the
incoherent excitation of electron-hole pairs. 
When an ion undergoes Auger neutralization close to a metal surface, energy and momentum are transferred from the ion to the surface that can 
excite not only electron-hole pairs (Auger electrons) but surface plasmons as well. This last mechanism is only possible if the associated frequency $\omega$ 
and wave vector parallel to the surface $q$ match those of the plasmon. The surface-plasmon assisted  channel 
for Auger neutralization was first introduced by Apell in the case of the direct Auger deexcitation process 
\cite{apellJPB1988} and later by Almulhem and Girardeau for studying neutralization of protons in Al \cite{almulhemSS1989}. 
Even though this mechanism has been assumed to be important for Auger neutralization, it has not been properly taken into account
\cite{almulhemSS1989,galosSS1999}.  The main reason for it is that surface plasmons and electron-hole pairs are coupled 
surface excitations (except at $q=0$) and the coupling intensities are not arbitrary, since they have to fulfill sum rules. 
Therefore surface plasmons and electron-hole pairs cannot be considered as different kinds of surface excitations and treated in an independent way.
In other words, any appropriate theory of the so-called plasmon-assisted neutralization channel should take into account the multielectron (many-body) nature of the system.
\newline

\subsection{General Formalism}
\label{secGeneralFormalism}

Inspired by the formulation in \cite{EFR} for the bulk problem, a general theory for multielectron Auger neutralization of ions at metal surfaces, based on a dielectric formalism, was developed by Monreal and Lorente \cite{lorenteMonrealSS1997,monrealLorentePRB1995,lorenteMonrealPRB1996}.   

The basic formula for the Auger transition rate
of Eq. (\ref{rate-general}) (denoted by $\frac{1}{\tau_{AN}}$ hereafter) is first written in an equivalent way as

\begin{equation}
\Gamma \equiv \frac {1}{\tau_{AN}} =2 \pi \sum_{i,j} |\langle f|\hat V|i
\rangle|^2 \delta(E_f-E_i)
\label{eq-tau0}
\end{equation}
with  $|i\rangle$
and  $|f
\rangle$ being the initial and final states of the system atom plus metal with energies $E_i$ and $E_f$ respectively and 
$\hat V$ is the Coulomb scattering potential. Now we separate the electron which neutralizes the ion core from the rest of the system of $N-1$ electrons.
 These $N-1$ electrons are initially in their 
ground state $|0\rangle$ and end up in an excited state $|n\rangle$ upon ion neutralization while the metal electron, which initially is in a state 
$|\vec{k_{or}}
\rangle$ of energy $E_{\vec{k}}$, ends up in an atomic state $|a\rangle$ of energy $E_a$. Under this approximation the initial and final states of the $N$- electron system are written as

\begin{eqnarray}
|i\rangle & = & |0\rangle \otimes |\vec {k}_{or}\rangle \nonumber \\
|f\rangle & = & |n\rangle \otimes |a\rangle
\label{if}
\end{eqnarray}
  In this equation, the subindex of $\vec{k}_{or}$ reminds us that states $|\vec{k}_{or}\rangle$ and $|a\rangle$ have to be
orthogonal, since they should be eigenstates of the same initial Hamiltonian and  $E_f-E_i=E_a+E_n-E_0-E_{\vec{k}}$.

The interaction potential $\hat V$ can be viewed as a density-density Coulomb interaction,

\begin{equation}
\hat V=\int d\vec{r}_1 \int d\vec{r}_2 \frac{\delta \hat n(\vec{r}_1) \hat \rho(\vec{r}_2)}{|\vec{r}_1-\vec{r}_2|}
\label{V},
\end{equation}
where $\delta \hat n(\vec{r}_1)$ is the density operator of the $N-1$ electron system and $\hat \rho(\vec{r}_2)$ is the 
density associated to the transition from $ |\vec{k}_{or}\rangle$ to $|a\rangle$. We will assume the ion to be outside the surface of a
semi-infinite metal described in the jellium model, the advantage being that  the system has translational invariance parallel to
the surface.  
 Then, it is convenient to Fourier-transform in
the coordinates $\vec {x}$ parallel to the surface and, using Eqs. (\ref{V}) and (\ref{if}), write down the matrix elements of Eq. (\ref{eq-tau0}) as

\begin{equation}
\langle f|\hat V|i\rangle=\int \frac{d \vec{q}}{(2 \pi)^2} \int dz_{1} \langle n|\delta \hat n(\vec{q}, z_1)|0\rangle \Phi(\vec{k}; \vec{q}, z_1),
\label{fVi}
\end{equation}
where we have defined

\begin{equation}
\Phi(\vec{k}; \vec{q}, z_{1})= \frac{2 \pi}{q} \langle a|e^{i \vec{q} \cdot \vec{x}_{2}} e^{-q|z_{1}-z_{2}|}|\vec{k}_{or}\rangle.
\label{potential}
\end{equation}
In this equation $\Phi(\vec{k}; \vec{q}, z_{1})$ is a potential depending on the initial state $|\vec{k}_{or}\rangle$,   
the wave vector parallel to the surface $\vec{q}$, the coordinate perpendicular to the surface $z_1$ and also implicitly on the
atomic position $z_a$ because the wave function of the atomic state is localized around that point. 
In Eq. (\ref{potential}) the internal integration implied in the matrix element is in the coordinates $\vec{x}_2$ and $z_2$.  

Now, by substituting Eq. (\ref{fVi}) into Eq. (\ref{eq-tau0}) and making use of the definition of the susceptibility of the many-electron system
$ \chi(q, \omega; z, z')$
in terms of the density operators, we obtain our final expression for the Auger rate as

\begin{eqnarray}
 \frac {1}{\tau_{AN}}(z_a) = & 2 & \sum_{k<k_F} \int_{0}^{\infty} d\omega \int \frac{d\vec{q}}{(2 \pi)^2} 
\int dz \int dz' Im(-\chi(q, \omega; z, z')) \nonumber  \\
& \times & \Phi(\vec{k}; \vec{q}, z) \Phi ^{*}(\vec{k}; \vec{q}, z') \delta(E_{a}(z_a)-E_{\vec{k}}+\omega).
\label{eq-taujelly}
\end{eqnarray}

The physical interpretation of the computationally involved Eq. (\ref{eq-taujelly}) is simple. One can consider the magnitude
$\Phi(\vec{k}; \vec{q}, z) e^{-i(E_{\vec{k}}-E_a)t}$, where  $\Phi(\vec{k}; \vec{q}, z)$ is given by Eq. (\ref{potential}), as an effective Coulomb potential 
caused by the transition of one electron
from the initial metallic state $|\vec{k}\rangle e^{- i E_{\vec{k}} t}$ to the final atomic state $|a\rangle e^{-i E_a t}$. This potential oscillates in time
with frequency $\omega=E_{\vec{k}}-E_a$ inducing fluctuations in the charge density of the electronic system.  The magnitude of these
fluctuations is determined by the surface susceptibility of the many-body electron system $\chi(q, \omega; z, z')$ and given by

\begin{equation}
\delta n(\vec{q}, \omega; z)= \int dz' \chi(q, \omega; z, z') \Phi(\vec{k}; \vec{q}, z').
\label{dn}
\end{equation}

The density given by Eq. (\ref{dn}) is a complex magnitude whose real part describes the screening charge and whose imaginary part is related to energy losses. 
Actually, the quantity $ Im (- \int dz \delta n(\vec{q}, \omega; z) \Phi^{*}(\vec{k}; \vec{q},z)) $ gives the rate at which the potential 
 $\Phi(\vec{k}; \vec{q}, z) e^{-i \omega t}$ produces
metal excitations of energy $\omega$ and momentum $\vec{q}$. When adding all possible contributions of all neutralizing states $|\vec{k}\rangle$,
 Eq. (\ref{eq-taujelly}) expresses the total rate at which metal excitations are produced in the Auger process. 
 \newline

The important physical magnitude in this theory is the surface susceptibility for interacting electrons (also called surface screened susceptibility)
$\chi(q, \omega; z, z')$, which can be obtained from the susceptibility $\chi_0(q, \omega; z, z')$ for non-interacting electrons by means of a self-consistent-field approximation. 

 The charge density induced in a system of interacting electrons by an external potential
$U_{ext}(\vec{r})e^{-i \omega t}$ is in general defined as

\begin{equation}
\delta n(\omega; \vec{r})= \int d\vec{r'} \chi(\omega; \vec{r}, \vec{r'}) U_{ext}(\vec{r'}).
\label{dn1}
\end{equation}
The self-consistent-field approximation introduces a self-consistent-field. This field is the sum of the external potential plus the potential produced
by the induced charge

\begin{equation}
U_{SCF}(\omega; \vec{r})= U_{ext}(\vec{r})+U_{ind}(\omega; \vec{r}),
\label{dn2}
\end{equation}
with

\begin{equation}
U_{ind}(\omega; \vec{r})=\int d\vec{r'} \; \frac{1}{|\vec{r}-\vec{r'}|} \delta n(\omega; \vec{r'}).
\label{dn3}
\end{equation}

Then, the approximation assumes that the interacting electrons respond to the external perturbation as if they were non-interacting electrons 
in the presence of the self-consistent-field:  

\begin{equation}
\delta n(\omega; \vec{r})= \int d\vec{r'} \; \chi_0(\omega; \vec{r}, \vec{r'}) U_{SCF}(\omega; \vec{r'}),
\label{dn4}
\end{equation}
where $\chi_0(\omega; \vec{r}, \vec{r'})$ is the susceptibility for non-interacting electrons. 
Thus, Eqs. (\ref{dn1})-(\ref{dn4}) imply that $\chi(\omega; \vec{r}, \vec{r'})$ has to fulfill the self-consistent equation

\begin{equation}
\chi(\omega; \vec{r}, \vec{r'})=\chi_0(\omega; \vec{r}, \vec{r'})+\int d\vec{r}_1 \int d\vec{r}_2 \; 
\chi_0(\omega; \vec{r}, \vec{r}_1) \frac{1}{|\vec{r}_1-\vec{r}_2|} \chi(\omega; \vec{r}_2, \vec{r'}).
\label{chir}
\end{equation}

 To obtain $\chi$ only a set of one-electron wave functions $\phi_{\vec{k}}(\vec{r})$ with energies $E_{\vec{k}}$ are needed, since from those
one first calculates $\chi_0$ by means of the equation \cite{griffin_Harris1976}

\begin{equation}
\chi_0(\omega; \vec{r}, \vec{r'})=2 \sum_{\vec{k},\vec{k'}}\frac{f_{\vec{k}}-f_{\vec{k'}}}{E_{\vec{k}}-E_{\vec{k'}}-\omega-i\eta}
\phi_{\vec{k}}(\vec{r})\phi_{\vec{k'}}^{*}(\vec{r})\phi_{\vec{k'}}(\vec{r'})\phi_{\vec{k}}^{*}(\vec{r'}),
\label{chi0r}
\end{equation} 
and then solves Eq. (\ref{chir}) self-consistently. In Eq. (\ref{chi0r}) $f_{\vec{k}}=\Theta(E_F-E_{\vec{k}})$ is the Fermi-Dirac
distribution function at zero temperature, $\eta$ is an infinitesimal and the factor of 2 comes from spin summation.

In the case of the semi-infinite jellium model, the wave functions and energies are of the form
$\phi_{\vec{k}}(\vec{r})=\phi_{k_z}(z) e^{i \vec{k}_{\parallel} \cdot \vec{x}} $ and 
$E_{\vec{k}}=\frac{1}{2}\vec{k}_{\parallel}^2 +E_{k_z}$ respectively. Then $\chi$ and $\chi_0$ only depend on the
difference of coordinates $\vec{x}-\vec{x'}$ and it is convenient to  Fourier-transform as

\begin{equation}
\chi_0(\omega; \vec{r}, \vec{r'})=\int \frac{d\vec{q}}{(2 \pi)^2} \; e^{i \vec{q} \cdot (\vec{x}-\vec{x}^{\prime})} \chi_0(q, \omega; z, z').
\end{equation}

 The self-consistent equation thus reads

\begin{equation}
\chi(q,\omega, z, z')=\chi_0(q, \omega, z, z')+\int dz_1 \int dz_2 \; \chi_0(q, \omega, z, z_1)R(q, \omega, z_1, z_2) \chi(q, \omega, z_2, z').
\label{chi}
\end{equation} 

There kernel $R$ appearing in the integral of the right-hand-side of Eq. (\ref{chi}) has been defined in the literature in two ways. 
The random phase approximation (RPA) 
exactly follows the derivation above and consequently $R$ is the Fourier-transformed Coulomb potential \cite{feibelmanPRB1975, eguiluz}

\begin{equation}
R_{RPA}(q, \omega, z_1, z_2)=\frac{2 \pi}{q} e^{-q|z_1-z_2|}.
\end{equation}

However, when using the eigenstates of a Lang-Kohn jellium surface \cite{langKohn} in Eq. (\ref{chi0r}), 
it has been argued that the exchange-correlation potential
used for calculating these eigenstates should also contribute to the self-consistent-field. Then, in the time-dependent local 
density approximation (TDLDA) one has \cite{liebschPRB}

\begin{equation}
R_{TDLDA}(q, \omega, z_1, z_2)=\frac{2 \pi}{q} e^{-q|z_1-z_2|}+\mu_{xc}^{\prime}(z_1) \delta(z_1-z_2),
\end{equation} 
where $\mu_{xc}^{\prime}$ is the derivative of the exchange-correlation potential. 

The screened susceptibility $\chi(q,\omega, z, z')$ obtained in the self-consistent-field approximation is a continuous
function of the spatial coordinates that 
smoothly connects the vacuum outside the surface with the volume of the metal. It contains the whole spectrum of metal excitations,
single-particle and collective modes, and the coupling among them.
 The use of $\chi_0$ instead of $\chi$ in Eq. (\ref{eq-taujelly}) recovers  
Eq. (\ref{rate-general}) when the one-electron wave functions $\phi_{\vec{k}}(\vec{r})$ are used in the
calculation of the matrix elements and  $V_{SC}$ is substituted by  the bare Coulomb interaction 
$1/|\vec{r}_2-\vec{r}_1|$. In this case collective modes are obviously not present in the formulation and therefore one should expect this approximation
to be good only in cases when the energies transferred in the Auger process are well above the energies of these modes. On the opposite limit, the
substitution of $V_{SC}$ by a static Thomas-Fermi screened Coulomb potential is only approximately valid when $\omega << \omega_p$, which is rarely 
the case for Auger neutralization of noble gas ions at metal surfaces.

\subsection{Jellium Model} 
\label{secJelliumModel}

Calculations of the Auger neutralization rate of slow He$^+$ scattered off aluminum and sodium surfaces using Eq. (\ref{eq-taujelly}) and the jellium model 
have been performed in Refs. \cite{lorenteMonrealSS1997,monrealLorentePRB1995,lorenteMonrealPRB1996}. These metals are prototypes of
free-electron-like metals and thus well described by a jellium. In these calculations the state $|\vec{k}_{or}\rangle$ neutralizing the ion is 
taken as a one-electron state of the jellium surface orthogonalized to the atomic state $|a\rangle$

\begin{equation}
|\vec{k}_{or}\rangle=|\vec{k}\rangle-\langle a|\vec{k}\rangle \;|a \rangle,
\label{opw}
\end{equation}
where the one-electron states $|\vec{k}\rangle$ are the same ones used in the calculation of $\chi$. These calculations are therefore fully consistent. One should note, however, that
the evaluation of Eq. (\ref{eq-taujelly}) is computationally very demanding since the self-consistent equation for $\chi(q, \omega, z, z')$ has to be solved for
many values of $q$ and $\omega$. Energy conservation limits the values of $\omega$ between $\omega_{min}=E_{bottom}-E_a$ and $\omega_{max}=E_F-E_a$,
$E_{bottom}$ being the bottom of the conduction band, but $q$ is unrestricted. 

As a first step, in \cite{monrealLorentePRB1995} the input metallic wave functions of the calculation
were those of a simple step-potential
barrier at the surface, which allowed one to perform a great deal of analytical work. Within this simple model, it was 
already shown that the excitation of surface plasmons
makes an important contribution to the Auger neutralization rate of He$^+$ on aluminum. 
In order to unravel
the contributions of collective modes and electron-hole pairs, it is convenient to look at the 
Auger rate per unit frequency and wave vector, $\gamma(q, \omega; z_a) $ which is defined from Eq. (\ref{eq-taujelly}) as,

\begin{equation}
 \frac {1}{\tau_{AN}}(z_a)= \int d\omega  \frac {1}{\tau_{AN}}(\omega,z_a) = \int d\omega \int dq \gamma(q,\omega; z_a)
\label{gammaq}.
\end{equation}

Fig.  \ref{figMonLor} shows $\gamma$ for $\omega=0.78 \omega_p$, for (a) $z_a=4 a.u.$ and (b) $z_a=-1 a.u.$, where distances are measured with respect to the jellium edge.
The strong peak at $q \simeq 0.2 k_F$ is due to the excitation of the surface plasmon and it gives almost the full contribution to the Auger rate at large distances between ion
and surface. In Fig. \ref{figMonLor}(b), the ion is inside the jellium, the
surface plasmon is still excited but the excitation of the continuum of electron-hole pairs is important as well.

In subsequent  works \cite{lorenteMonrealSS1997,lorenteMonrealPRB1996}, the Lang-Kohn jellium surface was considered and $\chi$ was calculated using TDLDA. 
One should note that in this description of the interacting electrons other collective modes appear in addition
to the surface plasmons which are called multipole surface plasmons. 
Fig.  \ref{fig_tau_w} shows $\frac{1}{\tau_{AN}}(\omega; z_a)$ (Eq. (\ref{gammaq})) for He$^+$ on Al  at a distance of $z_a= 5 $a.u. with respect to the jellium edge,
comparing the results when the interacting and non-interacting susceptibilities are used. 
It can be appreciated that, while the independent-particle calculation produces a smooth increase of $\frac{1}{\tau_{AN}} (\omega)$ with frequency, the interacting calculation
shows a sharp increase when the frequency reaches $\omega \simeq 0.65 \omega_p$. This is the region of frequencies were collective modes exist which
 practically give the full contribution to the rate.  It is  worth to note that in the region of small frequencies, 
$\omega < 0.65 \omega_p$ the non-interacting calculation overestimates the values of the rate because it does not take into account the efficient 
electronic screening of any perturbation in the quasi-static limit. 
As shown in Fig.  \ref{fig_tau_Al_jelly}, the overall effect in the total Auger neutralization rate is not so dramatic,  however, due to the compensation of 
the single particle and collective channels when integrating in $\omega$. Nevertheless one has to remember that such a compensation does not exist 
in other systems, such as H$^+$ on Al, where the energy transfer is always below $0.65 \omega_p$ and surface plasmons cannot be excited.

Fig. \ref{fig_tau_Na} shows the Auger rate for He$^+$ on Na as a function of the distance between the ion and the jellium edge, 
comparing the results of interacting and non-interacting 
calculations. In this case, the range of energy transfer is well above the energies of the surface modes and both calculations give very similar results for all energies and distances 
because the system responds as independent electron-hole pairs.
\newline

In spite of all this theoretical effort, the calculations of the Auger neutralization rate for He$^+$ on Al just presented were considered to be in poor agreement
with experiment. As shown in Fig.  \ref{fig_Hecht_SS}, the rates retrieved from the experiments in \cite{hechtSS1998}, using the classical concept
of image charge acceleration,  
were at least two orders of magnitude larger than the theoretical ones at the distances of about 
3-4 a.u. with respect to the image plane were ion neutralization was believed to take place. The reason for the discrepancy was sought in the neglect of the lowering
of the surface potential barrier by the presence of the strong ionic potential. These effects were first introduced by Propst \cite{propst1963} and later invoked in 
\cite{pennApellPRB1990} in connection to the spin-asymmetry of the emitted Auger electrons. Using a simple one-dimensional model, the effects were found important
in the calculations of the Auger rate presented in \cite{lorenteMonrealAlducinPRA1994}. A "piling up" of metallic charge on the ion was
considered in the calculations of neutralization rates of multiply charged ion inside metals \cite{diezmuinno}. 
A detailed investigation of the effect for the case of Auger neutralization of
He$^+$ at a jellium surface, was tackled by Cazalilla et al.
\cite{cazalillaPRB1998} who considered the 3D nature of the problem, including excited terms of the He atom. In this way they included not only the pure AN process of 
Fig.  2a but the processes of direct and indirect Auger deexcitation of Figs. 2c and 2d as well.
They found a huge effect at very large distances caused by the non-negligible probability of
tunneling of metal electrons to excited  terms of the He atom, but the effect decreases quickly with decreasing distance  and the difference found with calculations that do not
take these effects into account is not enough to bring theory in accord with experiment.  
Consequently none of the existing theories seemed
to be able to agree with the experiments and the reasons for the  discrepancy remained a mystery.
\newline

We have already shown that the excitation of surface plasmons is an important mechanism for Auger neutralization. Also, surface plasmons and electron-hole
pairs are coupled excitations so that surface plasmons decay into electron hole pairs.  Consequently, the excitation of surface plasmons in Auger neutralization can show up in the spectra
of the emitted electrons. First experimental evidence for plasmon excitation in the Auger neutralization of slow incident He$^+$ and Ne$^+$ ions on Al and Mg surfaces
was given by Baragiola and Dukes in 1996 \cite{baragiolaDukesPRL1996} and subsequently measured and analyzed in different systems by several authors  
\cite{niemannPRL1998,riccardiPRL2000,vanSomerenPRA2000,riccardiNIMB2001,baroneSS2001}.
On the theoretical side, calculations of the electron emission spectra by  Monreal  \cite{monrealSS1997} using the theory expounded in this subsection, showed 
that theory and experiment would agree if one admits that the incident ions are neutralized exciting plasmons near or even inside the jellium edge. Then, this was a first evidence that ion neutralization may occur near the metal surface. 
For a discussion of the surface plasmon-assisted neutralization mechanism and its connection to electron emission, we refer the reader to Ref. \cite{baragiolaMonrealBOOK} and references therein.

Other interesting many-body effects appear in the neutralization process of slow ions. An example is the so-called Fermi edge singularity, a final-state 
effect that manifests itself in the tails of the electron emission distributions, which decrease exponentially with the energy of the emitted electron
\cite{shakeupPRA2005,shakeupSS2007}. However, these effects have very little effect on the AN rate and will not be discussed here. 

\subsection{Corrugated Surface}
\label{secCorrugatedSurface}

Even though the jellium model for the metal surface provides insight into the impact that surface screening has in
the Auger neutralization rate, it is obvious that it cannot  describe real corrugated surfaces.
The need to go beyond the jellium model actually came from  experimental investigations, since precise measurements of
the very small surviving scattered ion fractions of He$^+$ in grazing collisions with Ag(111) \cite{wethekamPRL2003, monrealJPhys2003}
and Ag(110) surfaces revealed strong differences in the Auger neutralization probability of He$^+$ at different faces of the same metal
\cite{monrealJPhys2003,bandurinPRL2004}.
The only way in which a jellium can model
different crystallographic faces of the same metal is to withdraw the jellium edge with respect to the first atomic layer
 by half an interplanar spacing,
 as required by charge neutrality 
\cite{langKohn}, and in this way good agreement between theory and experiment was achieved in \cite{monrealJPhys2003,bandurinPRL2004}. 
The jellium model is, however, not suitable for describing the azimuthal orientation characteristics of ion neutralization seen in the experiments.

Although the influence of the crystal structure on the ion fractions of slow ions scattered from metals have been known for a long time
\cite{vanderweg1969, luitjensSS1979, taglauerPRL1980, winterZimnyZP1983,naermannSS1989},
rather simple and general models \cite{hentschke1986,snowdon1986,monrealGoldbergSS1989} cannot explain the details of the experimental results.
Therefore, a more realistic description of the metal surface is needed to account for observed crystal effects in the Auger rate.
Moreover a theory of the corrugated Auger rate contains information on the atomic structure and composition of the surface
and is then useful for investigating the "absence of matrix effects" 
in the Auger neutralization regime of LEIS.
\newline

To introduce corrugation in the theory of the Auger neutralization rate in a simple way,
the basic idea is to start from Eq. (\ref{if}) defining the initial and final states and consider that
the electron of the solid  neutralizing the ion has now to be described by a Bloch state 
whereas the rest of the electrons in the system are still described in the jellium model.
In this way we take into account the "local"  environment seen by the ion when it is neutralized even though we will neglect
band structure effects in the calculation of the screened susceptibility.  Hence the formula for the Auger transition rate
Eq. (\ref{eq-taujelly}) applies with the potential of Eq. (\ref{potential}) expressed in terms of the Bloch states. 

The Bloch states of wave function
$\phi_{\vec{k},n}(\vec{r})$ and energy $E_{\vec{k},n}$, where $n$ is the band index and
$\vec{k}$ is the wave vector restricted to the first Brillouin zone, 
are written in a linear combination of atomic orbitals (LCAO) basis 

\begin{equation}
\phi_{\vec{k},n}(\vec{r})= \frac{1}{\sqrt{N}} \sum_{\alpha} C_{\alpha}^{(n)}(\vec{k}) \sum_{\vec{R}}
e^{\imath \vec{k}\cdot\vec{R}} \varphi_{\alpha}(\vec{r}-\vec{R}), 
\label{bloch}
\end{equation}
where $\varphi_{\alpha}(\vec{r}-\vec{R})$ are atomic orbitals of symmetry $\alpha$
centered at the lattice points $\vec{R}$, N being the number of lattice points, and  $C_{\alpha}^{(n)}(\vec{k})$
are the coefficients of the linear combination. Then, the potential of Eq. (\ref{potential}) takes the form

\begin{equation}
\Phi( \vec{k},n; \vec{q},z)= \frac{1}{\sqrt{N}} \sum_{\alpha} C_{\alpha}^{(n)}(\vec{k}) \sum_{\vec{R}}
e^{\imath \vec{k}\cdot\vec{R}} \Phi_{\alpha,\vec{R}}(\vec{q},z),
\label{eq215}
\end{equation}
where we have defined

\begin{equation}
\Phi_{\alpha,\vec{R}}(\vec{q},z)\equiv \frac{2\pi}{q}
\langle a|e^{\imath \vec{q}\cdot \vec{x}_{2}}
e^{-q |z-z_{2}|} |\varphi_{\alpha}(\vec{r_{2}}-\vec{R})\rangle.
\label{potentialLCAO}
\end{equation}
The Auger transition rate, Eq. (\ref{eq-taujelly}), thus reads

\begin{eqnarray}
\frac {1}{\tau_{AN}}(\vec{R}_{a}) = & 2 &\sum_{\vec{k},n} \int_{0}^{\infty} d\omega \int \frac{d \vec{q}}
{(2\pi)^2}\int_{-\infty}^{\infty} dz \int_{-\infty}^{\infty} dz'    Im(-\chi(q,\omega;z,z'))    \nonumber \\
 & \times&  \Phi(\vec{k},n; \vec{q},z)
\Phi^{\ast}(\vec{k},n; \vec{q},z') \delta (\omega+E_{a}-E_{\vec{k},n}). 
\label{eq216}
\end{eqnarray}

If we now use the identities

\begin{equation}
\sum_{\vec{k},n} \int_{0}^{\infty} d\omega \delta(\omega+E_{a}-E_{\vec{k},n}) = \sum_{\vec{k},n}
\delta(\epsilon-E_{\vec{k},n}) \int_{0}^{\infty} d\omega \delta(\omega+E_{a}-\epsilon), 
\label{eq217}
\end{equation}
and

\begin{eqnarray}
& &\sum_{\vec{k},n} \delta(\epsilon-\epsilon_{\vec{k},n}) \frac{1}{N} \sum_{\alpha,\vec{R}} C_{\alpha}^{(n)}(\vec{k}) e^{\imath \vec{k}\cdot\vec{R}} \sum_{\alpha', \vec{R'}} C_{\alpha'}^{(n)\ast}(\vec{k})
e^{-\imath \vec{k}\cdot\vec{R'}}     \nonumber   \\
& &\equiv \sum_{\alpha,\alpha'} \sum_{\vec{R},\vec{R'}} \frac{1}{N} \sum_{\vec{k},n}
\delta(\epsilon-\epsilon_{\vec{k},n})C_{\alpha}^{(n)}(\vec{k})C_{\alpha'}^{(n)\ast}(\vec{k})e^{\imath
\vec{k}(\vec{R}-\vec{R'})}  \nonumber \\
& &\equiv \sum_{\alpha,\vec{R}} \sum_{\alpha',\vec{R'}}
\rho_{\alpha\vec{R},\alpha'\vec{R'}}(\epsilon),
\label{eq218}
\end{eqnarray}
where $\rho_{\alpha\vec{R},\alpha'\vec{R'}}(\epsilon)$ are the densities of states (DOS) projected in the local basis $\{\alpha,\vec{R}\}$,  
Eq. (\ref{eq216}) 
can be worked out finally yielding

\begin{eqnarray}
 \frac {1} {\tau_{AN}}(\vec{R_{a}}) = &2 &\sum_{\alpha,\vec{R}} \sum_{\alpha',\vec{R'}} \int_{-\infty}^{E_{F}} d\epsilon
\int_{0}^{\infty} d\omega \int \frac{d \vec {q}}{(2\pi)^2}\int_{-\infty}^{\infty} dz \int_{-\infty}^{\infty} dz'    Im(-\chi(q,\omega;z,z'))      \nonumber  \\  
& \times & \Phi_{\alpha,\vec{R}}(\vec{q},z)\
\Phi_{\alpha',\vec{R'}}^{\ast}(\vec{q},z') \ \rho_{\alpha\vec{R},\alpha'\vec{R'}}(\epsilon) \delta
(\omega+E_{a}(\vec{R_{a}})-\epsilon).
\label{eq219}
\end{eqnarray}
In this equation the upper limit of the energy $\epsilon$ is the Fermi energy because only electrons in occupied states can neutralize the ion. 
Also, energy conservation prevents electrons in bands below $E_a$ (such as core levels) to contribute directly to the Auger process.
Notice that in this formulation the rate depends on the position $\vec{R}_a$ of the ion with respect to the crystal unit cell and not only on
its coordinate perpendicular to the surface $z_a$ like in the jellium model.

The wave functions appearing in Eq. (\ref{potentialLCAO}) should be orthonormal.
To construct an orthonormal basis from an initial set of atomic orbitals
centered at different sites, $\psi_{\nu}$, we follow the  L\"owdin method \cite{lowdin}
in which the orthonormal basis is obtained as

\begin{equation}
\varphi_{\mu}= \sum_{\nu} (S^{-\frac{1}{2}})_{\mu\nu} \psi_{\nu},
\label{lowdin}
\end{equation}
 where $S$ is the overlap matrix having matrix elements $S_{\alpha\beta}=\langle\psi_{\alpha}|\psi_{\beta}\rangle$.

In all the calculations, we start with a set $\psi_{\nu}$ of Hartree-Fock atomic orbitals for He and the metal atoms 
expressed in  the gaussian basis of Ref. \cite{huzinaga}.  We include all atoms within a certain cut-off radius centered at the projectile position. 
This cut-off radius is chosen large enough to warrant that all important contributions to the AN-rate are considered. 
Densities of states are calculated ab initio using the FIREBALL code of Ref.\cite{fireball}.

The other ingredient in the theory of the Auger rate is the dielectric susceptibility
 $\chi(\vec{q},\omega;z,z') $.  A consistent treatment of this function in terms of Bloch wave functions
is not possible at present, mainly because it requires inclusion of a
large number of reciprocal lattice vectors in the surface plane. Moreover, as we said above, 
we need to evaluate $\chi$ numerically for many values of $\omega$ and $q$ 
and therefore the calculation has to be simplified. Hence, $\chi(\vec{q},\omega;z,z') $ is obtained for a jellium within 
the self-consistent-field approximation, with the jellium edge canonically placed at $\frac{1}{2}\,d$
above the first atomic layer in all the cases , $d$ being the interplanar distance, to ensure charge neutrality \cite{langKohn}.
Therefore, in this procedure corrugation enters in the AN rate through the basis set of atomic orbitals centered at different lattice points but 
we do not include corrugation effects in the surface susceptibility, apart from the position of the jellium edge.

\subsubsection{Results: Aluminum Surfaces}  
\label{secAlRates}

The present formulation for the corrugated Auger neutralization rate is first compared with the jellium model for He$^+$ on Al  \cite{valdesGoldbergPRB2005}. 
The calculation includes the 1s orbital of He and all orbitals of Al. However, we have checked that the results do not change 
if we only include the valence 3s and 3p orbitals. This is because the overlap of the core-levels of Al with He is small and so is their weight in the
orthogonal orbitals defined by Eq. (\ref{lowdin}). Figs. \ref{fig_tau_Al111_tcl_jell} and \ref{fig_tau_Al110_tc_jell}  
 show the results for Al(111) and Al(110), respectively. In each figure the rates for
different lateral positions of He are compared to the jellium results. One finds two types of behavior depending on the perpendicular
distance. At distances much larger than the jellium edge the corrugated rates are orders of magnitude larger than the jellium ones. This is due to the
different procedure for constructing orthogonal states:  L\" owdin's symmetrical orthogonalization  versus metal states orthogonalized to the ion state. 
These huge differences in the rates at large distances are, however, of little consequence, since their absolute values  are still very small in order to produce
an efficient neutralization of He$^+$. Efficient neutralization of slow He$^ +$ ions occurs near the jellium edge, where, interestingly,
the jellium and the corrugated rates cross.  One should remember that any measurable magnitude requires integration
along the trajectory. Consequently, the impact of corrugation in the neutralization probability of He$^ +$ at Al surfaces is not very dramatic and do not change
the conclusions reached  in the previous subsection.
Pointing to the same direction, notice how the corrugated rates are not too sensitive to lateral position. This is due to the fact that 	extit{ many} atoms of Al contribute 
in the first and also in the second atomic layers, due to
the large spatial extension of the atomic orbitals, as illustrated in Fig.  \ref{fig_tau_Al110t_vecs} where the contributions of different
neighbors to the total rate for He on Al(110) is shown. However, corrugation has to be taken into account for a detailed quantitative analysis of experiments, as we will see in subsection \ref{secComparisonAlxxx}. 

\subsubsection{Results: Noble Metal Surfaces} \label{secNobleRates}

The three noble metals Ag, Cu and Au have sp-bands extending about 10 eV below the Fermi level
and much narrower d-bands starting around 4 eV below the Fermi level in Ag and around 2 eV below the Fermi level
in Cu and Au and having a width of ca. 4 eV.  These localized  $d$ electrons can therefore contribute to Auger neutralization of noble gas 
ions and provide strongly corrugated Auger rates, as we will show in this subsection. 

In the calculation of the AN rates of He$^+$ on noble metals, we only use the 4s, 3d, 3p and 3s- orbitals of Cu, the 5s, 4d, 4p and 4s orbitals of Ag, and 
the 6s, 5d, 5p and 5s orbitals of Au.  Other orbitals are neglected because their overlap with He is too small to give any appreciable
contribution to Eq. (\ref{lowdin}), as was the case of the core orbitals of Al.
The dielectric susceptibility  $\chi $  will be evaluated by using the jellium model, with 
 suitable modifications to take into account that either s or d electrons can be excited in the Auger process.
From optical data \cite{ehrenreichPhilippPR62}, we know the number of electrons that contribute to the
optical properties of noble metals at each frequency $\omega$.
We can thus define an effective electronic density $n_{eff}(\omega)$  and an
effective value of the one-electron radius $r_s$ as  $r_s(\omega)=(\frac{3}{4\pi n_{eff}(\omega)})^{1/3}$.
 Then for each $\omega$, $\chi(\vec{q},\omega;z,z') $ is evaluated within the self-consistent-field approximation for a jellium surface described by that
$r_s(\omega)$. 

Fig. \ref{fig_rates_CuAgAu} shows the Auger rates for He on Cu(111), Ag(111) and Au(111) surfaces on-top position \cite{goeblValdesPRB2011,goeblNIMB2011}. 
The qualitative behavior of the three metals is very similar. The rate
presents a maximum at ca. 1 a.u. that is due to Auger neutralization of He$^+$ by the $d$ electrons of the atom on-top. The extended $sp$ electrons mainly contribute 
at distances larger than ca. 4 a.u. where they make the full rate. Moreover, the AN rates of the noble metals present a huge corrugation, which is absent in Al. 
This is illustrated 
in Fig. \ref{fig_rates_AuAl_tcl} which compares
AN rates of Au(111) and Al(111) respectively, assuming the He atom to be at different lateral positions with respect to the surface unit cell 
\cite{monrealGoeblNIMB2013} .
 We can appreciate that the rate of Al shows a weak dependence on lateral position compared to Au. 
 The reason is, again,
that the electrons contributing to the rate of Al are the extended 3$sp$ electrons while the
localized $d$ electrons are the important ones in the Au case at short distances, producing in addition a rate larger by a factor of 3-4 than that of Al.
In contrast, the long distance values of the rate are determined by the extended $sp$ electrons in all cases. 
 It is also interesting to note that in the case of Au, the on-top rate is the smallest one
close enough to the surface due to the strong decrease in overlap between the 1s electron
of He and the 5d electrons of Au.

The same physics seen at the (111) surfaces occurs at all surfaces of the noble metals. We illustrate this fact with a few examples. 
 The corrugated rate for He on Ag(110) is plotted in Fig. \ref{fig_rate_Ag110_z2_az35} as a function of the lateral distance $d_{\parallel}$ 
 along the [111] direction and 
 a fixed perpendicular distance of $z_a= 2$ a.u. with respect to the first atomic layer. While the contribution of s-electrons is rather flat, the total rate reflects the 
 spatial localization of the d-electrons. It is remarkable that the rate can change by a factor of 2 along the  azimuth.
 The corrugation of the Auger neutralization rate of  Cu(100) is illustrated  in Fig. \ref{fig_rate_Cu100_contour} as contour plots \cite{goeblMonrealNIMB2013}. 
 Notice that corrugation exist above the first atomic layer (lower panel) as well as  in the middle of the first and the second layers (upper panel).

\section{Energy Level Variation}
\label{secLevelshift}

The question of how well any theory of AN  is able to quantitatively reproduce the experiments is closely connected to the problem
of how the energy levels of atoms change in the proximity of a solid surface.
This has been at the origin of a historical controversy only solved recently, when 
consideration of energy level variation (level shift) of ions approaching solid surfaces has been the key point for advancing 
 the understanding of neutralization of ions at solid surfaces.

As mentioned in Sec. \ref{secExperimentalMethods}, measurements of the high-energy tails of the electron distributions
\cite{hagstrumPR1954} and measurements of energy gains of ions prior to neutralization \cite{hechtSS1998} showed changes in
the energy level of the incident ions of about 2 eV. From this value and making use of concepts of the classical image potential,
He$^+$ was assumed to be neutralized at distances of ca .4 a.u. from the image plane which required AN rates orders of
magnitude larger than theoretically predicted. 
However,  Merino \textit{et al.} \cite{merinoNIMB1997}, More \textit{et al.} \cite{morePRB1998}, and van Someren \textit{et al.} 
\cite{vanSomeren1PRA2000} pointed out that the He-1s  level
shift might be substantially reduced compared to the classical behavior for distances of some atomic units in front of the surface,
as a consequence  of a breaking down of the classical image-potential concept  at close distances. Actually, theoretical calculations of the He-1s level energy
shift showed reduced values or even negative shifts close to the surface as a result of chemical interactions with the surface
\cite{wangGarciaPRA2001,merinoNIMB1997, morePRB1998, cruzPRA2008}. Similar deviations from the classical behavior were also calculated for the
1s state of H \cite{merinoPRB1996, deutcherPRA1997},  for excited states of He in front of an Al surface \cite{makhmetovEL1994}, and predicted for
other systems \cite{carterJCP2004}.  A similar downward shift was  found for the ground state of Ar in front of a KCl(001) surface \cite{kimuraPRA2004}.  

Fig.\ref{fig_nivel_HeAl} shows the diabatic and adiabatic levels for He approaching Al(111) on a top position. (We refer the reader to Refs.\cite{merinoPRB1996,merinoNIMB1997,
morePRB1998} for details on the theoretical approach to this problem).
This is the relevant configuration to look at since scattering of He
is produced in the collision with a target atom.  
The diabatic level is taken from \cite{wangGarciaPRA2001} and the different adiabatic levels are obtained when the
hopping interactions between the 1s level of He and the different orbitals of Al are connected. 
At distances larger that ca. 7 a.u. from the first atomic layer we find the classical image potential behavior.  
As He approaches the surface it feels the
 hopping interaction with the 3sp electrons of Al which pushes the level down in energy. At  closer distances, 
 the interaction with the core 2sp levels of Al sets in. This is  strongly repulsive and promotes the level very quickly.
One can reveal in the figure  an upward level shift of ca. 2 eV  obtained at  a distance of 7 a.u.,
 the region of the image-shift, but also at distances of ca. 4 a.u. and 1.5 a.u..
Since the theoretical calculations of the Auger neutralization rate presented in section \ref{secJelliumModel}  predicted most probable distances 
of neutralization for slow ions of ca. 3 a.u. from the first atomic layer 
(1 a.u. from the jellium edge), it was suggested in \cite{morePRB1998} that experiments should be re-interpreted in terms of energy level variations beyond the classical image potential concept,  which is only valid in the limit of large distances, 
taking into account the chemical interactions that appear at closer distances from the surface.   
Comparison between theory and experiment to be presented in the following section will show that this is indeed the case.

\section{Comparison of Theory and Experiment}  
\label{secComparison}

Clear-cut experimental evidence that He$^+$ ions are Auger neutralized close to the surface in agreement with the theoretical
predictions was provided by the finding that small fractions of He$^+$ ions scattered from a metal surface under grazing conditions survive the
whole scattering event in their initial charge state \cite{wethekamPRL2003,monrealJPhys2003,bandurinPRL2004,valdesPRL2006}. The surviving ion fractions
would have been negligibly small (of the order of $10^{-8}$) for the Auger neutralization rates retrieved from experiments using the concept of classical image charges
\cite{hechtSS1998}.  
 The isotope effect for the surviving ion fractions demonstrates the existence of a
well-defined neutralization rate such that transient populations of
excited states during the neutralization process can be
neglected \cite{wethekamPRL2006} in accord with the results of Ref. \cite{morePRB1998}.
The picture was completed by measurements of shifts of the high-energy tails of Auger electron distributions \cite{lancasterPRB2003} and 
shifts of angular distributions for incident neutrals and ions for different energies (different distances of neutralization)
\cite{wethekamSSL2005,wethekamNIMB2007}, an experiment proposed by More \textit{et al.} \cite{morePRB1998}, that directly measured reduced (and even
negative)  energy shifts of the He-1s level close to the surface. 

Of particular interest are works that compare the neutralization of He$^+$ ions at different faces of the same metal,
 where pronounced variations of the surviving ion fractions are observed.  These variations are found under grazing scattering conditions 
 \cite{bandurinPRL2004,valdesPRB2007,wethekamValdesPRB2008} and, surprisingly, also in the
LEIS regime. The studies in \cite{primetzhoferPRL2008} revealed non-negligible differences in the ion fractions of
He$^+$ after scattering from Cu (110), (100) and polycrystalline surfaces,
demonstrating that the so-called ``matrix effects'' have indeed  to be
considered in the analysis of low-energy ion scattering data for studies on the composition of solid surfaces. 
In this section we will compare experimental measurements of ion fractions and energy gains with  theoretical calculations for these magnitudes using the Auger neutralization 
rates presented in section \ref{secMultielectron}.  
The comparison is made in subsections for grazing scattering and LEIS regime.
A unified picture emerges from this analysis: the calculated Auger neutralization rates are sufficiently accurate and, 
at  the distances to the surface where neutralization occurs, the distance-dependent position of the He 1s level plays a dominant role for obtaining good agreement with the measured values of the ion fractions.
 
\subsection{ Grazing Scattering} 
\label{secComparisonGrazing}

In this subsection we discuss Auger neutralization of He$^+$ on Ag(111) and Ag(110) surfaces exhibiting ion fractions that depend on the azimuthal direction
of the projectiles with respect to the surface unit cell.  Also we will discuss the case of 
 Al surfaces where, surprisingly, one also needs to invoke corrugation in the Auger rates. 

\subsubsection{Silver Surfaces.}
\label{secComparisonAgxxx}

 We first present a comparative study of Auger neutralization of He$^+$  ions on Ag(111) and Ag(110) surfaces at grazing incidence \cite{bandurinPRL2004}. 
 In the experiment, the angle of  incidence with respect to the surface is 3.5$^\circ$ and ions scattered
along a random direction are collected in the specular direction.  The number of surviving ions is always small and 
the remarkable and important feature of the experimental results shown if Fig. \ref{fig_HeAg_PRL} 
 is  an order of magnitude difference in the number of ions surviving 
Auger neutralization on Ag(110) as compared to Ag(111).  At first glance, this is surprising from the point of view of Auger neutralization, since the electronic
structure of these surfaces would not appear to present major differences. In order to comprehend this difference, 
we performed molecular dynamics simulation of scattered ion trajectories and then calculated the neutralization probability of scattered ions using theoretically
derived Auger neutralization rates.  Trajectories are calculated using the code KALYPSO \cite{Kalypso} in which Ziegler-Biersack-Littmark (ZBL) potentials 
\cite{ZBL} are used. As a first step, the Auger neutralization rates for He/Ag were calculated within the jellium model, with the jellium edge placed above the first atomic layer
by half the interplanar distance.  The  He-1s level was assumed to to be shifted up in energy by  a constant value of 2 eV. 
Some of the simulated trajectories are plotted in Fig. \ref{fig_traj_HeAg_PRL} where one can appreciate that, 
at the same incident energy, the trajectories followed by the ions at the Ag(111) or (110) surfaces do not present large differences and  
ions are deflected at approximately the same distance above the outermost atomic layer.
Therefore, the strong difference in ion survival is related to changes in atomic interlayer spacing, which leads to a different spill out of the electron density beyond the
surface and, hence, strong differences in the Auger neutralization rate. 
The good agreement between theory and experimental data shown in Fig. \ref{fig_HeAg_PRL} is achieved without any adjustable parameter.
\newline

The surviving ion fractions of He$^+$  on Ag(110) also exhibit a strong dependence on the azimuthal orientation of the surface 
with respect to the initial beam direction which cannot be accounted for in the jellium model
\cite{valdesPRL2006,valdesPRB2007} .
This is  shown in Figs. \ref{fig_ionfractionAg110_az1k} and \ref{fig_ionfractionAg110_az3k} for ion incident energies of 1 to 4 keV and the same angles of incidence and 
scattering as in Fig. \ref{fig_HeAg_PRL}.
In order to account for this dependence, the calculation of the corrugated Auger rates presented in subsection \ref{secCorrugatedSurface} were applied to this system 
in conjunction with  molecular dynamic simulations from KALYPSO, including in the code lattice vibrations at room temperature. Out of all the simulated trajectories, we select those that reach the detector. Then, for each trajectory
we calculate the ion survival probability along that trajectory as 

\begin{equation}
P_{i}^{+}=exp \{-\int_{t_{i}}^{t_{f}} \frac{dt}{\tau_{AN}} [\vec {R}_{a}(t)] \}
\label{eq-Pi}
\end{equation}
where $t_i$ and $t_f$ are the initial and final times in the simulation. Hence, the fraction of surviving ions to be compared to the experiment is given by

\begin{equation}
P^{+}=\frac{\sum_{i=1}^{N} P_{i}^{+}}{N}
\label{ionfraction}
\end{equation}
with $N$ being the number of trajectories that reach the detector. In the case of grazing scattering, we find that the inclusion 
of lattice vibrations is very important for obtaining $N$ and then the
theoretical ion fraction. 

 The deep minima in ion fraction at 0$^\circ$ and 90$^\circ$ are due to the fact that many particles penetrate the first
atomic layer and get completely neutralized. Out of these  symmetry directions, the 
scattered trajectories stay above the surface (see Fig. \ref{fig_traj_HeAg_PRL}). The apexes of these trajectories present a nearly Gaussian distribution, the
maximum of which we identify with the most probable distance of closest approach. For a random direction, the values of this distance are in the range of 0.7 -  2.3 a.u. 
for the incident energies of 1 - 4 keV used in the experiment. Hence, the projectiles reach distances where the contribution of 4$d$ electrons
to the total Auger neutralization rate is important at all incident energies. This is clearly demonstrated in Fig. \ref{fig_ionfractionAg110_az1k} where the contribution of the 
5$s$ electrons to the ion fraction 
is shown separately:  one can appreciate that they underestimate the ion fraction by one order of magnitude. In contrast, the full calculation including s and d electrons
reproduce quantitatively the experiment at all incident energies. It is worth to note  the small variations of  the ion fractions with azimuth 
out of the symmetry directions which partially reflects the weak dependence 
of the Auger neutralization rate on lateral position in the unit cell around a surface atom (see the contour plots in Fig. \ref{fig_rate_Cu100_contour}). 

 It is also interesting to compare the theoretical ion fractions obtained 
within the jellium model shown in Fig. \ref{fig_HeAg_PRL}, to the ones we get using the corrugated rates, for random directions. This is presented in Fig. \ref{fig_ionfraction_LCAOvsJelly}
together with the experimental results, demonstrating the accuracy of the LCAO approximation to the calculation of the Auger neutralization rate. 

The quantitative agreement between theory and experiment found for  scattering of He$^+$ on Ag(110) is not fortuitous. A similar analysis of the azimuthal dependence
of the fraction of ions surviving scattering off Ag(111) reveals an equally good agreement, as shown in Fig. \ref{fig_ionfractionAg111_az}. 
\newline

Another important output of the calculations presented so far is the necessity of a correct description of the surface dielectric screening. Even though in the present
formulation screening by the $d$ electrons is treated in an effective way, the self-consistent-field evaluation of $\chi$ ensuring that the jellium edge is placed at half the interplanar
distance as required by charge neutrality is essential to yield  quantitative agreement  between theory and experiment. Fig. \ref{fig_tauAg110_displacedjelly}
shows the Auger neutralization rate of He$^+$ on-top and on-center positions with respect to the Ag(110) unit cell, calculated when placing the jellium edge at half the
interplanar distance of the (110) planes (as it should) and at half  the interplanar distance of the (111) planes. 
The differences of 30-35 $\%$ found for the on-top position are enough to destroy
 the quantitative agreement with the experiment shown in Figs. \ref{fig_ionfractionAg110_az1k} and  \ref{fig_ionfractionAg110_az3k}.
 
 \subsubsection{Aluminum Surfaces}
 \label{secComparisonAlxxx}

In this subsection we present results on ion fractions and energy level shifts for He$^+$ scattered at Al(111), Al(100) and Al(110) surfaces at 
random directions and with very low normal
incident energies \cite{wethekamValdesPRB2008}.  The experimental technique was briefly described in subsection  \ref{secExperimentalMethods} and measures polar
angular distributions of scattered projectiles from which the energy gain (or energy level shift) can be obtained via Eq. (\ref{Egain}).
 Ion fractions are also measured in the same experimental set-up.  Details can be found in \cite{wethekamValdesPRB2008} and references therein.
 \newline
  
 Theoretical calculations of ion fractions and polar angular distributions are performed using 3D molecular dynamics simulations of trajectories
 followed by neutral atoms He$^0$ and ions He$^+$.  Correlated  thermal
displacements are included in the simulations within the Debye model at T=300 K, as in Ref. \cite{wethekamNIMB2007}.
 For the calculation of trajectories for He$^0$ atoms we made use of
a Moliere potential with a screening length modified by O'Connor and Biersack (OCB) \cite{oconnor}. This choice is different from Refs.
\cite{wethekamSSL2005, wethekamNIMB2007}, where an interaction potential derived from data for rainbow scattering under axial 
surface channeling \cite{schullerPRA} was used. Although being very sensitive to the interaction potential, the rainbow data does not
yield enough information for an unequivocal derivation of the potential and, without further input, very different potentials can
be constructed from the data.  
The potential for ions $V_{+}(\vec{r})$ is constructed from the potential for neutrals  $V_{0}(\vec{r})$ and the theoretical energy level  shift of He. 
We should remember that the  level shift is defined as the change in the ionization potential of He due to its interaction with the metal:

\begin{equation}
\Delta E_{1s}(\vec {r}) = E_{1s}(\vec{r}) - E_{1s}(\infty) = V_{0}(\vec{ r}) - V_{+}(\vec {r}) . 
\label{eqLevelPot} 
\end{equation}

To construct $E_{1s}(\vec {r})$,  the energy shift of Fig. \ref{fig_nivel_HeAl} calculated for the on-top position is evaluated as a
function of distance to the closest target atom (with slight modifications to ensure continuity). 

The curves in Fig. \ref{fig_levelshift_Alxxx}(a) depict the level shift averaged parallel to Al(111), Al(100), and
Al(110). The energy shifts coincide for large distances, which is the correct behavior due to the small
differences of the image plane positions, and show the expected relative behavior close to the surface. This averaged value is the one that has to be related to
the energy gain measured in the present experiments. At very grazing angles of incidence and random azimuthal directions, ions spend a
long time traveling parallel to the surface and therefore probe many different lateral positions within the unit cell. 
With respect to the AN rates, we use our calculations for the corrugated rates of He at Al surfaces presented in subsection \ref{secAlRates},
considering the on-top  and the hollow (on-center) rates, in order to enlighten the effects of corrugation.  The averaged potentials 
$V_{0}(\vec{ r})$ and  $ V_{+}(\vec {r})$ are displayed in Fig. \ref{fig_levelshift_Alxxx}(b).

Fig. \ref{fig_ionfractionAlxxx} shows the measured surviving ion fractions for He$^+$ ions scattered
from Al(111), Al(100)  and Al(110),  compared to simulations based on the on-top AN
rates as well as AN rates for the hollow position.
As in the case of Ag surfaces, the order of magnitude and relative dependence of the ion fractions with crystal face is reproduced by the simulations.
 However, the absolute magnitude of the ion fractions depend on lateral position within the unit cell: the on-top rates give ion fractions much
larger than in the experiment, while use of the hollow rates yields much better agreement, especially in the case
of the open Al(110) surface. This is to be expected because at random directions an ion is most of the time above hollow site positions in the
unit cell.

The results of Fig. \ref{fig_ionfractionAlxxx} clearly show the extreme sensitivity of the ion fractions to the values of the AN rate.
Due to the exponential dependence of the ion fractions with AN rate, an increase of the
rates by 30\% reduces the ion fractions by factors of about 5 to 10. Therefore, the fact that the agreement between theory and
experiment is systematically improved by using the hollow position rates, shows the importance of a proper description of the face dependence of the AN rate
beyond the simple "jellium edge" concept, the same conclusion reached for Ag in Fig. \ref{fig_ionfraction_LCAOvsJelly}.

Fig. \ref{fig_polardistributions} compares experimental and simulated polar angular distributions.
The distributions in panel (a) for incident ions, which correspond to an incident normal energy of about 0.8 eV, are clearly shifted
towards larger outgoing angles compared to the distributions for incident atoms. Via Eq. (\ref{Egain}) this is related to a
positive normal energy gain due to neutralization at large distances from the surface where the energy shift is positive.
Panel (b) shows data for a larger incident normal energy of about 17 eV where shifts of the distributions for incident ions and neutrals
are absent. The normal energy gain is close to zero which means neutralization at about 3 a.u. in front of
the surface where the energy shift changes sign (see Fig. \ref{fig_nivel_HeAl}). The overall agreement of measured and simulated
distributions including the tails is good, which shows that the simulations catch the main ingredients of the physics involved. In
addition it can be concluded that the density of surface defects is negligibly small \cite{pfandzelterPRB}. 

Measured and simulated  normal energy gains as function of incident normal energy for scattering of
He$^+$ ions from Al(111), Al(100) and Al(110) surface for the on-top rates are compared in Fig. \ref{fig_energygain_Alxxx}(a) and
Fig. \ref{fig_energygain_Alxxx}(b) shows the comparison of the experiments and simulations based on the hollow site rates.
The differences between the normal energy gain for different faces can be understood from an energy level shift with a weaker face
dependence than the AN rate.  As the AN rate for Al(111) is larger than for Al(110) (see subsection \ref{secAlRates}), the incident normal energy that results in the
same distance of neutralization is larger for Al(111) than for Al(110). Therefore, for Al(111), the downward shift of the level is
seen at larger incident normal energies. Measured and simulated normal energy gains agree on a quantitative level which shows that
the calculated level shifts are accurate. Only the simulations for Al(110) for large normal energies, 
where also the relative effects get very small, show small deviations from the experimental data.
This is a result of our implementation of the energy level shift for the three faces, based on the on-top level shift for 
Al(111) evaluated as function to the closest target atom. This ansatz will be least valid for Al(110), as it represents the most open surface with the smallest AN
rate and therefore closest distance of neutralization for incident ions. However, in the region before the zero crossing for the energy gain, where the
effects of the downward shift are most pronounced, the agreement is very good. Also, the fact that the measured energy gains are smaller
than -1 eV is consistent with our interpretation of the energy gain as average of level shifts parallel to the surface, plotted in Fig. \ref{fig_levelshift_Alxxx}(a). 
\newline

The good overall agreement of experimental data and theoretical predictions for Ag and Al shown in this subsection, obtained without
adjustable parameters,  
 demonstrates that a detailed microscopic understanding of He-metal surface interactions has finally been achieved.
It is rooted on two basic physical concepts:  i) modification of the atomic energy levels beyond the simple image-potential concept due to the close interaction with the metal surface and ii)
the many-body character of the Coulomb 
electron-electron interactions causing the Auger neutralization process, in and at metal surfaces.
We can unequivocally conclude that neutralization of slow He$^+$ ions at metal surfaces takes place at distances between 2 a.u. and 4 a.u. from the first atomic layer, for
perpendicular energies smaller than 20 eV. At these close distances to the image plane, the classical image-potential concept breaks down and
the close interactions between He and the metal atoms have to be considered in the analysis of experiments.

\subsection{ LEIS Regime.} 
\label{secComparisonLEIS}

To proceed one step further in the understanding of noble gas-metal surface interactions, 
we investigate in this subsection how the electronic properties of the target influence Auger neutralization of He$^+$ ions in low energy ion scattering, 
where single scattering events between He and a target atom prevail.  Important differences  with respect to grazing scattering are 
i) the turning points of the trajectories are smaller than 1 a.u. from the first atomic layer and ii) except for very low incident energies, 
more than 10$\%$ of the incident ions survive neutralization.
Here we will compare calculations of ion fractions for the different surfaces of the 
three noble metals with experimental results. Ion fractions are in general presented as a function 
of $\frac{1}{v_{\perp}}=\frac{1}{v_{\perp,in}}+\frac{1}{v_{\perp, out}}$, where $v_{\perp,in}$ and $v_{\perp,out}$ are the components of the velocity perpendicular to the surface for the
ingoing and outgoing trajectories, respectively. From Eq. (\ref{eq-P+}), a fit of the data to a single exponential retrieves
the characteristic velocity for the ion-surface combination under study.  

Most of the experiments presented here for the (100) and (110) surfaces were performed in double alignment geometry, 
which corresponds to normal incidence and exit in [001] and 
[1 $\bar{1}$ 2] azimuth direction, respectively, which ensures that collected particles have been scattered from the first atomic layer. 
We refer the reader to Refs. \cite{primetzhoferPRL2008,draxler_Vacuum2004} and references therein for experimental details.
In the simulations we use the molecular dynamics simulation code KALYPSO \cite{Kalypso} and calculate ion survival 
probabilities using the corrugated Auger rates in Eq. (\ref{eq-Pi}).  

We start by analyzing the case of Cu, where the presence of physical matrix-effects in LEIS was first observed \cite{primetzhoferPRL2008}.  
Three different crystal orientations are investigated: (110), (100), and
polycrystalline Cu. As a first step, the AN rates were calculated with a constant upward  shift of the He level of 2 eV, similar to the case of Ag 
investigated in grazing collisions in subsection \ref{secComparisonAgxxx}.   
However, unlike that case, for Cu we only find qualitative agreement between theory and experiment: the relative
differences in ion fractions between the different faces are reproduced, but the rates have to be increased by more than 30$\%$ to get a quantitative agreement.
Uncertainties in the external parameters (DOS, effective number of electrons, work function) do not have a
 sufficiently high impact on the resulting AN rate to explain this discrepancy. However, it turns out that the Auger rate reacts very sensitively to an energy shift of the He 1s level. 
 Unfortunately, detailed calculations of the distance dependent level shift for He interacting with noble metal surfaces are not available nowadays. Therefore, in the calculations
 to be presented below, the level position is independent of distance and chosen as a fitting parameter. 
 
In Fig. \ref{fig_LEIS_Cuxxx} ion fractions for three crystal orientations of Cu are compared \cite{goeblValdesPRB2011}. In the simulations,
the He level was set to - 20.5 eV with respect to the Fermi level which results in almost perfect agreement
with the experimental data for the polycrystalline surface. As a reference, the standard upward shift of 2 eV will correspond to 
a level placed - 17.5 eV below the Fermi level. Then, we have to invoke a 	extit{ downward} level shift of -1 eV in this case.  As can be seen in the figure, the data for
the (100) and (110) surfaces do not fit equally well and we need to use other values of the level shift.
For the case of Cu(110)  \cite{goeblMonrealNIMB2013}, the data shown in Fig. \ref{fig_LEIS_Cu110_azimuth} fit for a He level at -19.5 eV below the Fermi level and it is important to note that the
same value is valid for two different azimuthal directions at the surface shown in the figure. 
To fit the data for Cu(100) at random directions the level has to be at -20.0 eV with respect to the Fermi level. 
Thus, we find a difference of 1 eV among the three faces at most, which seems reasonable in view of the findings of subsection \ref{secComparisonAlxxx} for He on Al.
However, we have to keep in mind that in LEIS the turning points of the trajectories are smaller than 0.6 a.u. from the first atomic layer, which 
is the region close to the surface were the interaction of He with the metal atoms is very strong  (Fig.\ref{fig_nivel_HeAl}).
Then, we would expect the level shift to be basically influenced by one target atom and, consequently, 
a unique value independent of the azimuthal direction, and also independent of whether the
scattering has taken place in the first or in the second atomic layers.  This is the case for Cu(110) already noted in  Fig. \ref{fig_LEIS_Cu110_azimuth} 
and the same
behavior is found for Ag(110) in \cite{goeblNIMB2011}. 
 Fig. \ref{fig_LEIS_Cu100_ionsneutrals_azimuth} shows azimuthal scans for ion and neutral fractions of He$^+$ scattered from Cu(100) at 2 keV incident energy.
One should note here that 0$^\circ$ corresponds to particles exiting along the [001] direction after scattering with a Cu atom of the first atomic layer, while for
45$^\circ$ particles are scattered from first and second atomic layers. In the simulations a single value of the level position is used
and the excellent agreement between theory and experiment for both neutral and ion yields supports our interpretation. The slight disagreement
in the neutral yields at 45$^\circ$ can be corrected when subtracting the background contribution to the experimental data, see Ref.  \cite{goeblMonrealNIMB2013} for details. 
\newline

 The  profound influence of the position of the He level on the calculated AN rates of the noble metals was not found for Al.
 The origin for  this phenomenon has been recently examined in \cite{monrealGoeblNIMB2013}.
 Figs. \ref{fig_LEIS_rates_AgAu} (a) and (b) show the AN rates of He$^+$  on Au and Ag, respectively,
 for several values of the position of the He-1s level with respect to the Fermi level.  Here the rates are presented in a linear scale to better reveal the changes. 
 The values of the level position near -18eV correspond to the standard  upward level shift of 2eV.
 We note that, for Au and Ag (and also for Cu, not shown here) the rates increase notably as the He- level goes down in energies. The reason is the following.
 The value of $E_a$, measured with respect to the Fermi level,  determines the range of energies transferred to the metal,
 $\omega_{min}=\epsilon_{bottom}-E_a$, $\omega_{max}=-E_a$, where $\epsilon_{bottom} $ is the energy at the bottom of the conduction band.   
 This range moves toward higher energies when the level goes down. In our approximation for the screened susceptibility  $\chi(\omega) $ 
 we increase the density of the electron gas,  $n_{eff}(\omega)$, with increasing excitation energy, which also causes an increase in the plasma frequency.  
 As stressed in section \ref{secMultielectron}, when the plasmons of the metal can be excited they make an important contribution to the rate.
Therefore our approximation makes the metal to screen very efficiently at high frequencies.
Thus, the rates can change by 30\% when the level changes by 2 eV. This is not the case for free-electron metals of similar band width. 

 Figs. \ref{fig_LEIS_fractions_AgAu} (a) and (b) show the comparison of experimental \cite{primetzhoferPRB2009,primetzhoferNIMB2009}
and calculated ion fractions of He$^+$ 
 scattered from polycrystalline Ag and polycrystalline Au, 
 respectively, for different values of the energy level position of He with respect to the Fermi level.
The polycrystalline samples, were approximated as a surface with randomly oriented (111) domains \cite{chatain2004}, and,
consequently, the ion fractions were obtained as an average over trajectories scattered from a (111) surface with normal incidence and arbitrary azimuth exit directions.
The order of magnitude of the ion fractions is well reproduced by  theory and almost perfect agreement with experiment can be obtained by fitting the 
values of the level position, as in the case of Cu.  In the light of our discussion above, given our approximate treatment of the dielectric susceptibility of noble metal surfaces, 
we cannot assess how accurate these values are. We deem them to be not too unrealistic because, as commented before, these values are independent
of azimuth and experimental ion yields for scattering along  symmetry directions have contributions from the first and the second atomic layers. If band 
 structure effects were very important in the calculation of $\chi(\omega)$, they would produce significant differences in the rates for the first and second 
 atomic layers. Then these differences had to be somehow compensated by differences in the level shifts. We do not see these differences in the level 
 shift, since our approximations yield  good agreement with experiments for both Ag and Cu surfaces. With respect to the magnitude of the shifts, we 
note that,
while the value that fits the experimental results for He/ polycrystalline Au is -1.5 eV, 
which is similar to the value for He/ polycrystalline Cu quoted above,  
in the case of polycrystalline Ag
 we need to invoke a level shift of 4 eV.
This large and positive value of the shift could be an indication that these ions probe the region of distances to a surface atom where the He level is being 
strongly  promoted, since the experimental incident energies are near the threshold for collision induced neutralization and reionization processes 
in the system He/Ag. 
Since an exact treatment of the surface screened susceptibility is not feasible nowadays, we conclude  that first principles theoretical calculations of the level shift variation of ions in front of noble metal surfaces are of prime importance to improve our understanding of ion surface interactions.

\section{Mutielectron Theory of Auger Ionization}
\label{secAugerIonization}

The detailed understanding achieved for AN processes allows us to address the inverse problem of AI.
This is motivated by the fact that in the grazing scattering regime ionization of neutral atoms is experimentally seen even at
low perpendicular energies, with a threshold in parallel energy of 5 keV for the system He/Al.
Earlier experimental results on the ionization of neutral atoms
\cite{wethekamNIMB2007,winterAI, wethekamAI} seemed to agree with theoretical estimates \cite{zimnyMiskovicNIMB1991} for the
existence of a kinetic energy threshold below which AI is forbidden, but a detailed theory have been lacking until recently.

\subsection{General Formalism}
\label{secFormalismAI}

The theory for Auger ionization follows closely that of Auger neutralization expounded in subsection 
\ref{secGeneralFormalism}, since they are inverse processes. For AI we follow a similar approach, however, take into account the
kinetic energy of the projectile.  We will start by using   
the jellium model because it has the advantage that the surface is translationally invariant and this makes the problem
of Auger ionization  mathematically simpler \cite{wethekamValdesAIPRB2008,wethekamWinterValdesAIPRB2009} . 
Since we are concerned with grazing incident experiments, 
we assume a projectile motion parallel to the surface with velocity ${\vec v}$, at a perpendicular distance 
$z_a$.  Working in the restframe of the solid, we write the initial state of the electron bound to the atom  as 

\begin{equation}
|\Phi_a \rangle=\phi_a({\vec x}-{\vec v}t,z-z_a)e^{i{\vec x}\cdot {\vec v}} e^{-i(E_a+{1 \over 2}v^2)t},
\label{Phi-atom}
\end{equation}
with $\phi_a({\vec x},z)$ being the wave function of the bound electron in the restframe of the ion, $({\vec x},z)$ are the electron coordinates
parallel and perpendicular to the surface, respectively, and $E_a$ the electron binding energy. Eq.(\ref{Phi-atom}) shows the translation factor
$e^{i{\vec x}\cdot {\vec v}}$ as well as the addition of the kinetic  energy $v^2/2$ to the potential energy of the projectile.
In its final state, this electron is in the empty part of the conduction band of the metal and is  described by

\begin{equation}
|{\vec k}\rangle=\frac{1}{\sqrt{2\pi}}e^{i{\vec k}_{\parallel}\cdot{\vec x}}\phi_{k_z}(z) 
e^{-i E_{\vec k} t},
\label{Phi-metal}
\end{equation}
representing an electron of wave vector ${\vec k}=({\vec k}_{\parallel},k_z)$ and energy $E_{\bf k}=k^2/2$. Simultaneously, the electron-electron 
Coulomb interaction $\hat V$ causes the excitation of the metal from its many-body ground state $|0\rangle$ , of energy $E_0$, to a state $|n\rangle$, of energy $E_n$,
 with matrix elements having exactly the form of Eq. (\ref{fVi}).  Due to the translational invariance of the metal surface, 
the time-dependence of the matrix elements can be completely 
factored out as $e^{-i(E_0-E_n+E_a-v^2/2-E_{\vec k}+({\vec k}_{\parallel}-{\vec q})
\cdot {\vec v}) t}$, which allows us to write down the Fermi's golden-rule for the 
process as
\begin{eqnarray}
\frac{1}{\tau_{AI}}=2\pi\sum_{n} \sum_{k>k_F}  & \int_{0}^{\infty} d\omega 
|\langle f|\hat V|i \rangle|^2  \delta(\omega-(E_n-E_0))   \nonumber   \\
&\times  \delta(-\omega+E_a-v^2/2-E_{\vec k}+({\vec k}_{\parallel}-{\vec q}) \cdot {\bf v}),
\end{eqnarray}
 where
$\langle f|\hat V|i\rangle$ denotes now only the spatial part of Eq. (\ref{fVi}). 
Then, the sum over excited states $|n\rangle$ 
can be again related to the imaginary part of the dielectric susceptibility of the metal surface 
$\chi(q, \omega; z, z')$. Also, it is possible to perform the Galilean transformation
${\bf k} \rightarrow {\vec k}+{\vec v}$, yielding our final expression for the
 Auger ionization rate as 

\begin{eqnarray}
{1\over \tau_{AI}}({\vec v},z_a,E_a)&  =&  2\sum_{|{\vec k}+{\vec v}|> k_F} 
 \int_0^\infty d\omega
 \int {d{\vec q}\over(2\pi)^2}
 \int dz  \int dz'
 [-Im\chi(q, \omega; z, z')] 
\nonumber \\
& & \times   \Phi({\vec k};{\vec q},z) \Phi^\ast({\vec k};{\vec q},z')  \delta(E_a-E_k-\omega-{\vec q}\cdot{\vec v}), 
\label{eq-tauAI}
\end{eqnarray}
where the potential $\Phi({\vec k};{\vec q},z)$ has the same expression as for AN, given by Eq. (\ref{potential}).

In Eq.(\ref{eq-tauAI}), energy conservation implies that the frequency 
and the wave vector are related via the Doppler relation $\omega'=\omega+{\vec q}\cdot{\vec v}$.
As a consequence, different from previous assumptions \cite{winterAI, wethekamAI, zimnyMiskovicNIMB1991}
there is not a clear-cut threshold for the kinetic energy of the projectile below which the
AI process is energetically forbidden. Rather, the  $\delta$-function of Eq.(\ref{eq-tauAI}) can be used to
give the cosine of the angle between vectors $\bf q$ and $\bf v$ and therefore imposes the constraint

\begin{equation}
-1\leq \frac{E_a-E_{\vec k}-\omega}{qv} \leq 1.
\label{cos}
\end{equation}

 Also, the fact that the bound electron has to be excited to an empty state of the
shifted Fermi sphere implies the restriction in energies $E_{\vec k}$

\begin{equation}
\frac{1}{2}(k_F-v)^2 \leq E_{\vec k}
\label{Ekmin}
\end{equation}

Then Eqs. (\ref{cos}) and (\ref{Ekmin}), together with the condition
$\omega\geq 0$,  imply that:

\begin{equation}
q \geq \frac{\frac{1}{2}(k_F-v)^2-E_a}{v}.
\label{qmin}
\end{equation}

Therefore, according to Eq. (\ref{qmin}), if $v$ is small or $E_a$ has a large and negative value,
the allowed values of the parallel wave vector $q$ will be much larger than $k_F$.
The surface response function strongly disfavors excitations of such large 
wave vectors and, consequently, the efficiency of the Auger ionization process can be very 
small in these cases. This is illustrated in Fig. \ref{fig_AI_rate_vs_v}, where we plot the Auger ionization
rate of He on Al as a function of the projectile velocity, for two values of the 
energy level of He. Instead of a threshold like behavior, the AI rate shows an
overall exponential increase with velocity. We also find  
the nearly exponential increase with $E_a$ shown in Fig. \ref{fig_AI_rate_vs_E}.
However, the increase is steeper for the smaller values of velocity and energy level, 
as expected from our discussion above. 
With respect to the dependence of the AI rate with distance to the surface, it is nearly exponentially decreasing with distance, as for AN.
Fig. \ref{fig_AI_rate_vs_v} shows results for two distances illustrating that the spatial  decay length depends mainly
on $E_a$ and it is not so much dependent on velocity, since, for a given value of $E_a$ the curves for different $z_a$ are nearly parallel.
Note also in this figure that the AI rates can get large values. As a reference, the largest values of the calculated  AN rates 
for He/Al are  0.01-0.02 a.u. at the distances shown in Fig. \ref{fig_AI_rate_vs_v}.  Consequently, we expect efficient Auger ionization of
 neutral He when atoms  
with a velocity of about 0.25 a.u. approach the Al surface closer than about 1 a.u., where  the calculated values of $E_a$ presented in
Fig. \ref{fig_nivel_HeAl} show level shifts greater than 10 eV.  We will show next that this is indeed the case.

\subsection{Comparison to experiments} 
\label{secComparisonAI}

In the experiments, a beam of $^4$He$^0$ atoms with energies of 1-11 keV under grazing angles of incidence is scattered along a random direction
from a flat and atomically clean Al(111) surface. Using the experimental set-up described in Ref. \cite{wethekamValdesPRB2008}, 
polar angular distributions of the scattered neutral and
ion projectiles are recorded. Also, we performed 3D molecular dynamic simulations of the trajectories using the 
 same interaction potential as for AN described in subsection 
\ref{secComparisonAlxxx}. Since ionized projectiles can be Auger-neutralized in their way out from the surface, we need to include in the simulations the rate for AN as well.
We will use the AN rates calculated within the LCAO method which give more precise results. Since the main difference between the jellium and the
LCAO calculations of the AN rate is due to differences in the orthogonalization procedure, we consistently correct the rates for AI by the same factor that brings the AN rate from
the jellium to the LCAO value.  From the simulations we can also extract the fractions of projectiles that were never ionized and those  that were ionized at least once, as well
as their distribution with respect to the polar exit angle, gaining further insight into the physics of the AI process. 

Comparison of measured and calculated ion fractions are shown in Fig. \ref{fig_AI_ionfraction} (a). In the simulations we find that incident neutrals can reach 
distances to the surface where the energy level of He is substantially promoted and efficient Auger loss is present, the ionization efficiency can reach 35$\%$ at 
the highest energies and angles of incidence shown in this figure.
We note that the simulations reproduce the
experimental trends on a quantitative level. The largest discrepancies occur at the larges values of energy and angle of incidence, where the projectiles 
can get  near to the surface, and the interaction potentials are more uncertain. 
Another source on uncertainty is related to the accuracy of the Auger neutralization rates, 
which already lead to the larger discrepancy between theory and experiment in the survival regime for the (111) surface 
(see Fig. \ref{fig_ionfractionAlxxx}). 
When we multiply the AN rate by a factor of 1.2  we get a perfect agreement between theory and experiment in the survival regime and this also tends to improve
the agreement in the ionization regime, as shown in Fig. \ref{fig_AI_ionfraction} (b). 

We note that the theory gives too low ion fractions for the smaller angles of incidence.
This is due to the existence of surface defects whose effects are more pronounced for the longer trajectories. Further evidence of it is given in Fig. \ref{fig_AI_exp_polar} (a)
where we present polar angular distributions for neutral and ionized projectiles, with a kinetic energy of 7.5 keV and several angles on incidence, 
the corresponding simulated distributions are shown in Fig. \ref{fig_AI_exp_polar} (b). For the two smallest angles of incidence 
the distributions for ions are broad and have a long tail extending to small exit angles, so this indicates that surface defects play an important role. 
However, for the largest angles of incidence, the discrepancy between experimental
and simulated distributions for ions are most probably due to  deficiencies in the interaction potentials, which are more important at the small distances of approach reached for
a He atom. Fig. \ref{fig_AI_exp_polar} (c) shows simulated distributions of projectiles that have never been ionized and projectiles that have been ionized at least once.
Ionized projectiles have been scattered toward larger exit angles also indicating that the ionization processes has been produced in a close collision between 
He and a target atom. Further evidence for it is given in Fig. \ref{fig_AI_proj_zmin} which shows the distributions of projectiles versus the minimum distance to a target atom.
The distributions are normalized so that they give the calculated ion fractions.
Here we can appreciate that these incident neutral atoms that reach distances shorter than ca. 1.0 a.u. are ionized with appreciable probability. 
At these distances, the promotion of the He-level is large enough to make the Auger process an efficient ionization mechanism 
for kinetic energies larger than 5 keV. Since the interaction potentials and the lattice vibrations control the distances of closest approach, they are crucial magnitudes,
together with ionization and neutralization rates, to give ion fractions and polar distributions.

Given that, in the range of distances between He and the Al surface where the AI process occurs, the He-level is in resonance with the conduction band of Al,
the contribution of the one-electron resonant processes cannot be ruled out. Actually, when they are introduced by means of  a simple model 
\cite{wethekamWinterValdesAIPRB2009}, one finds that they are as
important as Auger processes. Fig. \ref{fig_AI_ionfraction_AR} summarizes these findings. The remarkable agreement achieved between theory and experiment should not
be overemphasized in view of the limitations and approximations of the theory. A realistic theory should be based on improved calculations of level shifts and
interaction potentials, thermal displacements and transition rates including resonant and Auger processes for a quantitative description of charge transfer.

\section{Conclusions and Outlook}

In spite of its apparent simplicity, a microscopic understanding of Auger neutralization and ionization processes of ions at metal surfaces has only been
achieved recently. The key physical concepts involved are: i) modification of the atomic energy levels beyond the classical image potential approximation 
that occurs near the surface and ii) the many body character of the electron-electron Coulomb interaction requires a detailed description of the screening properties of the surface. Calculations of the Auger transition rates using these concepts have provided quantitative agreement with experimental
measurements of ion fractions for He interacting with metal surfaces under grazing scattering conditions and in the Auger neutralization
regime of LEIS. Measured energy gains for the He/Al system are also in accord with theoretical predictions. 

The theoretical methods presented here
should be equally applicable to the analysis of charge transfer at semiconductor surfaces, since experiments are usually performed at room temperature.
With suitable modifications these methods can be applied to the characterization of oxidized surfaces, 
a problem of renewed interest nowadays \cite{uhmAPL2009,uhmAPL2011,kurnsteinerSS2013,leeJP2013}.
Further progress in this field
requires, first, realistic "ab initio" calculations for the interaction energies of different atomic terms in the proximity of the surface and, second, a detailed evaluation of the surface response function by improving on the simple jellium model with the 
inclusion of the band structure of the surface.

Ion-surface interactions are very sensitive to the electronic properties of the target. Hence one should take advantage of previous knowledge
 and use atoms/ions as a tool
to investigate and characterize new materials such as graphene, nanoparticles, nanostructured, organic materials, etc, that exhibit new interesting properties and
are promising for technological applications.

\section*{Acknowledgments}
\label{secAck}

Much of the work presented here would have been not possible without the benefit of discussions with many colleagues and the frequent disagreement among us.
I want to especially express my gratitude to Fernando Flores, Pedro Echenique and Werner Heiland, who first introduced and guided me in the field of particle-surface interactions, 
to my fellow theorists  Jaime Merino, Nicol\'as Lorente 
and Diego Vald\'es, for contributing actively to the development of the theory and to the calculations of energy level variation and 
Auger neutralization rates shown in this work, and to the experimentalists whose investigations have been decisive to establish the present 
new scenario for analyzing Auger charge exchange processes at metal surfaces:  
Ra\'ul Baragiola, Vladimir Esaulov, Stefan Wethekam and Helmut Winter, Dominik Goebl and Peter Bauer.
I am also indebted to Jos\'e Manuel Benavides for drawing figures 1 and 2 of this manuscript.
 This work has been funded by the Spanish
Ministerio de Ciencia e Innovaci\'on, Project No. FIS2011-26516.

\newpage

Fig. 1:
Schematic energy diagram for the one-electron resonant processes.
\newline

Fig. 2:
Schematic energy diagram for the two-electron Auger processes. (a) Auger neutralization, (b) Auger ionization, (c) direct Auger
deexcitation, (d) indirect Auger deexcitation.
\newline

Fig. \ref{figHeMetal}:
Schematic energy diagram for interaction of
He with a high work function metal surface. $W$: work function; blue shaded area: occupied
states of conduction band; brown curves: energy levels of He
as function of distance from the surface for states indicated. Green
arrow: resonant neutralization (RN), blue arrows: Auger
neutralization (AN).
Reprinted with permission from \cite{monrealGoeblNIMB2013}. \copyright (2013) by Elsevier.
\newline

Fig. \ref{figWinterJP}:
Schematic  diagram of the experimental method for measuring energy gains. 
Reproduced with permission from H. Winter, J. Phys.: Condens. Matter \textbf{5} (1993) p. A295 \cite{winterJPCM1993}.
\copyright (1993) by IOP Publishing.
\newline

 Fig. \ref{figMonLor}:
The Auger neutralization rate of He$^+$ on a jellium representative of Al as a function of parallel momentum transfer, for fixed value of the
energy transfer $\omega=0.78 \omega_p$ for (a) $z_a= 4$ a.u. and (b) $z_a= -1 $a.u. with respect to the jellium edge.
Reprinted with permission from R. Monreal, N. Lorente, Phys. Rev. B \textbf{52} (1995) p. 4760 \cite{monrealLorentePRB1995}. 
\copyright (1995) by the American Physical Society. 
\newline

 Fig. \ref{fig_tau_w}:
The Auger neutralization rate of He$^+$ on a jellium representative of Al as a function of the energy transfer $\omega$, for $z_a= 5$ a.u. and with respect to the jellium edge. Crosses: full interacting calculation, squares: non-interacting calculation. 
Reprinted with permission from  N. Lorente, R. Monreal, Phys. Rev. B \textbf{53} (1996) p. 9622 \cite{lorenteMonrealPRB1996}. 
\copyright (1996) by the American Physical Society. 
 \newline

Fig. \ref{fig_tau_Al_jelly}:
The Auger neutralization rate of He$^+$ on a jellium representative of Al as a function of distance with respect to the jellium edge.
Continuous line: full interacting calculation, dashed-line: non-interacting calculation. 
Reprinted with permission from  N. Lorente, R. Monreal, Phys. Rev. B \textbf{53} (1996) p. 9622 \cite{lorenteMonrealPRB1996}. 
\copyright (1996) by the American Physical Society. 
\newline

Fig. \ref{fig_tau_Na}:
The Auger neutralization rate of He$^+$ on a jellium representative of Na as a function of distance with respect to the jellium edge.
Continuous line: full interacting calculation, dashed-line: non-interacting calculation. 
Reprinted with permission from \cite{lorenteMonrealSS1997}. \copyright (1997) by Elsevier.
\newline

Fig. \ref{fig_Hecht_SS}:
The AN, AD, and RI transition rates obtained/used in the simulation of experimental data as function of distance from
the image plane. Dotted curve: calculation of the Auger rate in\cite{lorenteMonrealSS1997}. 
Reprinted with permission from \cite{hechtSS1998}. \copyright (1998) by Elsevier. 
\newline

Fig. \ref{fig_tau_Al111_tcl_jell}:
The Auger neutralization rate of He$^+$ on Al(111) as a function of distance with respect to the first atomic layer compared to the jellium rate (blue line).
Several positions of He within the surface unit cell are considered:  on-top (black triangles), center (red dots) and bridge (blue asterisks). 
Reprinted with permission from  Diego Vald\'es 	\textit{et al.},
Phys Rev. B \textbf{71} (2005) p. 245417 \cite{valdesGoldbergPRB2005}.
\copyright (2005) by the American Physical Society. 
\newline

Fig. \ref{fig_tau_Al110_tc_jell}:
The Auger neutralization rate of He$^+$ on Al(110) as a function of distance with respect to the first atomic layer compared to the jellium rate (blue line).
Several positions of He within the surface unit cell are considered:  on-top (black squares), and center (red dots). 
Reprinted with permission from  Diego Vald\'es 	\textit{et al.},
Phys Rev. B \textbf{71} (2005) p. 245417 \cite{valdesGoldbergPRB2005}.
\copyright (2005) by the American Physical Society. 
\newline

Fig. \ref{fig_tau_Al110t_vecs}:
The total Auger neutralization rate of He$^+$ approaching the Al(110) surface on-top of an atom of the first atomic layer (black squares)
is decomposed in the contributions of neighboring atoms of Al:  on-top atom (red dots), first neighbors (blue up triangles ) and second neighbors (black triangles) in the 
first atomic layer and first neighbors (blue down triangles) on the second atomic layer.
Adapted with permission from  Diego Vald\'es  \textit{et al.},
Phys Rev. B \textbf{71} (2005) p. 245417 \cite{valdesGoldbergPRB2005}.
\copyright (2005) by the American Physical Society. 
\newline

Fig. \ref{fig_rates_CuAgAu}:
The Auger neutralization rate of He$^+$ on-top of an atom of the Cu(111) (a), Ag(111) (b), and Au(111) (c)  surfaces as a function of distance 
with respect to the first atomic layer (black squares) is decomposed in the contributions of  s-electrons (blue triangles) and d-electrons (green diamonds). 
The contribution of the atom on-top is shown as red dots. 
Reprinted with permission from  D. Goebl \textit{et al.},
Phys. Rev. B \textbf{84} (2011) p. 165428 \cite{goeblValdesPRB2011}.
\copyright (2011) by the American Physical Society. 
\newline

Fig. \ref{fig_rates_AuAl_tcl}:
The Auger neutralization rates of He$^+$ on Au(111) (a), and Al(111) (b) assuming the He atom to be at the following
lateral positions within the (111) unit cell: on-top, the two non-equivalent center positions
and in the mid point between two neighbor atoms (Pos 1).
Reprinted with permission from \cite{monrealGoeblNIMB2013}. \copyright (2013) by Elsevier.
\newline

Fig. \ref{fig_rate_Ag110_z2_az35}:
The Auger neutralization rate of He$^+$ on Ag(110) as a function of the lateral distance to a surface atom along the azimuthal
direction [111], at fixed perpendicular distance  $z_a=2$ a.u.,  is shown by the black line. The blue dashed line
represents the contribution of s-electrons and the thin straight line is the jellium result.
\newline

Fig. \ref{fig_rate_Cu100_contour}:
Contour plots of the Auger neutralization rate of He$^+$ on Cu(100) at $z_a= -2.4 $ a.u (upper graph) and $z_a=1.0$ a.u. (lower graph). 
z-distances are given with respect to the first atomic layer.
Reprinted with permission from \cite{goeblMonrealNIMB2013}. \copyright (2013) by Elsevier. 
\newline

Fig. \ref{fig_nivel_HeAl}:
The energy of He 1s-level on Al as a function of distance with respect to the first atomic layer.
Reprinted with permission from  Diego Vald\'es 	\textit{et al.},
Phys Rev. B \textbf{71} (2005) p. 245417 \cite{valdesGoldbergPRB2005}.
\copyright (2005) by the American Physical Society. 
\newline

Fig. \ref{fig_HeAg_PRL}:
The experimental ion fractions for grazing scattering of He$^+$ ions on Ag(111) and (110) surfaces as a function of the incident energy are compared to
theoretical predictions using the jellium model.
Reprinted with permission from  Yu Bandurin \textit{et al.},
Phys. Rev. Lett. \textbf{92} (2004) p. 017601 \cite{bandurinPRL2004}.
\copyright (2004) by the American Physical Society. 
\newline

Fig. \ref{fig_traj_HeAg_PRL}:
Calculated  trajectories for scattering of He on Ag(110) and (111) surfaces for the incident energies indicated in the figure. The position of the jellium
edges for both surfaces are indicated by the lines at the right.
Reprinted with permission from  Yu Bandurin \textit{et al.},
Phys. Rev. Lett. \textbf{92} (2004) 017601 \cite{bandurinPRL2004}.
\copyright (2004) by the American Physical Society. 
\newline

Fig. \ref{fig_ionfractionAg110_az1k}:
 Experimental ion fractions for scattering of He ions on Ag(110) as a function of azimuth for incident energies of 1 keV (a) and 2 keV (b).  
The experimental data are shown by blue squares. The results of theory including only s electrons are shown by green diamonds and those
including s and d electrons by open triangles. 
Reprinted with permission from  Diego Vald\'es \textit{et al.},
Phys. Rev. Lett. \textbf{97} (2006) 047601 \cite{valdesPRL2006}.
\copyright (2006) by the American Physical Society.  
\newline

Fig. \ref{fig_ionfractionAg110_az3k}:
 Experimental ion fractions for scattering of He ions on Ag(110) as a function of azimuth for incident energies of 3 keV (a) and 4 keV (b).  
The experimental data are shown by black dots and the results of theory including s and d electrons by open triangles.
Reprinted with permission from  Diego Vald\'es \textit{et al.}
Phys. Rev. B  \textbf{75} (2007) 165404 \cite{valdesPRB2007}.
\copyright (2007) by the American Physical Society.    
\newline

Fig. \ref{fig_ionfraction_LCAOvsJelly}:
Comparison between LCAO (continuous line) and jellium approaches (dotted line) for the calculation of the ion fraction for He$^+$ on Ag(110) at
random directions. Experimental points are also shown as circles with error bars.
Reprinted with permission from  Diego Vald\'es \textit{et al.},
Phys. Rev. Lett. \textbf{97} (2006) 047601 \cite{valdesPRL2006}.
\copyright (2006) by the American Physical Society.  
\newline

Fig. \ref{fig_ionfractionAg111_az}:
The experimental ion fractions for scattering of He ions on Ag(111) as a function of azimuth and 4 keV of incident energy (closed squares) 
are compared with theoretical calculations, using the  LCAO model including s and d-electrons (open circles).
Reprinted with permission from  Diego Vald\'es \textit{et al.},
Phys. Rev. B  \textbf{75} (2007) 165404 \cite{valdesPRB2007}.
\copyright (2007) by the American Physical Society.     
\newline

Fig. \ref{fig_tauAg110_displacedjelly}:
Auger neutralization rate for He$^+$ on Ag(110)  with $\chi$ calculated using the jellium edge 
placed at half the interplanar distance of the (100) for the on-top (black squares) and on-center (black dots) positions and 
and using the interplanar distance of the (111) planes for the on-top (blue triangles up) and
on-center (blue triangles down) positions.
\newline

Fig. \ref{fig_levelshift_Alxxx}:
Panel a: He-1s energy shift as a
function of distance from Al(111) surface calculated for on-top position (green dots) and values
used in the simulations (green dash-dotted curve). Black solid, red dashed, and blue dotted curves: level shifts averaged parallel to
surface for Al(111), Al(100), and Al(110). Panel b: potential energy averaged parallel to surface as function of distance for He$^0$
(thin curves) and He$^+$ (thick curves) in front of Al(111) (black solid curves), Al(100) (red dashed curves), and Al(110) (blue dotted curves).
Reprinted with permission from  S. Wethekam \textit{et al.},
Phys. Rev. B  \textbf{78} (2008) 75423 \cite{wethekamValdesPRB2008}.
\copyright (2008) by the American Physical Society.     
\newline

Fig. \ref{fig_ionfractionAlxxx}:
The experimental ion fraction for scattering of He$^+$ on Al(111), Al(110) and Al(100) surfaces are compared to theoretical results using the calculated position dependent values of the rates indicated in the figure.
Reprinted with permission from  S. Wethekam \textit{et al.},
Phys. Rev. B  \textbf{78} (2008) 75423 \cite{wethekamValdesPRB2008}.
\copyright (2008) by the American Physical Society.     
\newline

Fig. \ref{fig_polardistributions}:
Polar angular distributions for scattering of 1 keV (panel a) and 10 keV (panel b) He$^+$ ions (full symbols) and He$^0$ atoms (open symbols) from Al(111). Black dots (red squares) show experimental data (simulation for on-top rates).
Reprinted with permission from  S. Wethekam \textit{et al.},
Phys. Rev. B  \textbf{78} (2008) 75423 \cite{wethekamValdesPRB2008}.
\copyright (2008) by the American Physical Society.     
\newline

Fig. \ref{fig_energygain_Alxxx}:
Experimental normal energy gains  as function of incident normal energy for He$^+$ scattered from Al(111) (black open squares), Al(100) (red
open dots), and Al(110) (blue open squares) compared to simulations using on-top AN rates (panel a) for Al(111) (black full squares),
Al(100) (red full dots), and Al(110) (blue full upward triangles) as well as AN rates for the hollow position (panel b) for Al(111) (black full
diamonds) and Al(110) (blue full downward triangles).
Reprinted with permission from  S. Wethekam \textit{et al.},
Phys. Rev. B  \textbf{78} (2008) 75423 \cite{wethekamValdesPRB2008}.
\copyright (2008) by the American Physical Society.      
\newline

Fig. \ref{fig_LEIS_Cuxxx}:
Experimental (open symbols) and theoretical (full symbols) ion fractions for He$^+$ scattered from  Cu(100) (squares), Cu(110) (circles and polycrystalline Cu (triangles).  
The straight lines represent single exponential fits to the data yielding the characteristics velocities indicated in the figure.
\newline

Fig. \ref{fig_LEIS_Cu110_azimuth}:
 Experimental (open symbols) and theoretical (full symbols) ion fractions for He$^+$ scattered from Cu(110) 
 in [1 $\bar{1}$ 2] (black squares) and [1 $\bar{1}$ 0] (red circles) directions.   
 The straight lines represent single exponential fits to the data yielding the characteristics velocities indicated in the figure.
Reprinted with permission from \cite{goeblMonrealNIMB2013}. \copyright (2013) by Elsevier. 
\newline

Fig. \ref{fig_LEIS_Cu100_ionsneutrals_azimuth}:
 Experimental (open symbols) and theoretical (full symbols)  neutral (dots) and ion (squares) yields for He$^+$ scattered from Cu(100) as a function of
azimuth.   Experimental data without background subtraction. 
Reprinted with permission from \cite{goeblMonrealNIMB2013}. \copyright (2013) by Elsevier.   
\newline

Fig. \ref{fig_LEIS_rates_AgAu}:
The AN rates for He$^+$ on Au(111) (a) and Ag(111) (b) on-top position as a function of the distance to the first atomic layer, for the values of the He- level position 
specified in the figure.
Reprinted with permission from \cite{monrealGoeblNIMB2013}. \copyright (2013) by Elsevier.  
\newline

Fig. \ref{fig_LEIS_fractions_AgAu}:
 Experimental and theoretical ion fractions for He$^+$ on polycrystalline Ag (a) and Au (b) for different values of the He-level position.
Reprinted with permission from \cite{monrealGoeblNIMB2013}. \copyright (2013) by Elsevier.
\newline

Fig. \ref{fig_AI_rate_vs_v}:
The Auger ionization rate of $^4$He$^0$ on Al(111) is shown as a function of the projectile velocity for the values of the level shift 
and the distance to the first atomic layer indicated in the figure.
Reprinted with permission from  S. Wethekam \textit{et al.},
Phys. Rev. B  \textbf{79} (2009) 195408 \cite{wethekamWinterValdesAIPRB2009}.
\copyright (2009) by the American Physical Society.     
\newline

Fig. \ref{fig_AI_rate_vs_E}:
The Auger ionization rate of $^4$He$^0$ on Al(111) is shown as a function of the He level shift  for a fixed distance of 1.21 a.u.
with respect to the first atomic layer and different values of the kinetic  energy indicated in the figure.
The inset shows the calculated He- level shift as a function of the distance. The conduction band of Al is indicated by the shaded area.
Reprinted with permission from  S. Wethekam \textit{et al.},
Phys. Rev. B  \textbf{78} (2008) 033105 \cite{wethekamValdesAIPRB2008}.
\copyright (2008) by the American Physical Society.     
\newline

Fig. \ref{fig_AI_ionfraction}:
Measured (full symbols) and simulated (open symbols with lines) ion fractions as a function of the kinetic energy of the projectiles
for scattering of $^4$He$^0$ atoms from Al(111) 
along random directions under grazing angles of incidence indicated in the figure.  Simulations including AN and AI processes with the original 
values of the rates (panel a) and with the AN rates multiplied by a factor of 1.2 (panel b). The inset shows a simple sketch of the experimental setup.
Reprinted with permission from  S. Wethekam \textit{et al.},
Phys. Rev. B  \textbf{79} (2009) 195408 \cite{wethekamWinterValdesAIPRB2009}.
\copyright (2009) by the American Physical Society.      
\newline

Fig. \ref{fig_AI_exp_polar}:
Panel a: experimental angular distributions as function of polar exit angle for scattering of 7.5 keV $^4$He$^0$ atoms from Al(111) under angles of incidence as indicated. 
Open black circles (full red circles): outgoing atoms `0' (ions `+'). Panel b: simulated angular distributions for the same conditions 
based on AI and AN as charge transfer mechanisms. 
Panel c (same conditions): Normalized angular distributions for projectiles that have never been ionized (blue open triangles) and projectiles that have been 
ionized at least once (magenta full triangles with solid curve). Ion survival probability $P_{AN}^{+,out}$ on the
outgoing trajectory (full green squares)  as function of the exit angle. All distributions are normalized to 1. Data is offset by multiples of 1.1.
Reprinted with permission from  S. Wethekam \textit{et al.},
Phys. Rev. B  \textbf{79} (2009) 195408 \cite{wethekamWinterValdesAIPRB2009}.
\copyright (2009) by the American Physical Society.     
\newline

Fig. \ref{fig_AI_proj_zmin}:
Simulated distributions of projectiles that have been never ionized (blue, open triangles), of projectiles that have been ionized at least once (red, full triangles) and
of all projectiles that reach the detector versus the minimum distance to a target atom. The incident energy of He is 7.5 keV and the incidence angles are indicated in the figure.
Reprinted with permission from  S. Wethekam \textit{et al.},
Phys. Rev. B  \textbf{79} (2009) 195408 \cite{wethekamWinterValdesAIPRB2009}.
\copyright (2009) by the American Physical Society.     
\newline

Fig. \ref{fig_AI_ionfraction_AR}:
Measured (full symbols) and simulated (open symbols with lines) ion fractions as a function of the kinetic energy of the projectiles
for scattering of $^4$He$^0$ atoms from Al(111) 
along random directions under grazing angles of incidence indicated in the figure.  Simulations including AN, AI and resonant processes with the original 
values of the rates (panel a) and with the AN rates multiplied by a factor of 1.2 (panel b). 
Reprinted with permission from  S. Wethekam \textit{et al.},
Phys. Rev. B  \textbf{79} (2009) 195408 \cite{wethekamWinterValdesAIPRB2009}.
\copyright (2009) by the American Physical Society.      
\newline

\newpage

\begin{figure}[htbp]
\centering
\includegraphics[width=7cm]{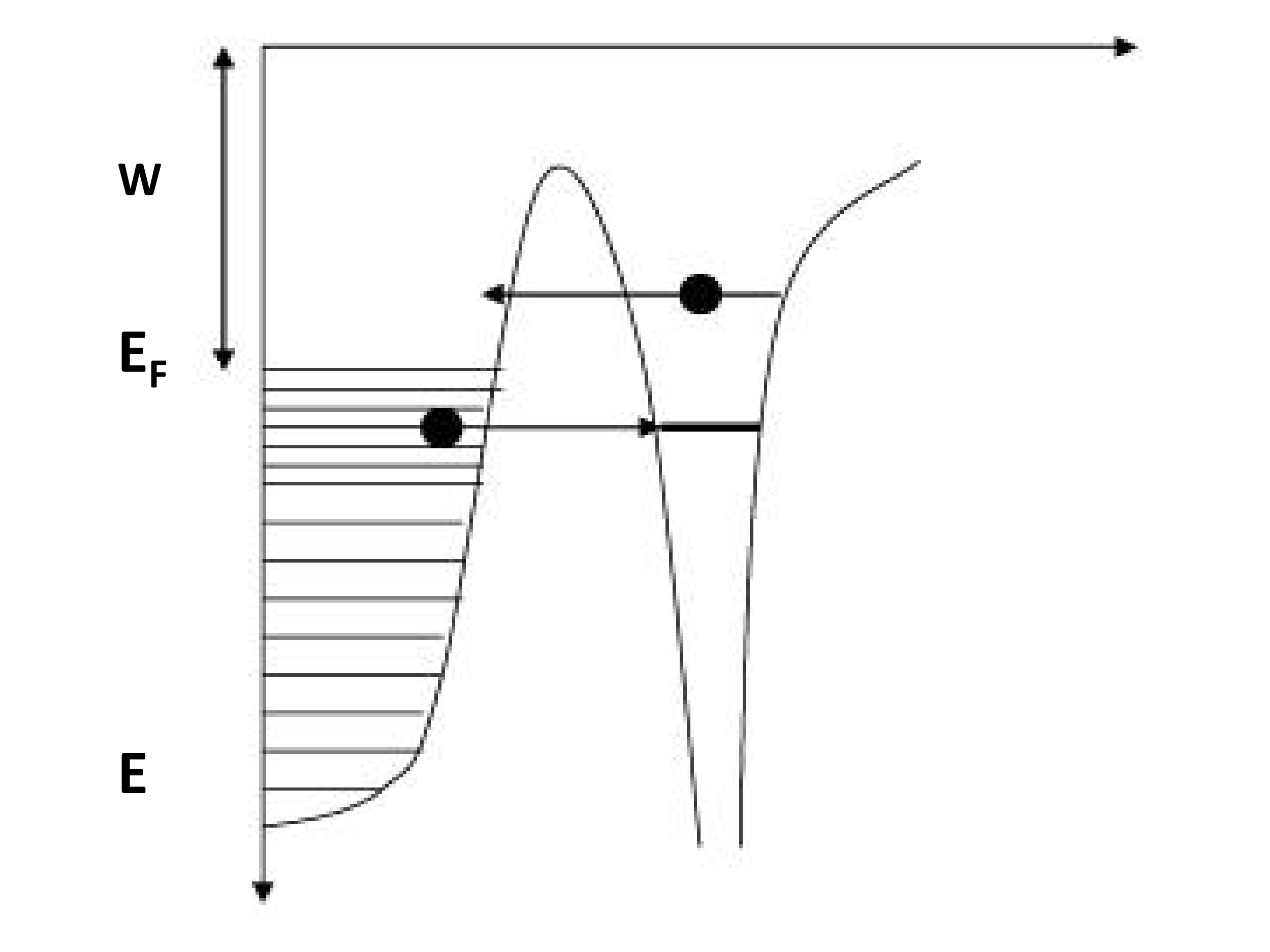}
\caption{}
\label{fig1}
\end{figure}

\begin{figure}[htbp]
\centering
\includegraphics[width=7cm]{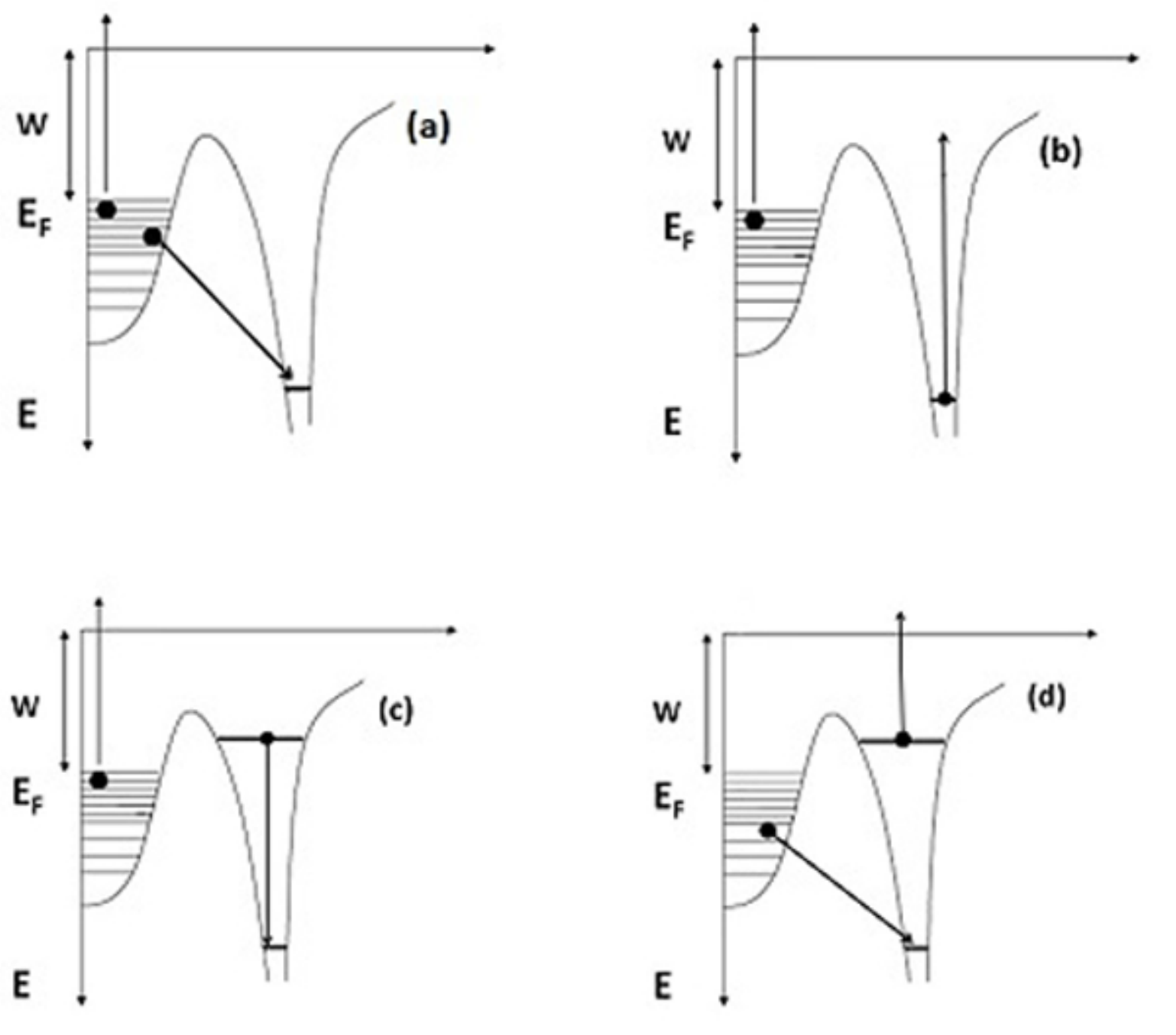}
\caption{}
\label{fig2}
\end{figure}

\newpage

\begin{figure}[htbp]
\centering
\includegraphics[width=7cm]{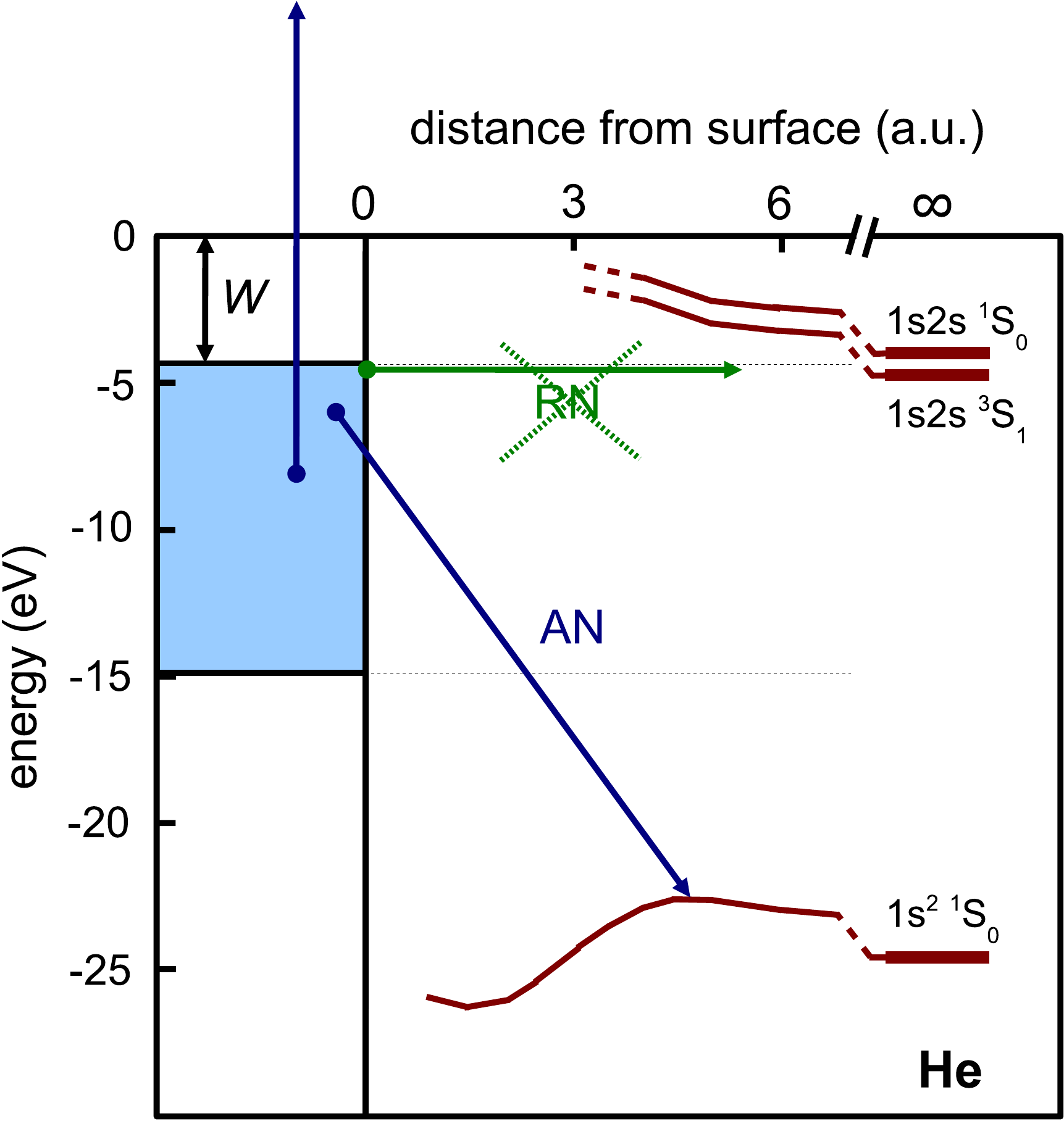}
\caption{}
\label{figHeMetal}
\end{figure}

\begin{figure}[htbp]
\centering
\includegraphics[width=7cm]{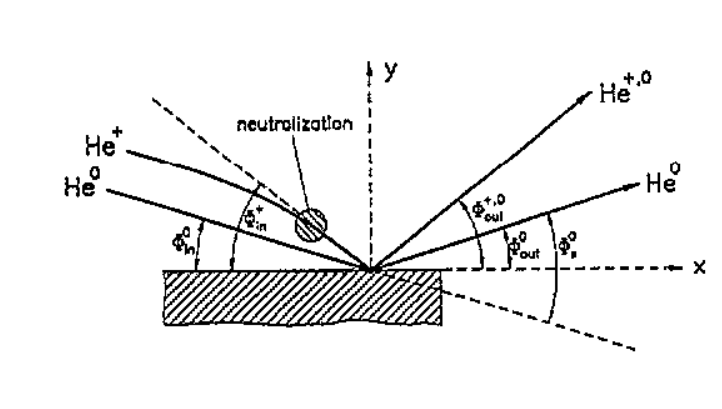}
\caption{}
\label{figWinterJP}
\end{figure}

\newpage

\begin{figure}[htbp]
\centering
\includegraphics[width=7cm]{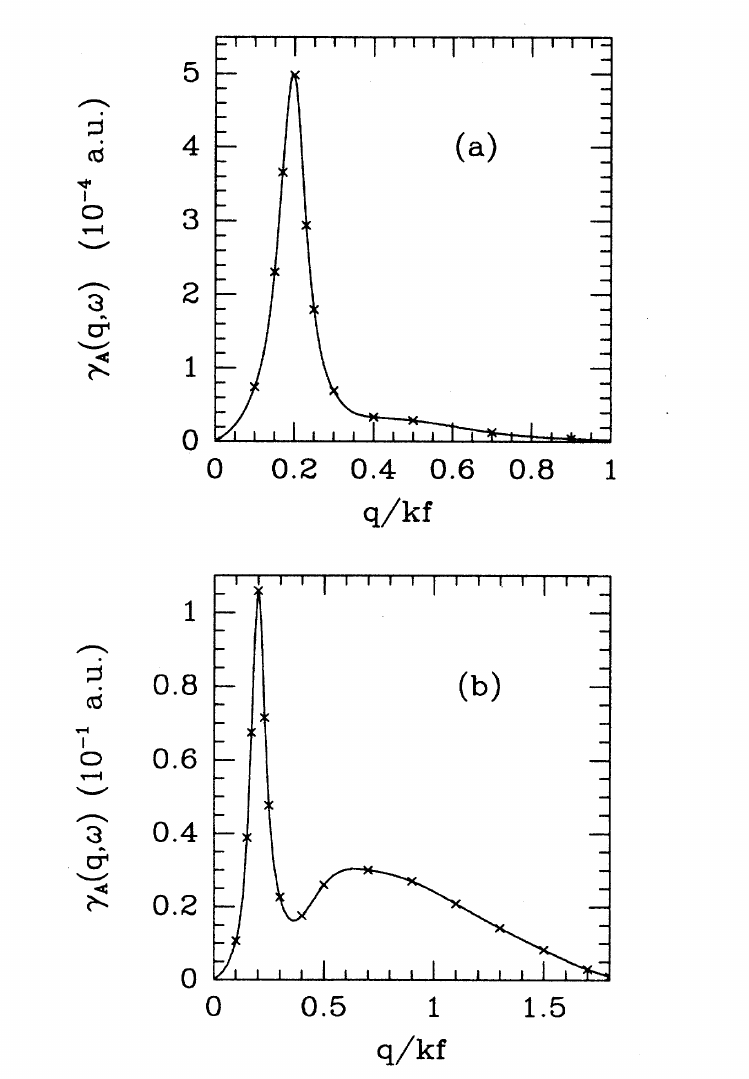}
\caption{}
 \label{figMonLor}
\end{figure}

\newpage

\begin{figure}[htbp]
\centering
\includegraphics[width=7cm]{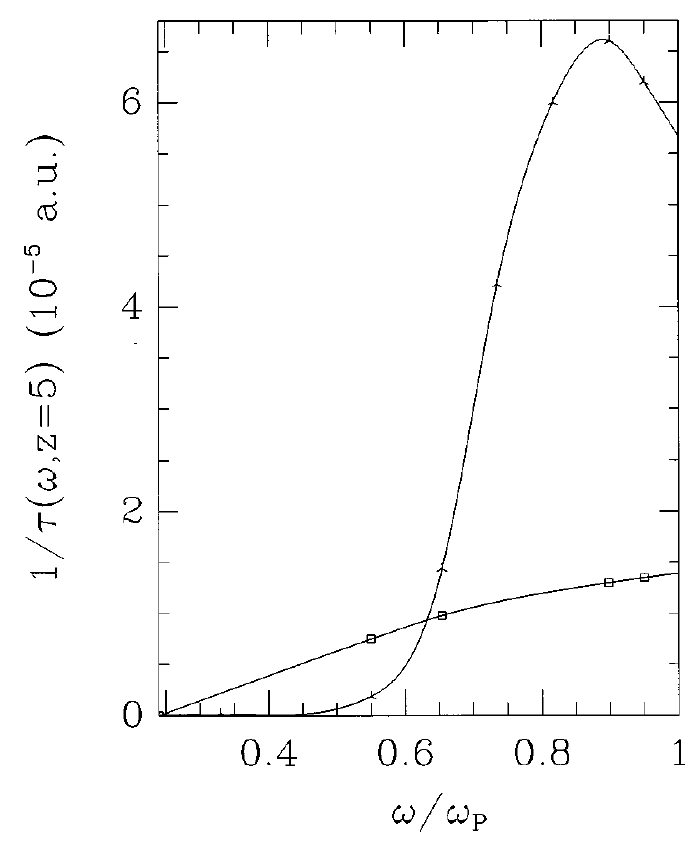}
\caption{}
\label{fig_tau_w}
\end{figure}

\begin{figure}[htbp]
\centering
\includegraphics[width=7cm]{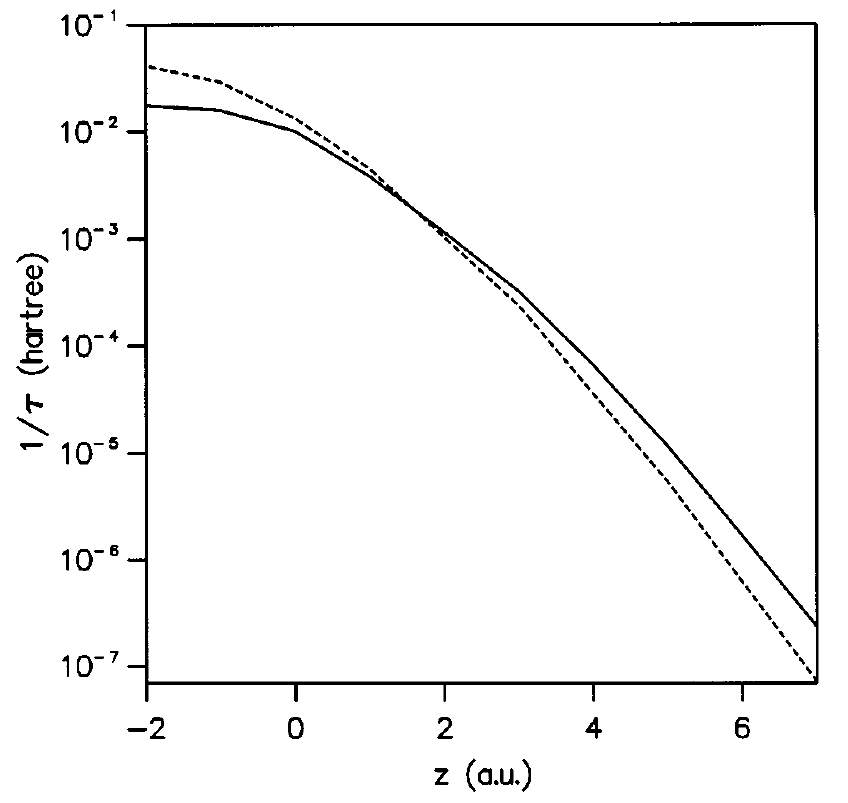}
\caption{}
\label{fig_tau_Al_jelly}
\end{figure}

\newpage

\begin{figure}[htbp]
\centering
\includegraphics[width=7cm]{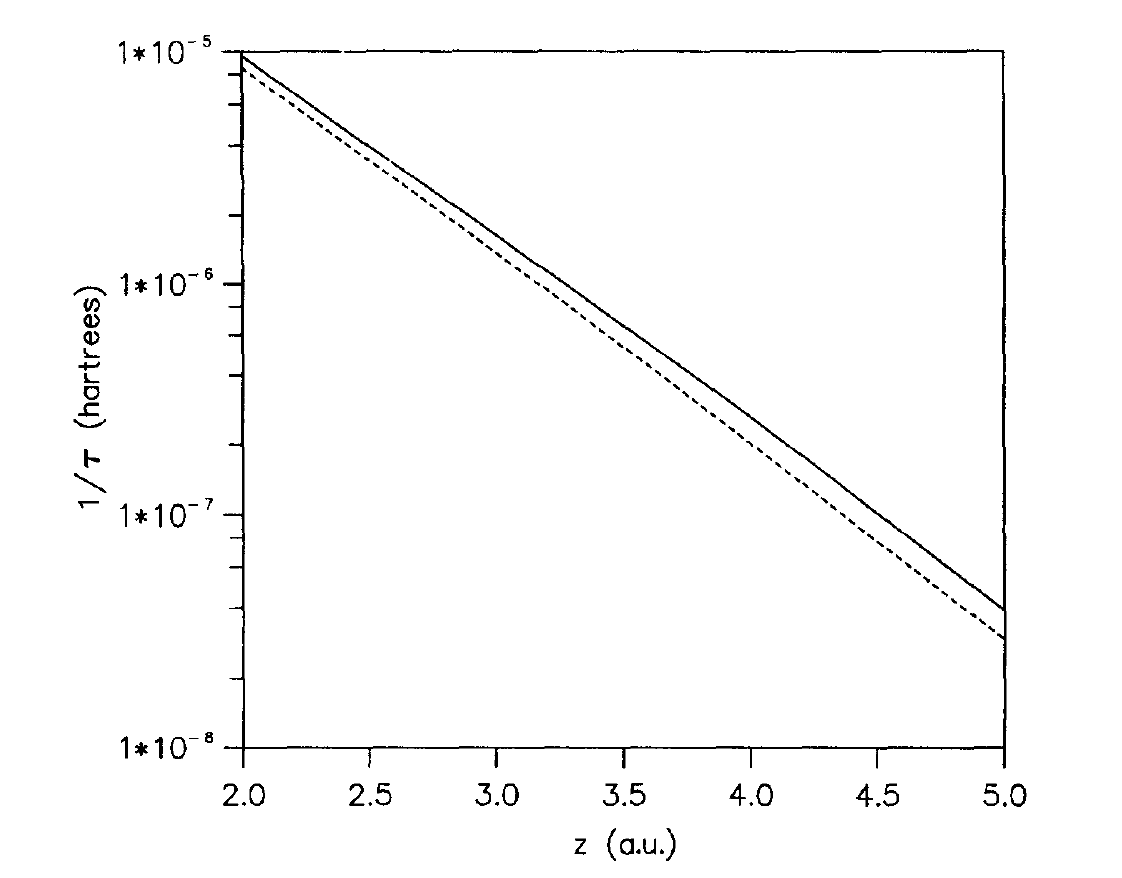}
\caption{}
\label{fig_tau_Na}
\end{figure}

\begin{figure}[htbp]
\centering
\includegraphics[width=7cm]{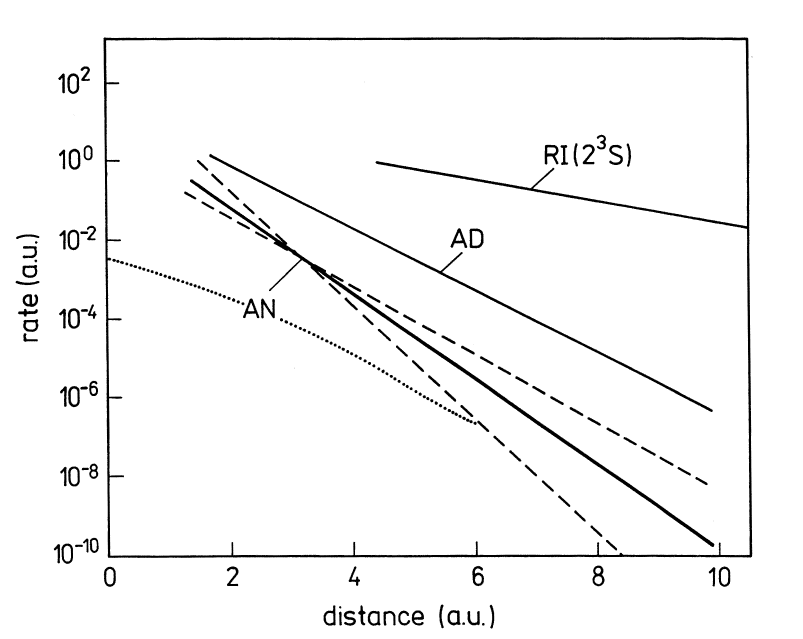}
\caption{}
\label{fig_Hecht_SS}
\end{figure}

\newpage

\begin{figure}[htbp]
\centering
\includegraphics[width=7cm]{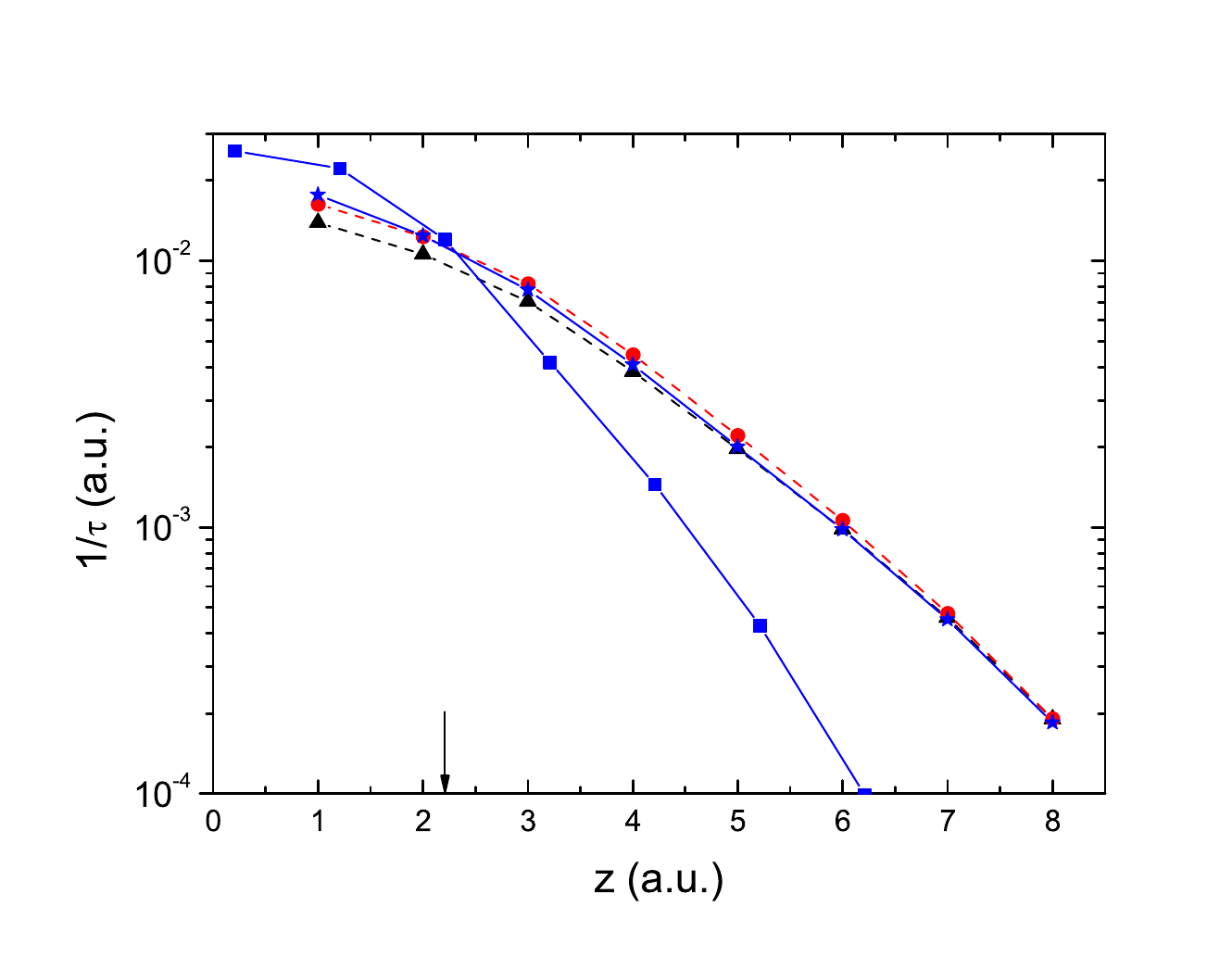}
\caption{}
\label{fig_tau_Al111_tcl_jell}
\end{figure}

\begin{figure}[htbp]
\centering
\includegraphics[width=7cm]{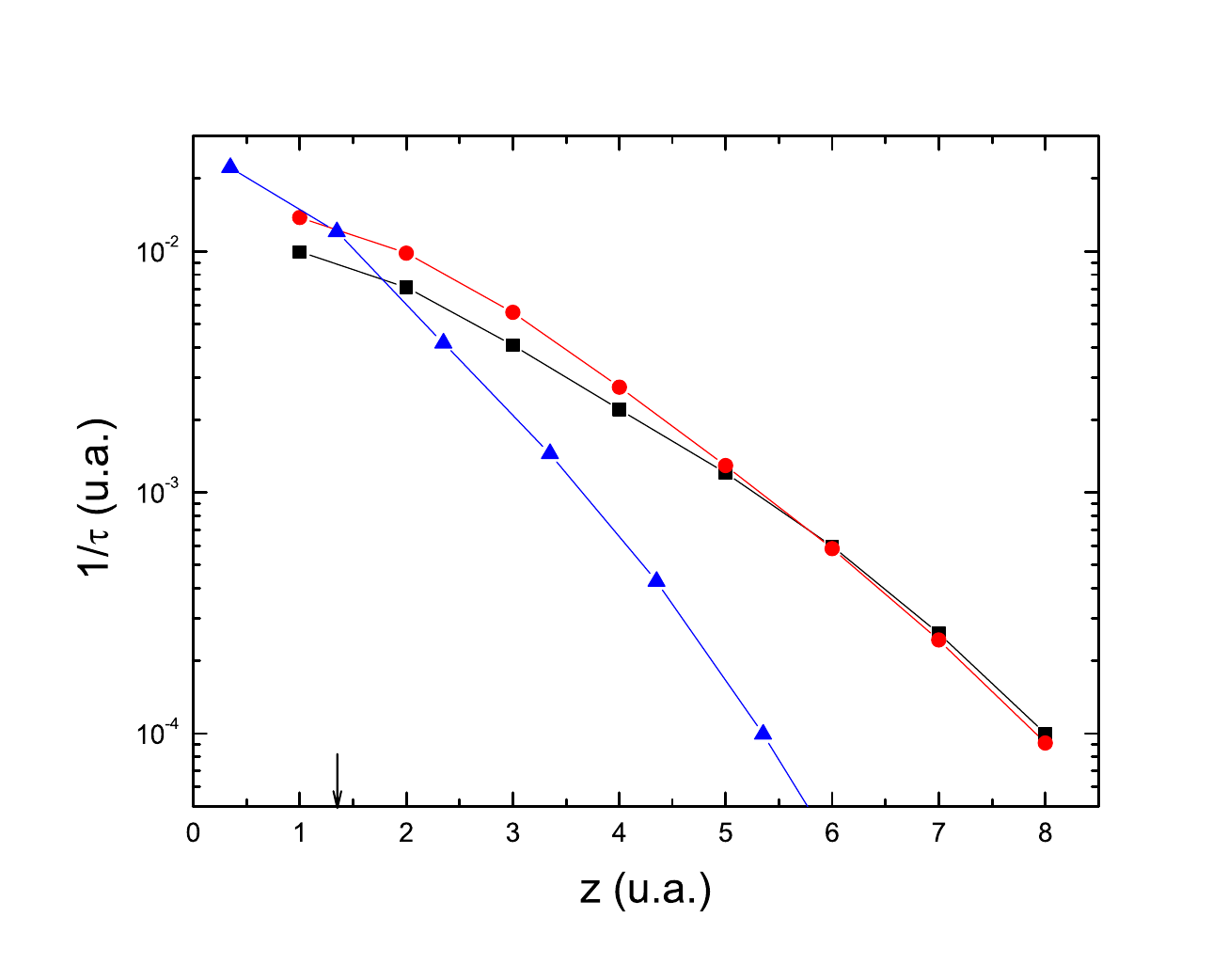}
\caption{}
\label{fig_tau_Al110_tc_jell}
\end{figure}

\newpage

\begin{figure}[htbp]
\centering
\includegraphics[width=7cm]{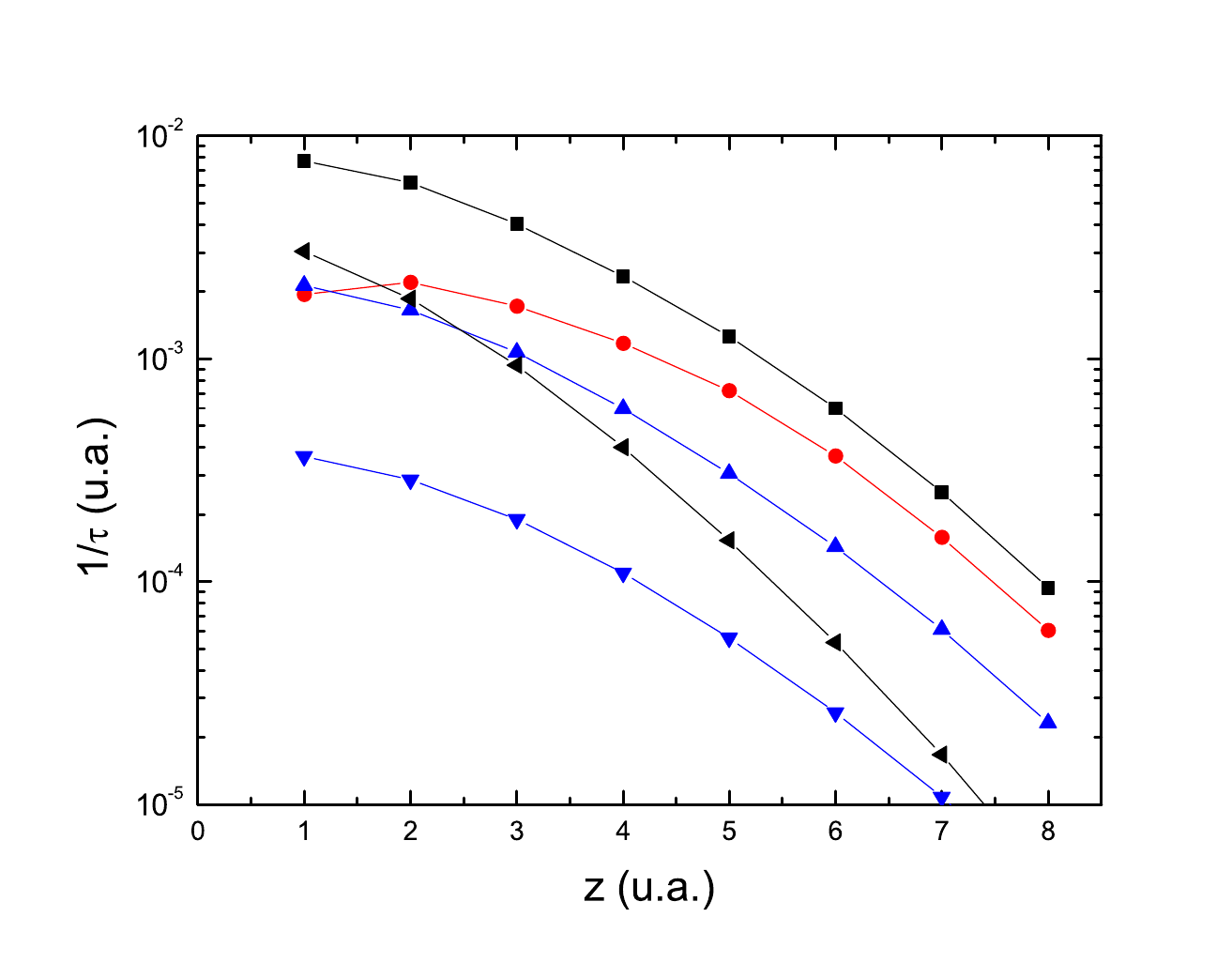}
\caption{}
\label{fig_tau_Al110t_vecs}
\end{figure}

\newpage

\begin{figure}[htbp]
\centering
\includegraphics[width=7cm]{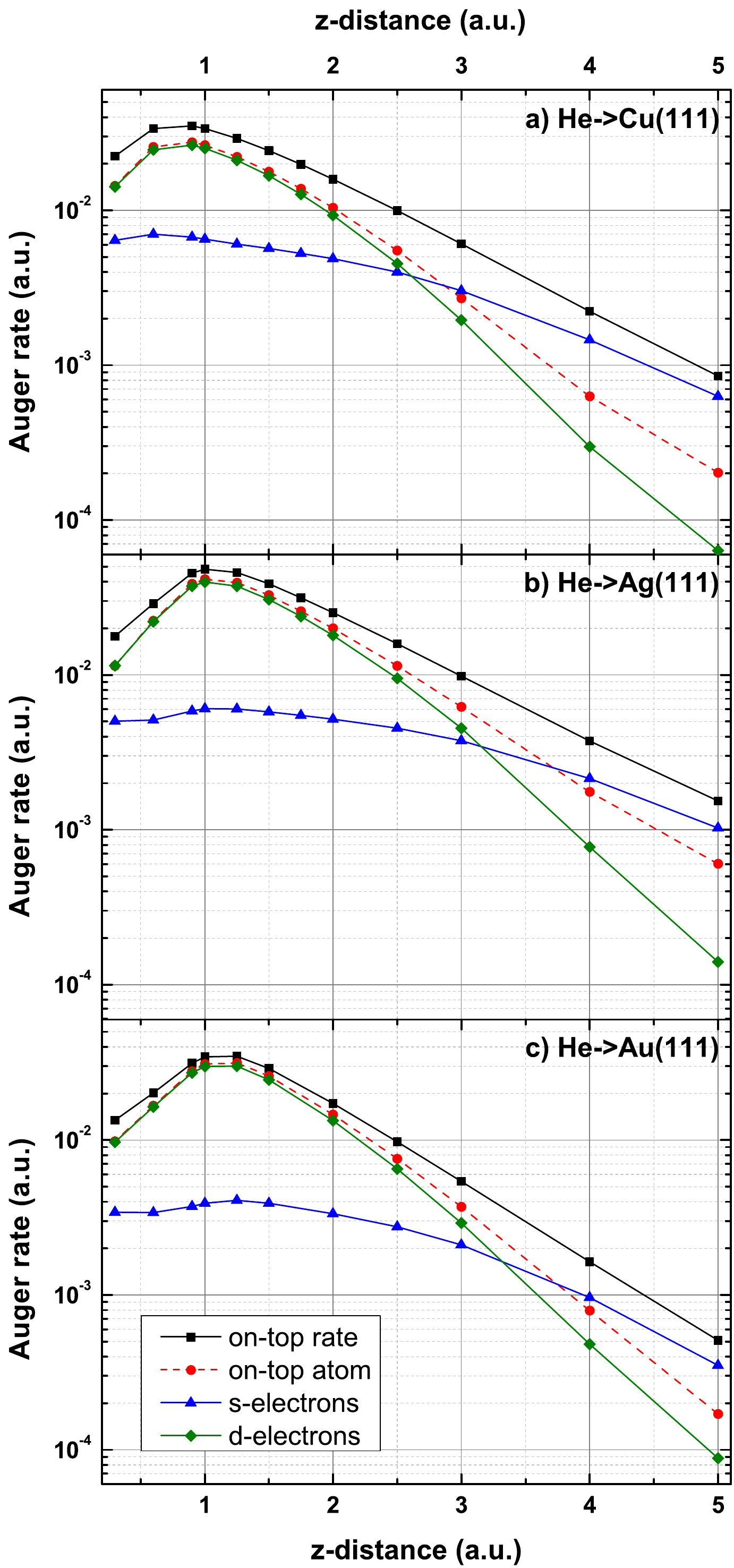}
\caption{}
\label{fig_rates_CuAgAu}
\end{figure}

\newpage

\begin{figure}[htbp]
\centering
\includegraphics[width=7cm]{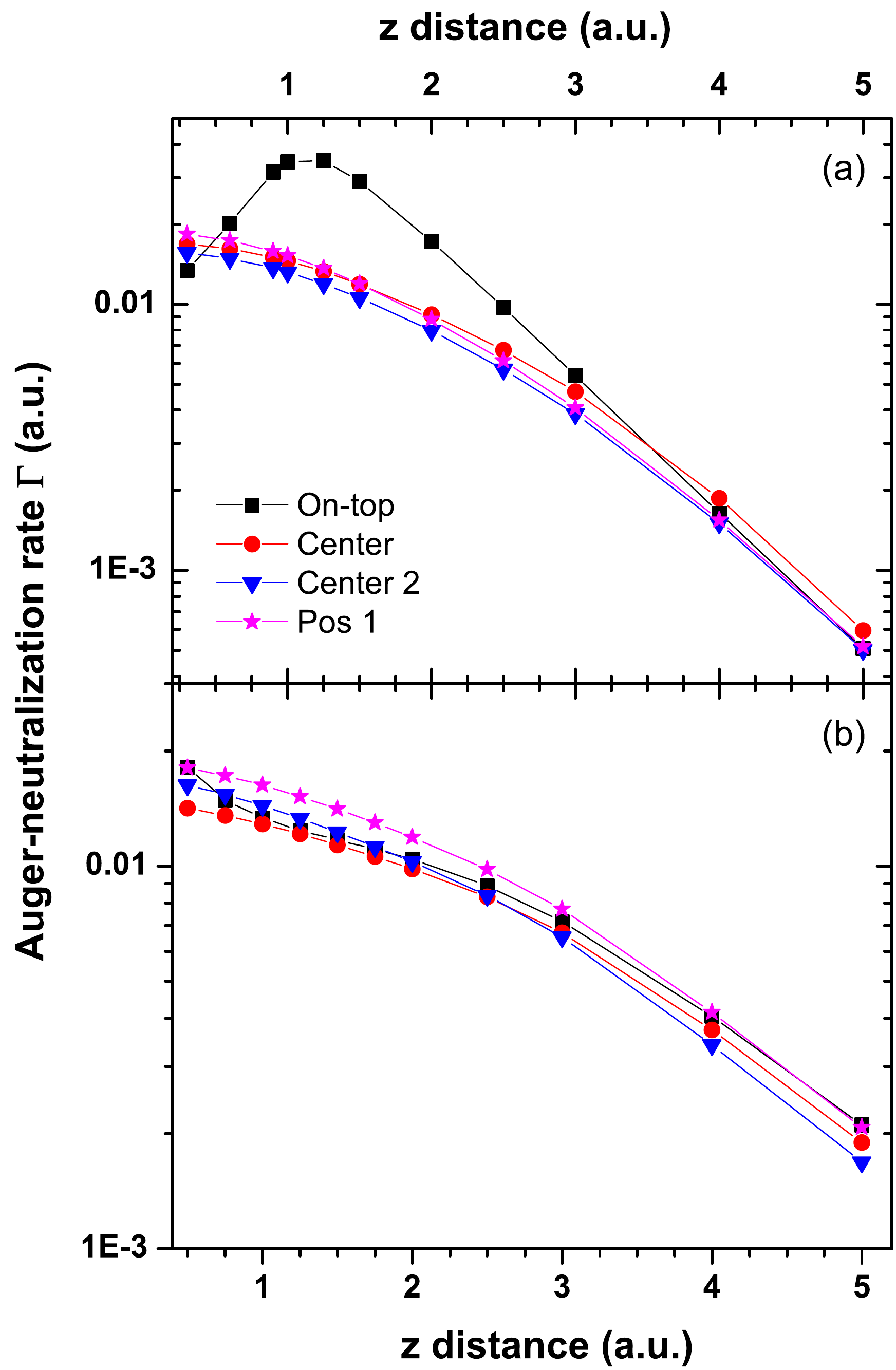}
\caption{}
\label{fig_rates_AuAl_tcl}
\end{figure}

\newpage

\begin{figure}[htbp]
\centering
\includegraphics[width=7cm]{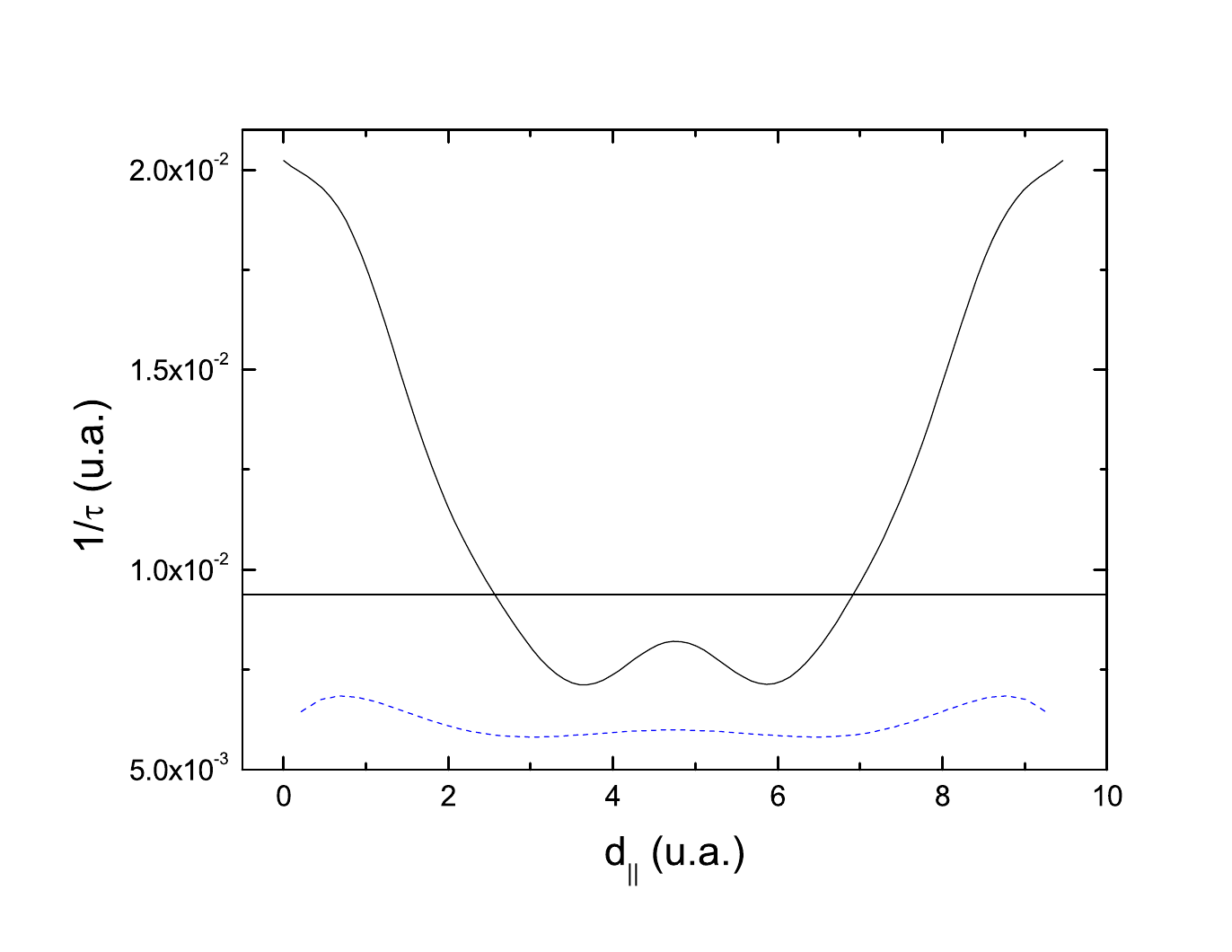}
\caption{}
\label{fig_rate_Ag110_z2_az35}
\end{figure}

\newpage

\begin{figure}[htbp]
\centering
\includegraphics[width=7cm]{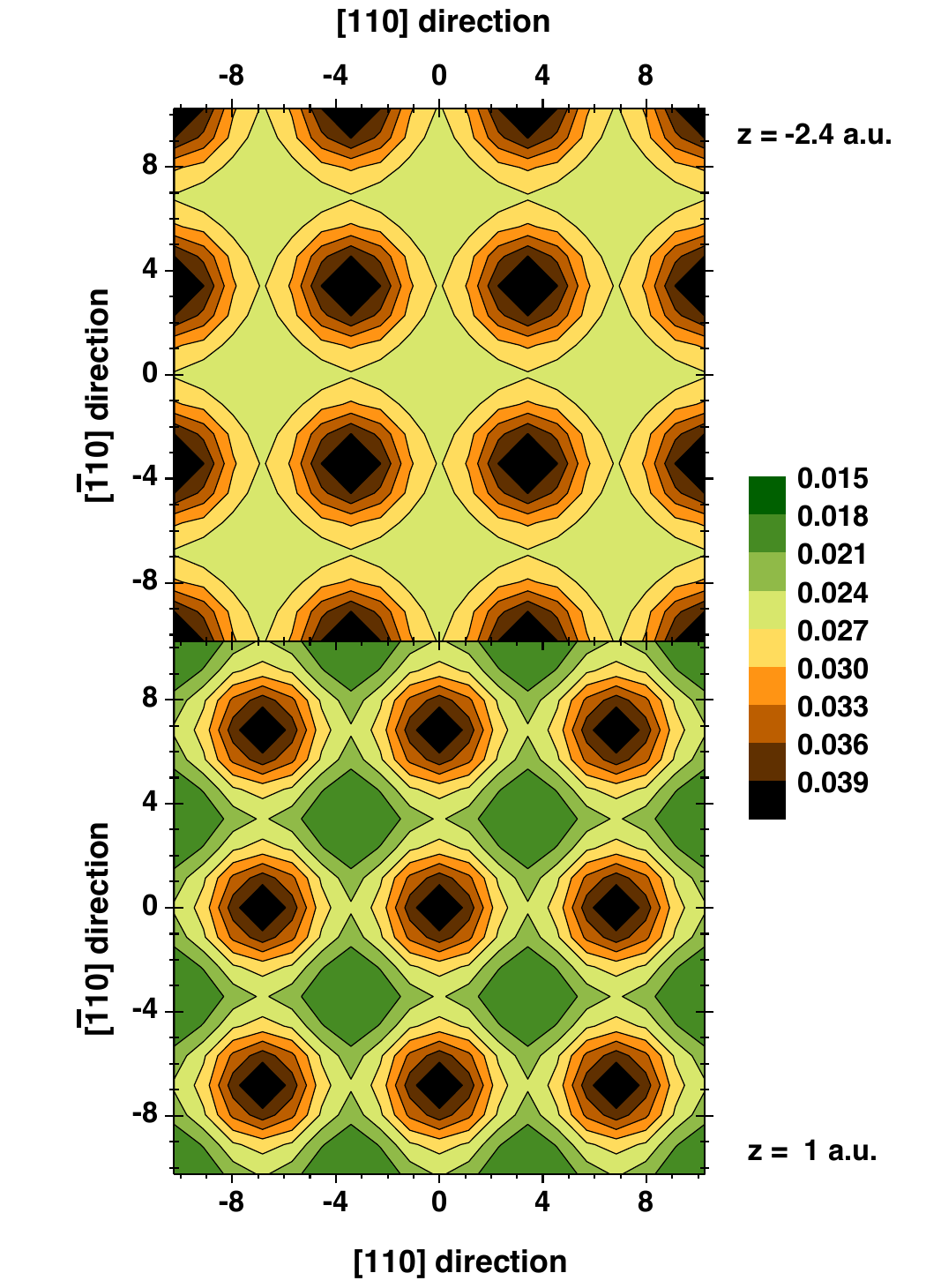}
\caption{}
\label{fig_rate_Cu100_contour}
\end{figure}

\begin{figure}[htbp]
\centering
\includegraphics[width=7cm]{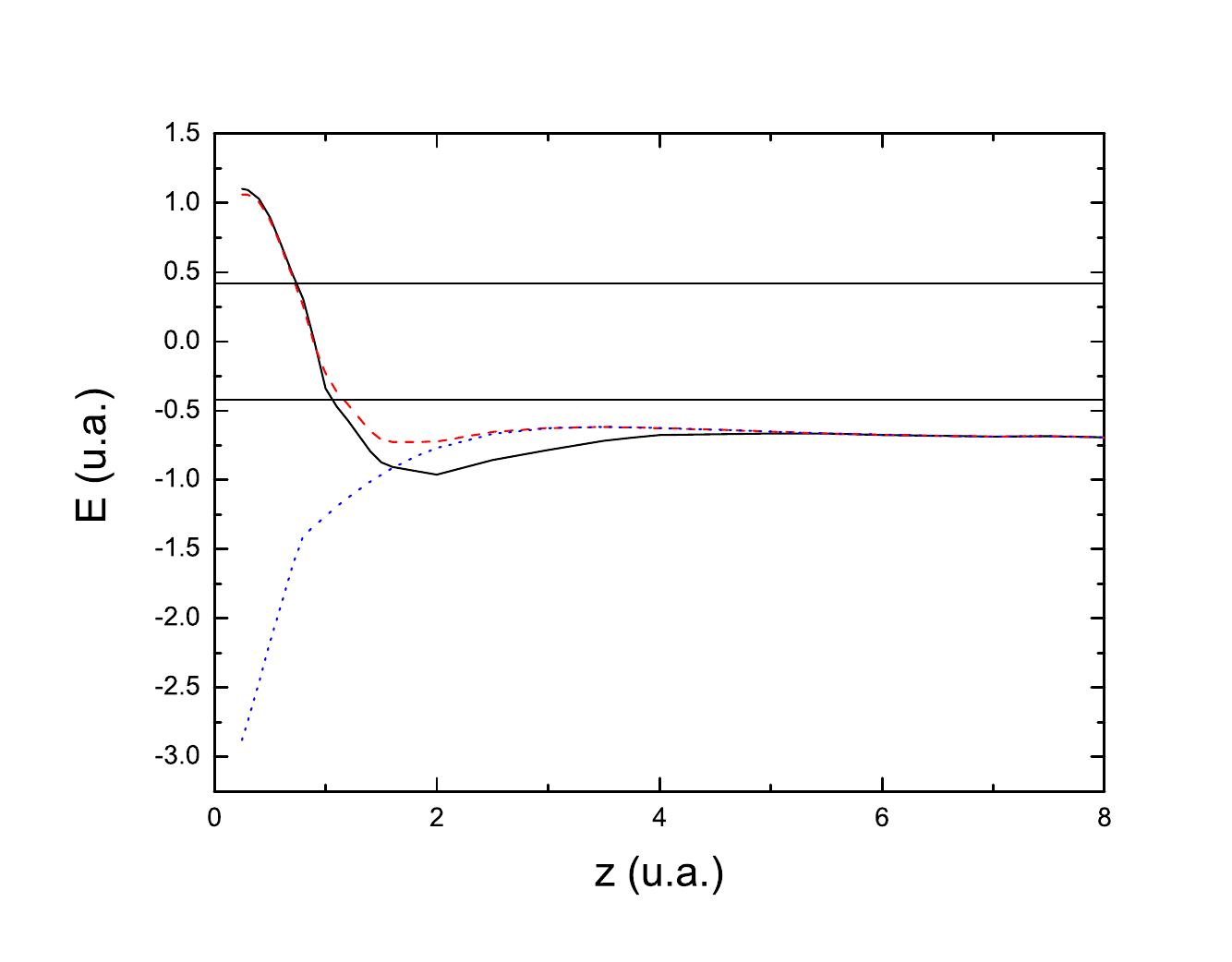}
\caption{} 
\label{fig_nivel_HeAl}
\end{figure}

\newpage


\begin{figure}[htbp]
\centering
\includegraphics[width=7cm]{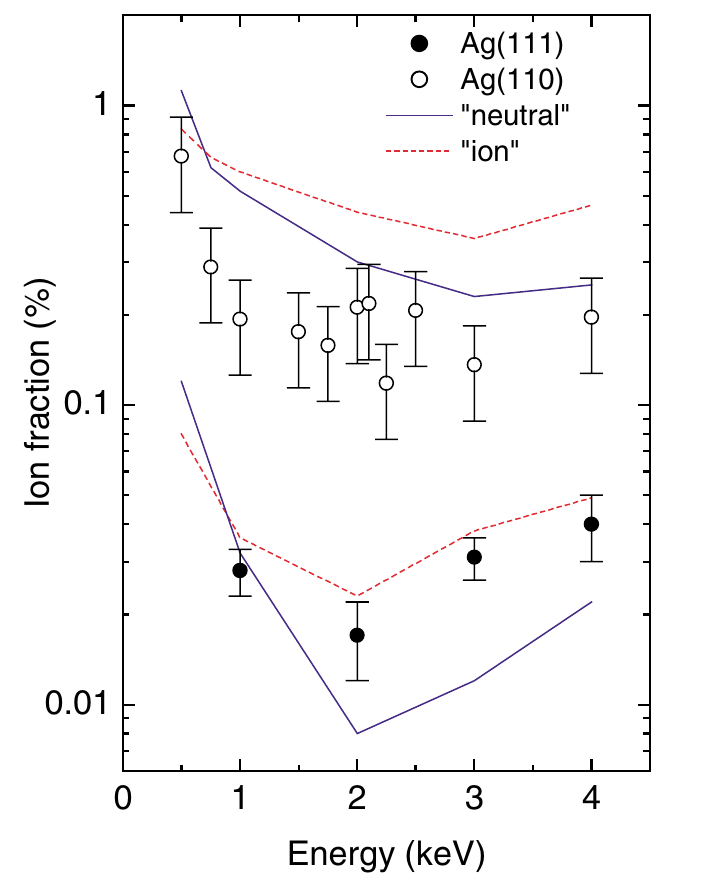}
\caption{}
\label{fig_HeAg_PRL}
\end{figure}

\begin{figure}[htbp]
\centering
\includegraphics[width=7cm]{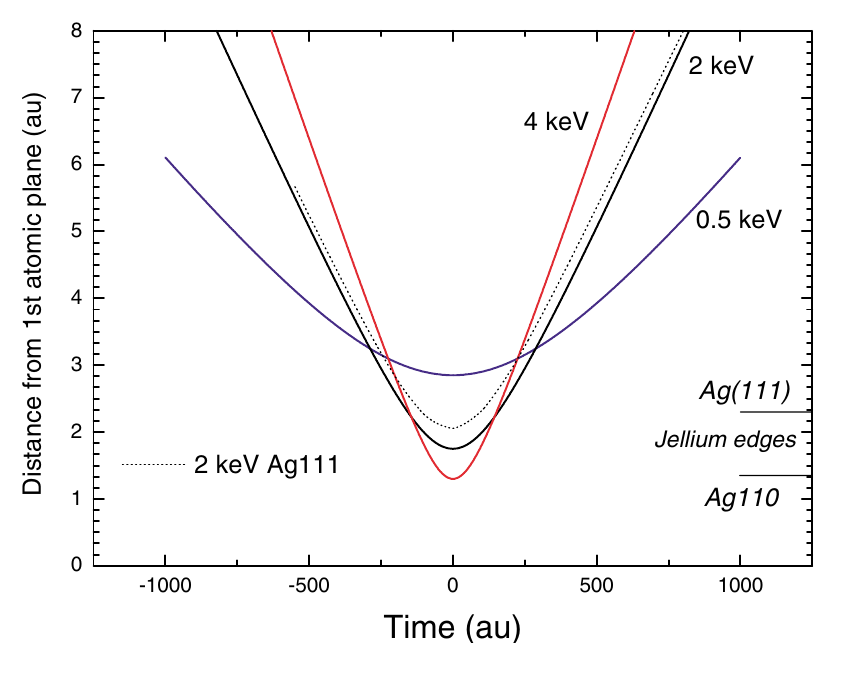}
\caption{}
\label{fig_traj_HeAg_PRL}
\end{figure}

\newpage

\begin{figure}[htbp]
\centering
\includegraphics[width=7cm]{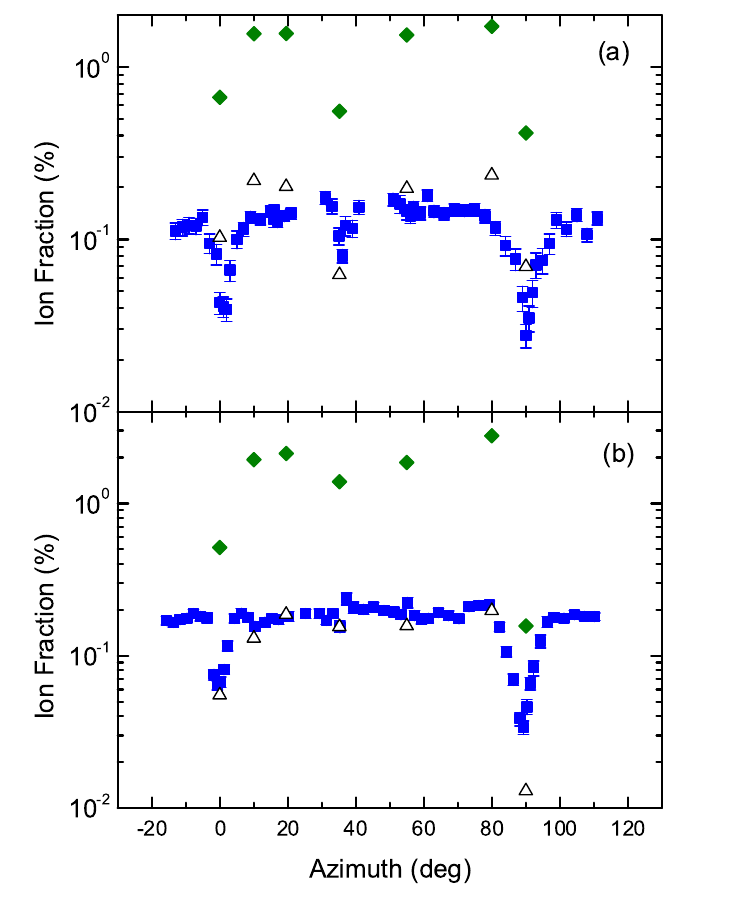}
\caption{} 
\label{fig_ionfractionAg110_az1k}
\end{figure}

\begin{figure}[htbp]
\centering
\includegraphics[width=8cm]{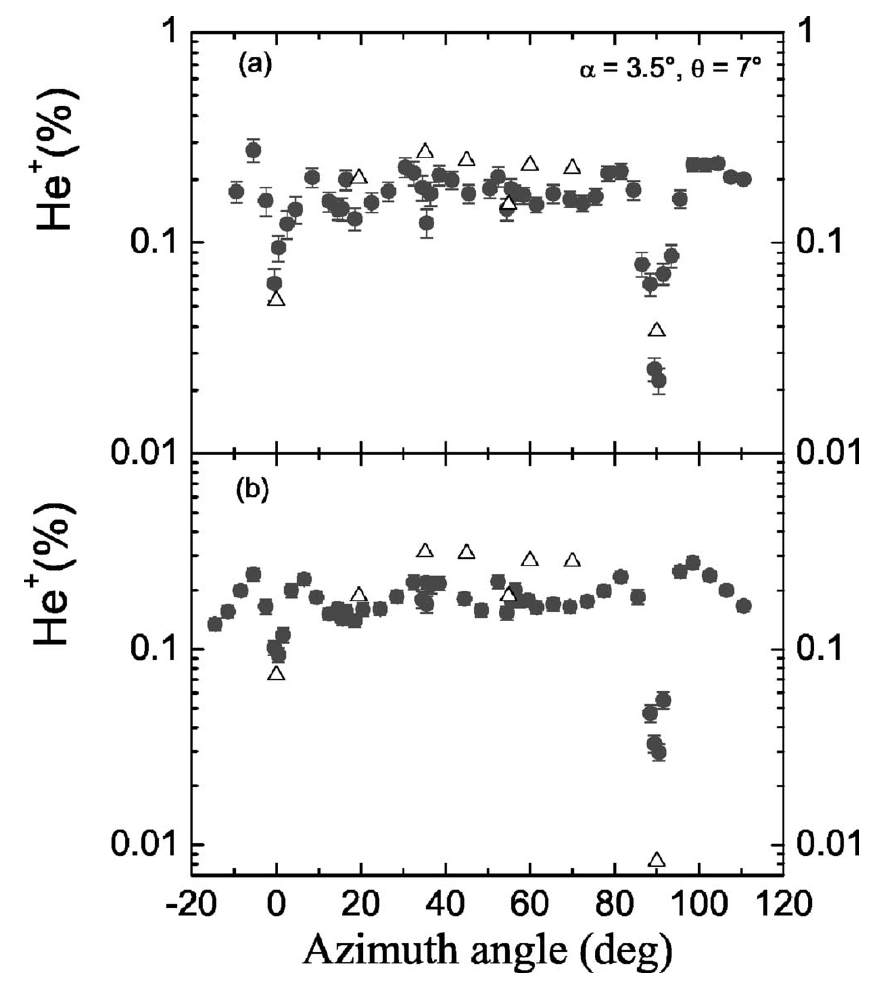}
\caption{} 
\label{fig_ionfractionAg110_az3k}
\end{figure}

\newpage

\begin{figure}[htbp]
\centering
\includegraphics[width=8cm]{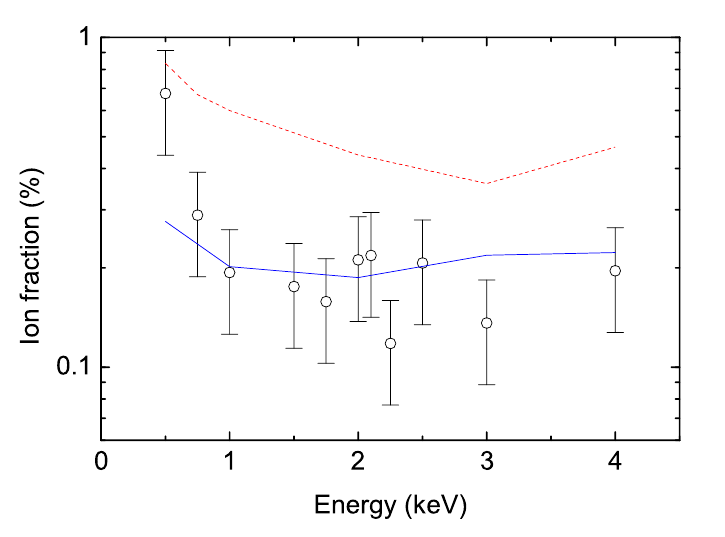}
\caption{} 
\label{fig_ionfraction_LCAOvsJelly}
\end{figure}

\begin{figure}[htbp]
\centering
\includegraphics[width=8.5cm]{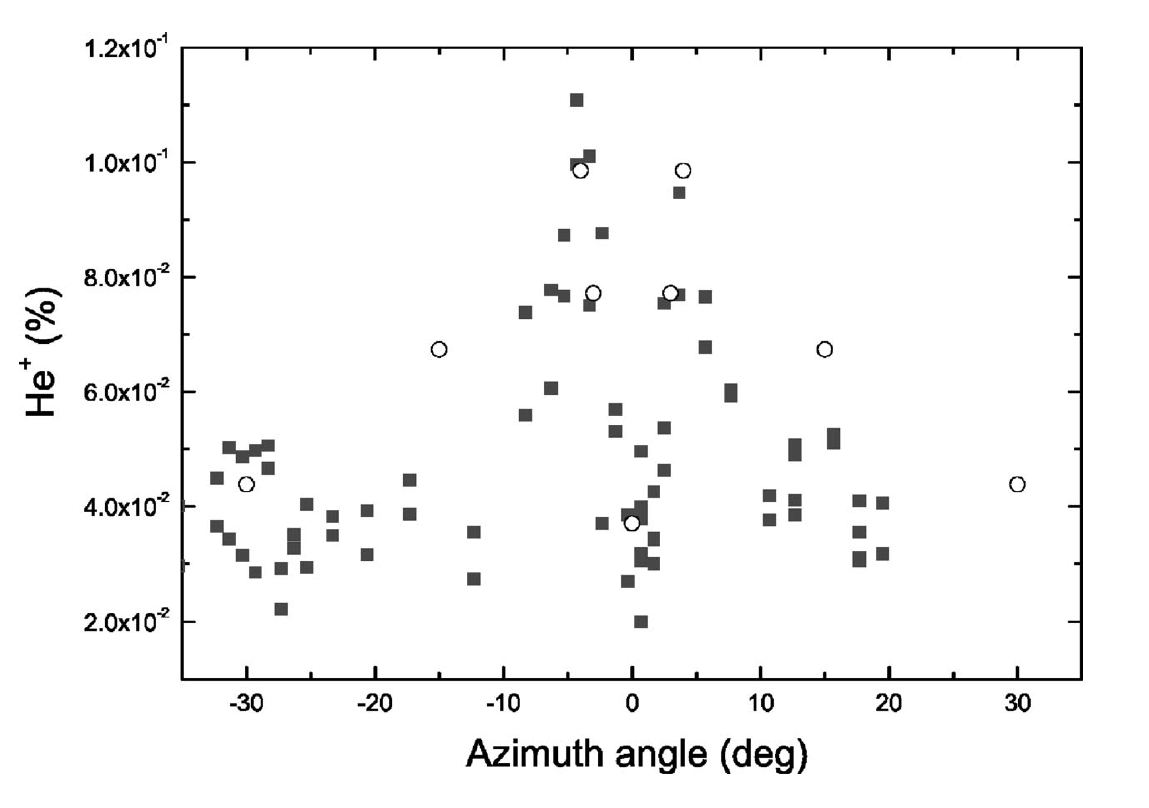}
\caption{} 
\label{fig_ionfractionAg111_az}
\end{figure}

\newpage

\begin{figure}[htbp]
\centering
\includegraphics[width=7cm]{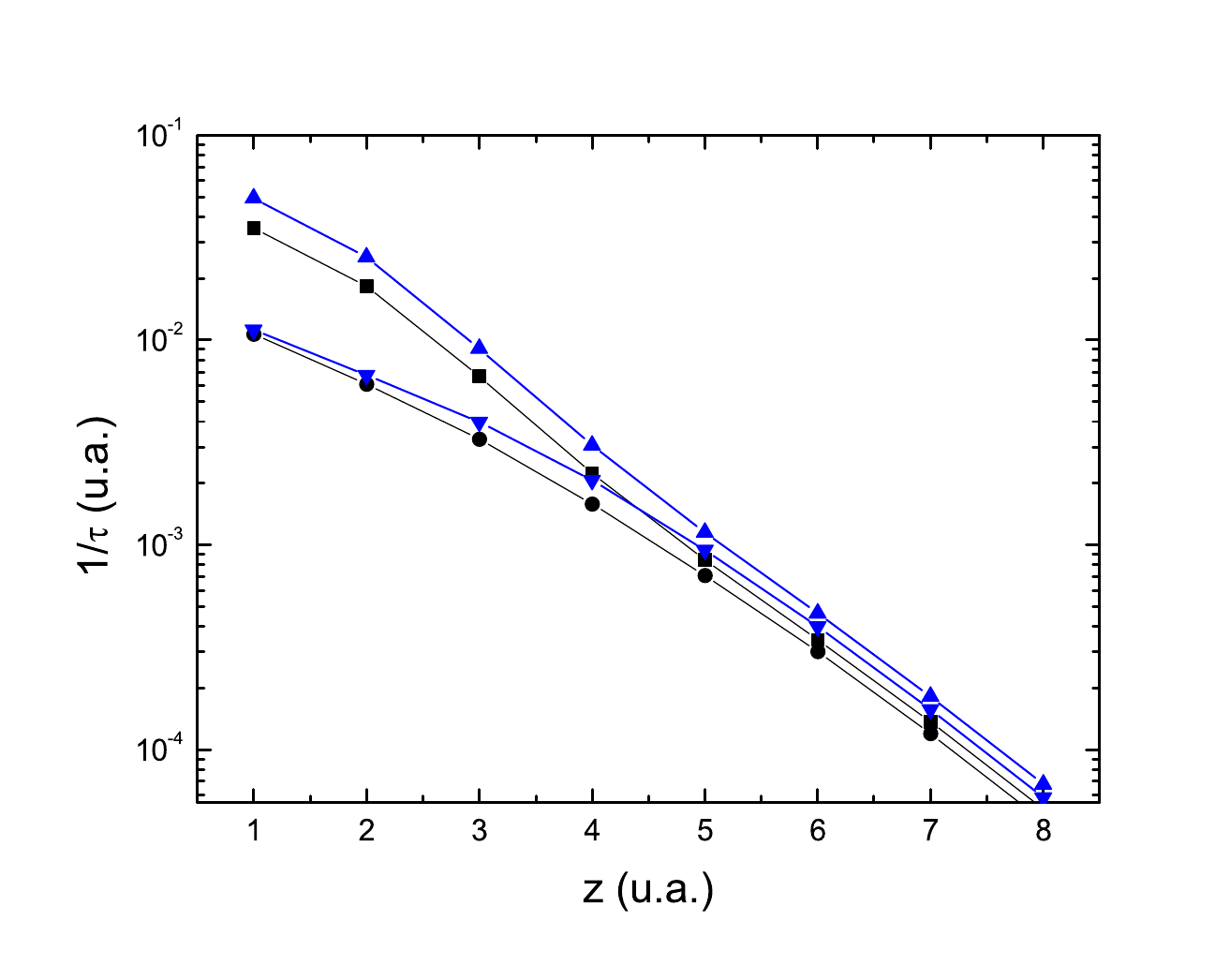}
\caption{} 
\label{fig_tauAg110_displacedjelly}
\end{figure}

\newpage


\begin{figure}[htbp]
\centering
\includegraphics[width=7cm]{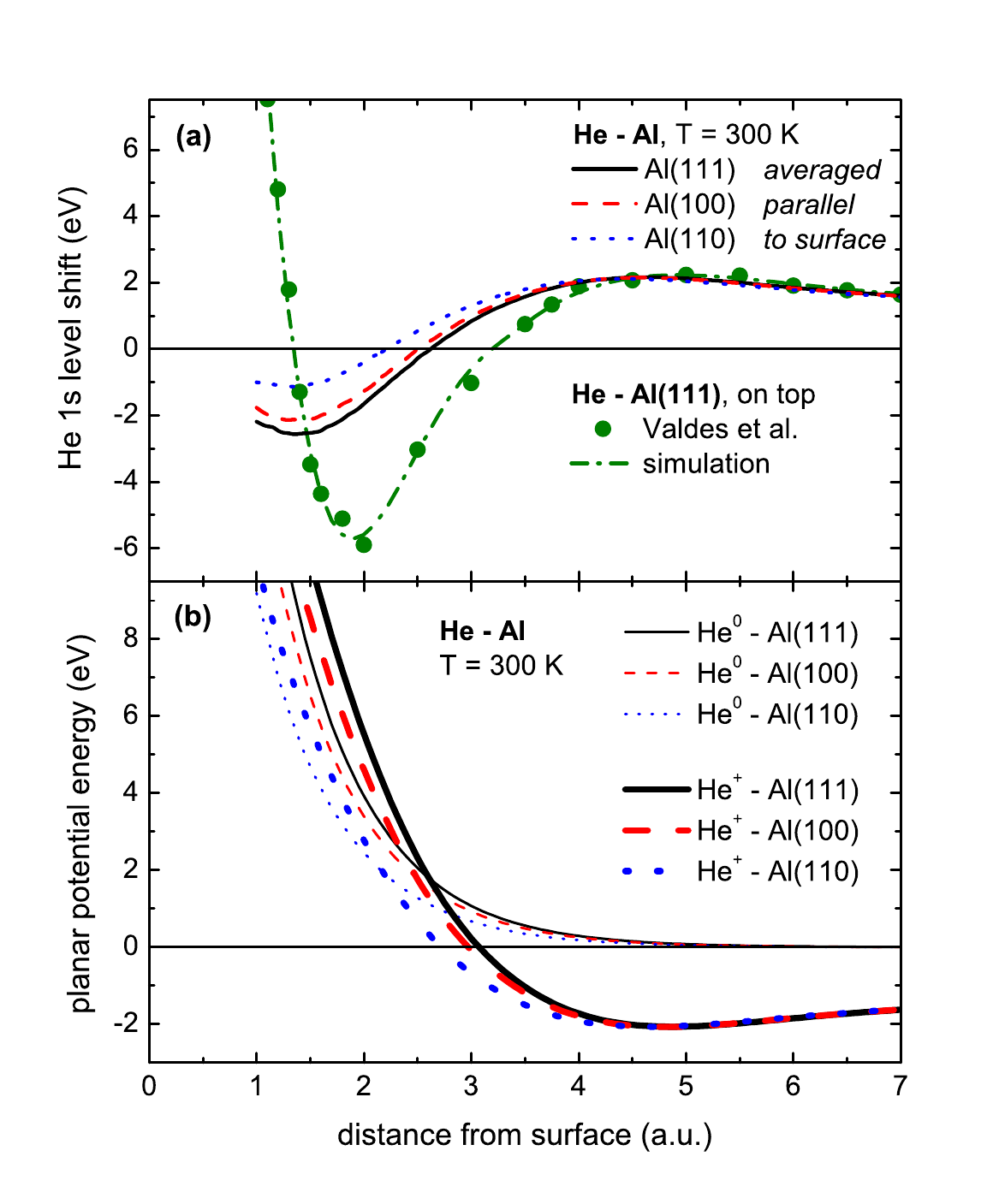}
\caption{} 
\label{fig_levelshift_Alxxx}
\end{figure}

\newpage

\begin{figure}[htbp]
\centering
\includegraphics[width=8cm]{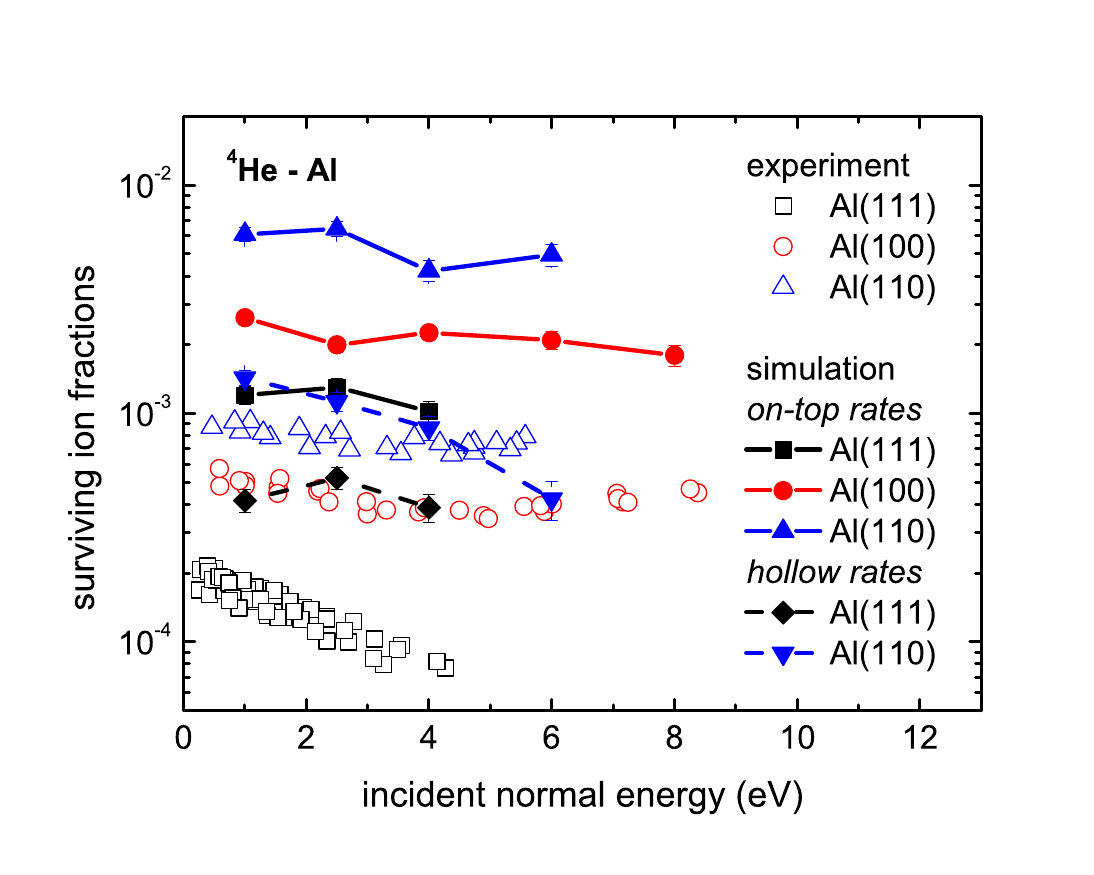}
\caption{} 
\label{fig_ionfractionAlxxx}
\end{figure}

\newpage

\begin{figure}[htbp]
\centering
\includegraphics[width=9cm]{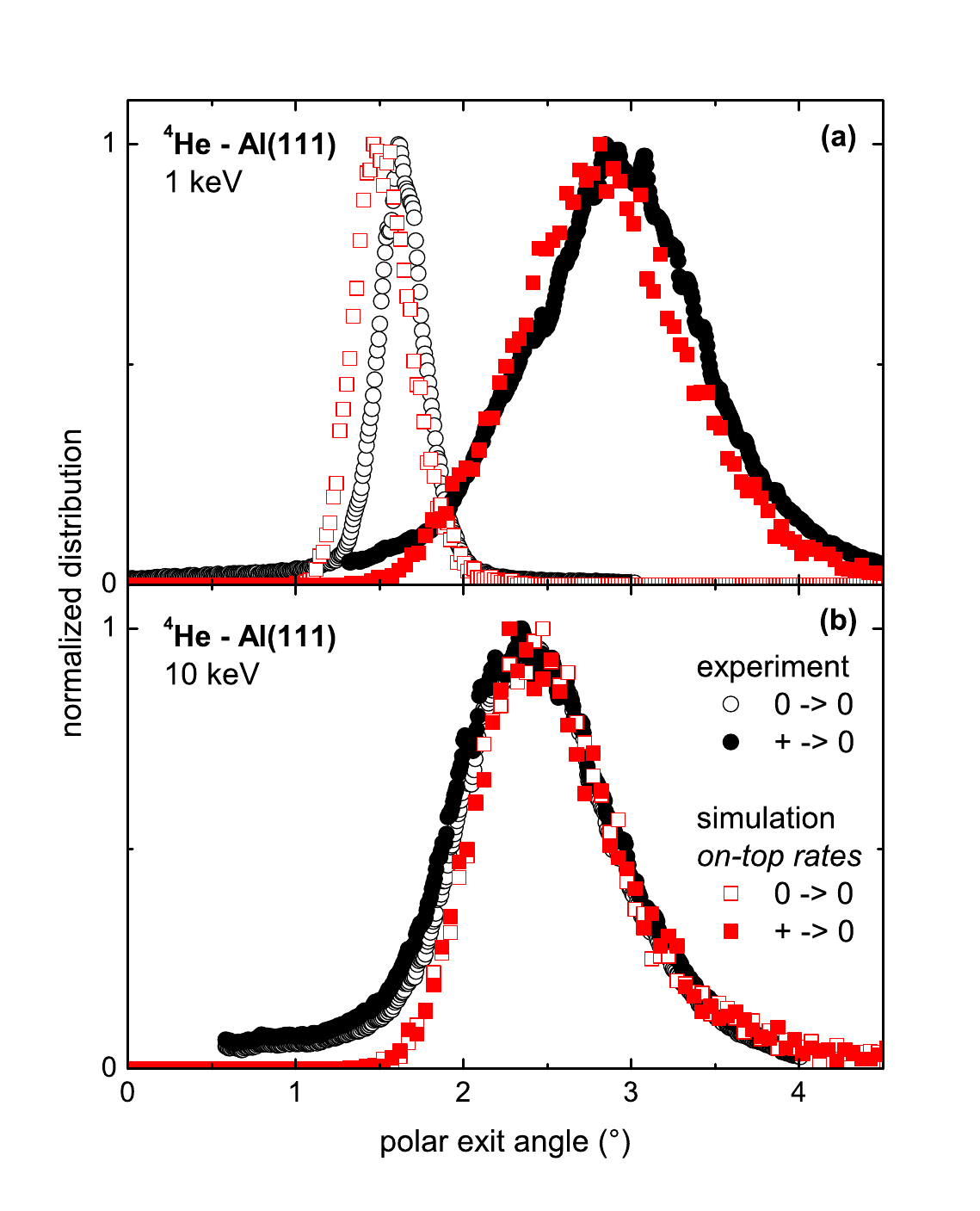}
\caption{} 
\label{fig_polardistributions}
\end{figure}

\newpage

\begin{figure}[htbp]
\centering
\includegraphics[width=9cm]{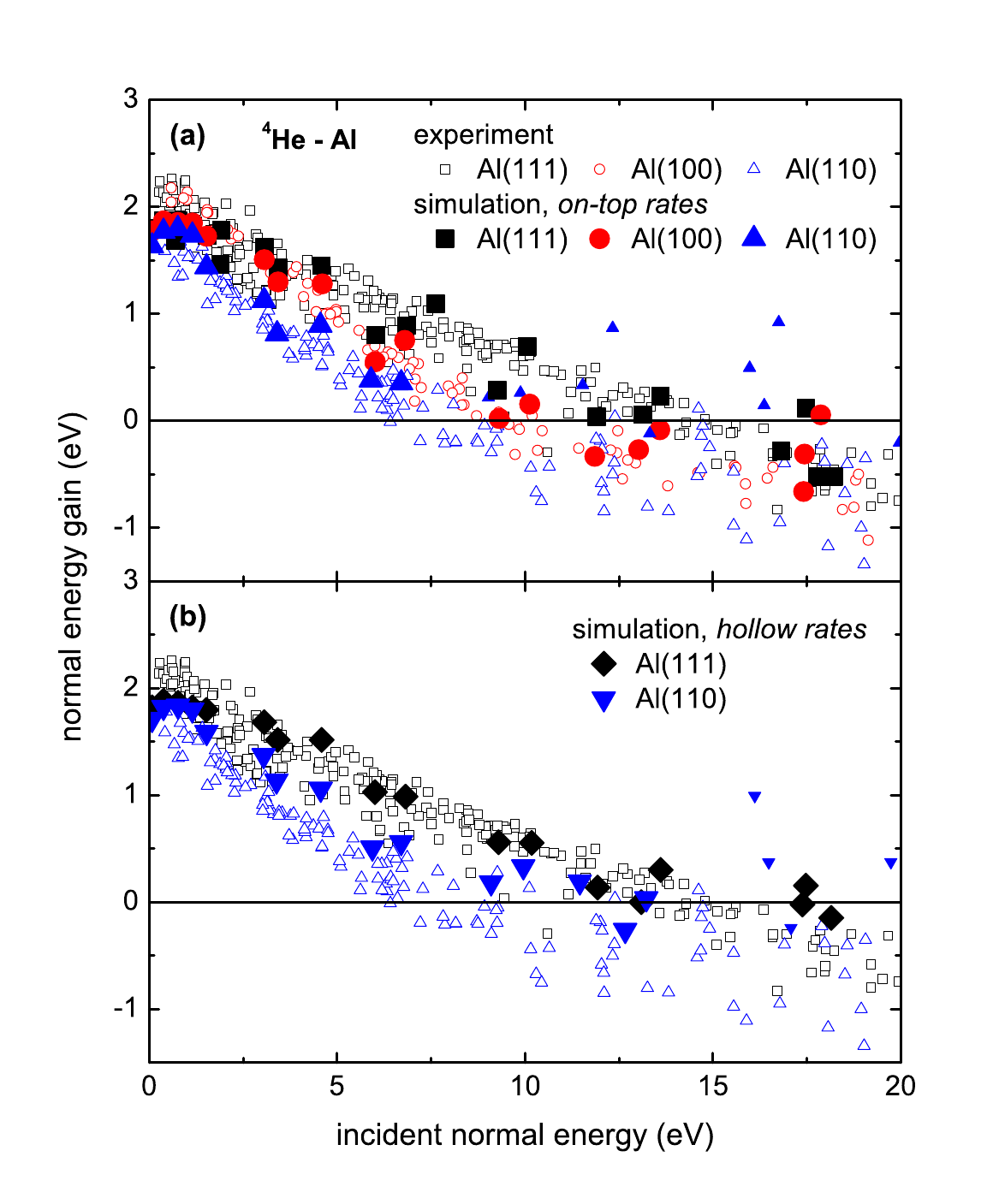}
\caption{} 
\label{fig_energygain_Alxxx}
\end{figure}

\newpage

\begin{figure}[htbp]
\centering
\includegraphics[width=10cm]{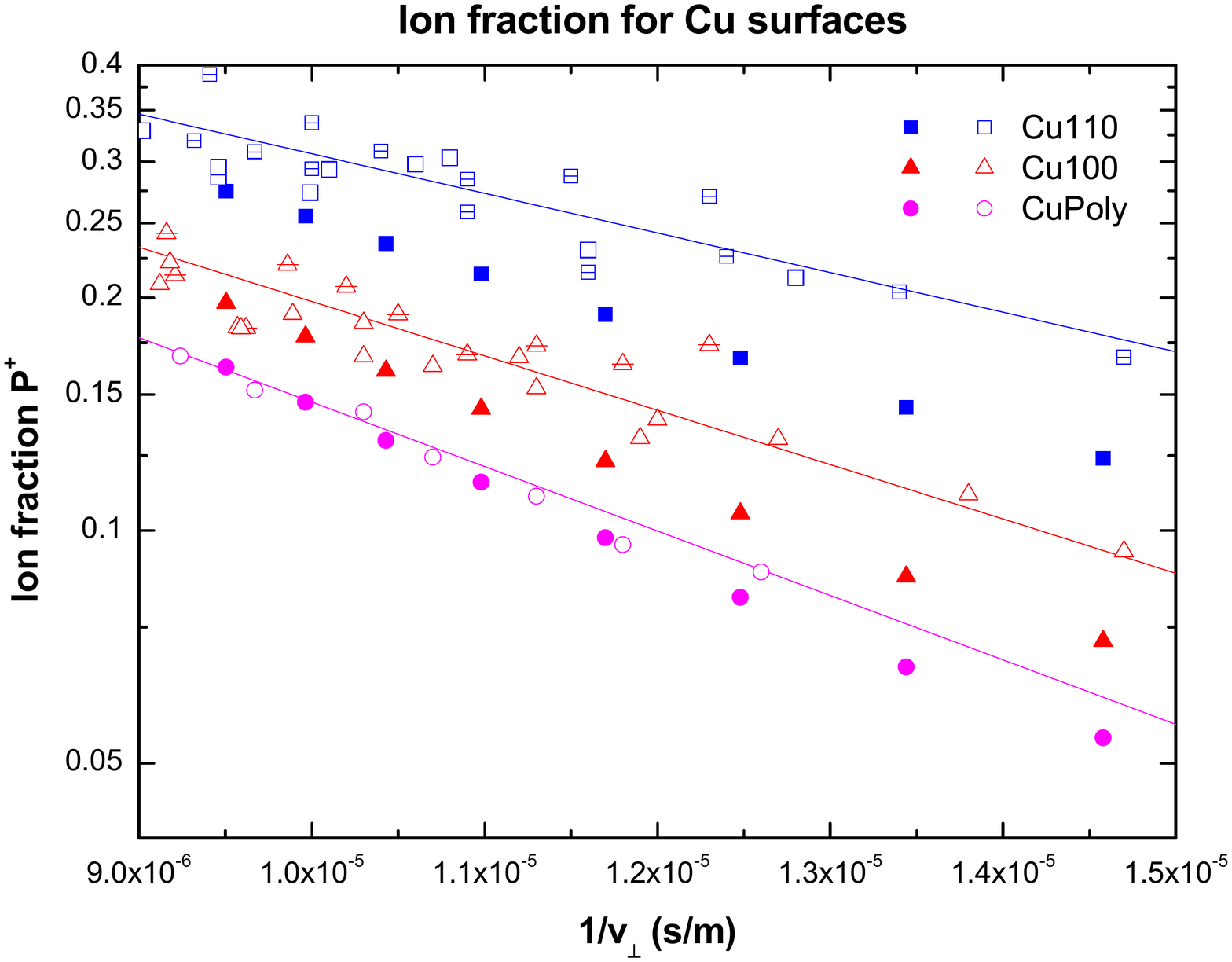}
\caption{} 
\label{fig_LEIS_Cuxxx}
\end{figure}

\newpage

\begin{figure}[htbp]
\centering
\includegraphics[width=9cm]{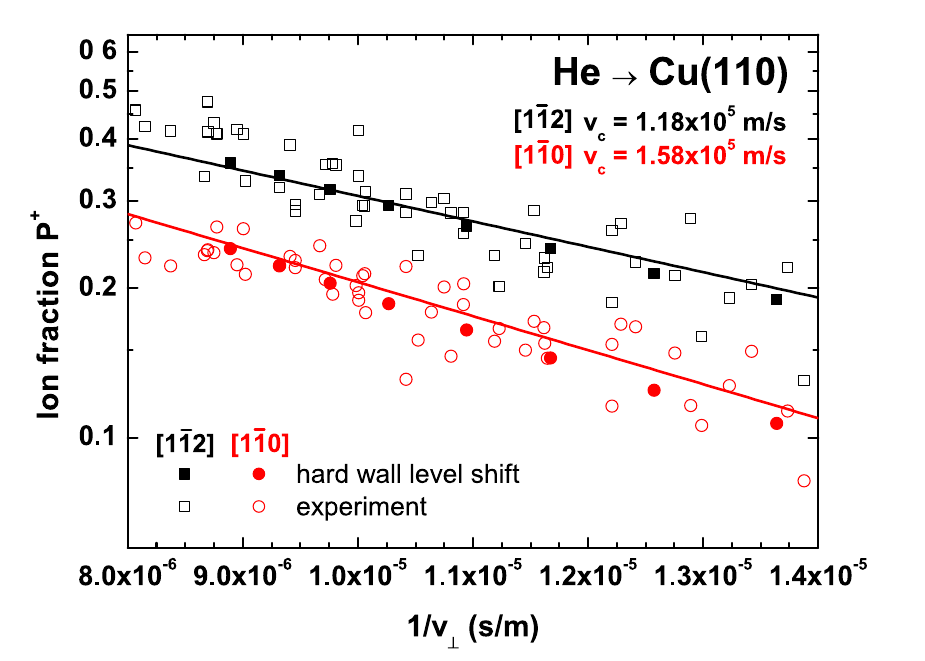}
\caption{} 
\label{fig_LEIS_Cu110_azimuth}
\end{figure}

\newpage

\begin{figure}[htbp]
\centering
\includegraphics[width=9cm]{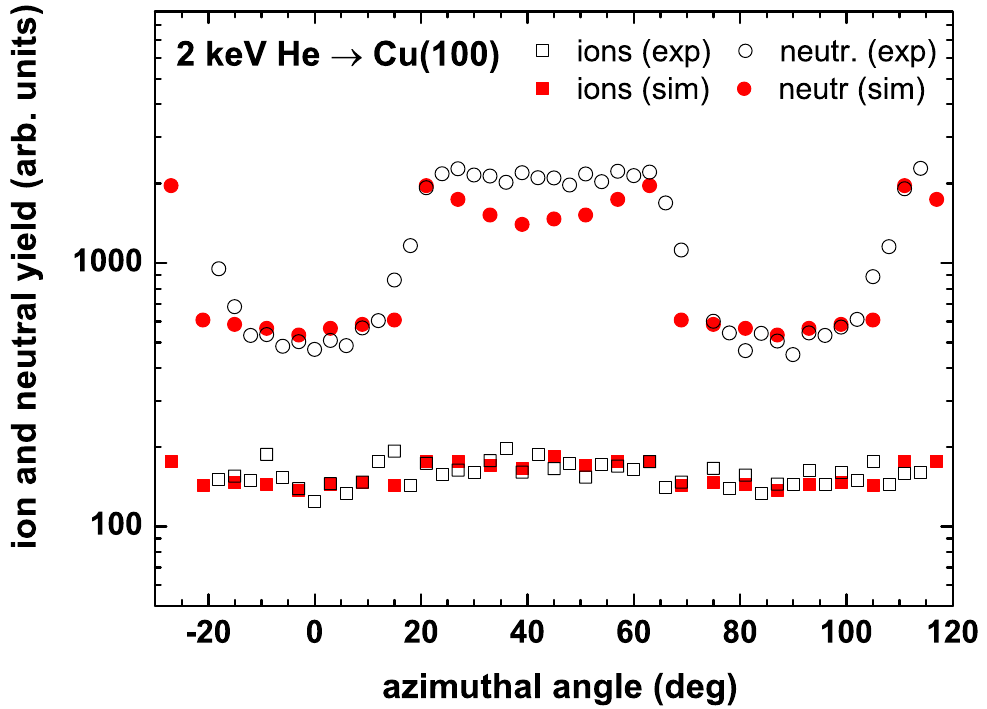}
\caption{} 
\label{fig_LEIS_Cu100_ionsneutrals_azimuth}
\end{figure}

\newpage

\begin{figure}[htbp]
\centering
\includegraphics[width=9cm]{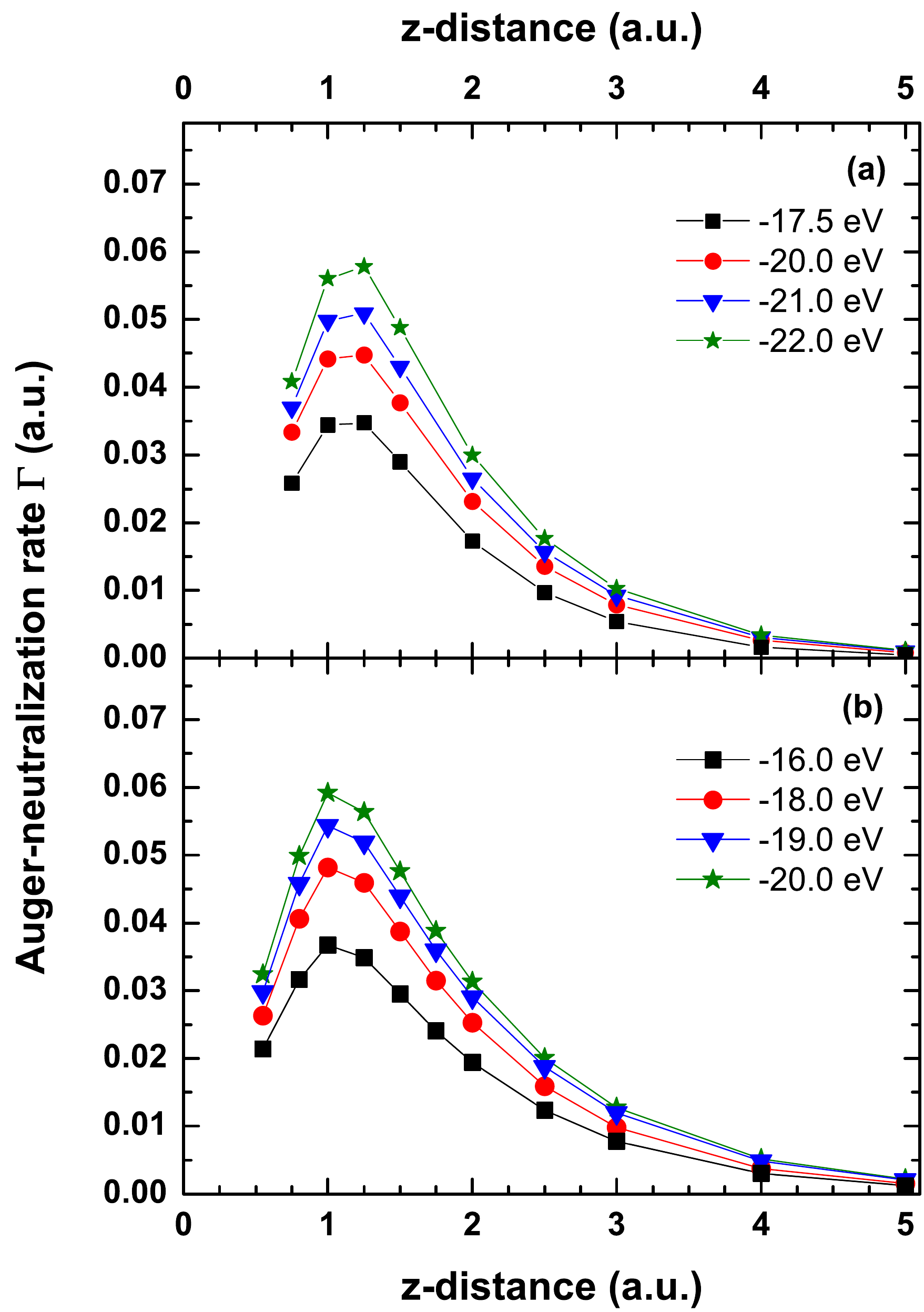}
\caption{} 
\label{fig_LEIS_rates_AgAu}
\end{figure}

\newpage

\begin{figure}[htbp]
\centering
\includegraphics[width=9cm]{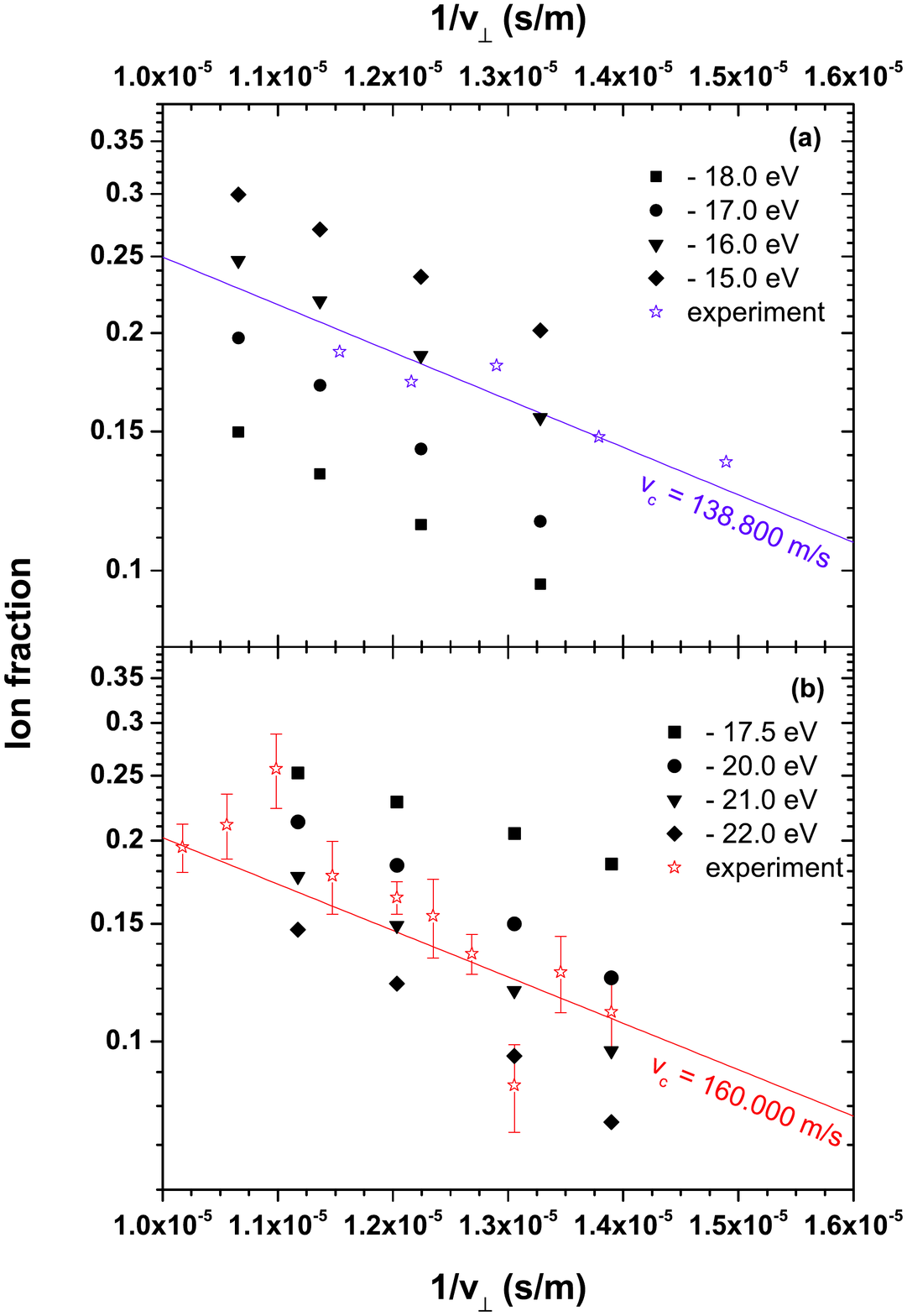}
\caption{} 
\label{fig_LEIS_fractions_AgAu}
\end{figure}

\newpage

\begin{figure}[htbp]
\centering
\includegraphics[width=9cm]{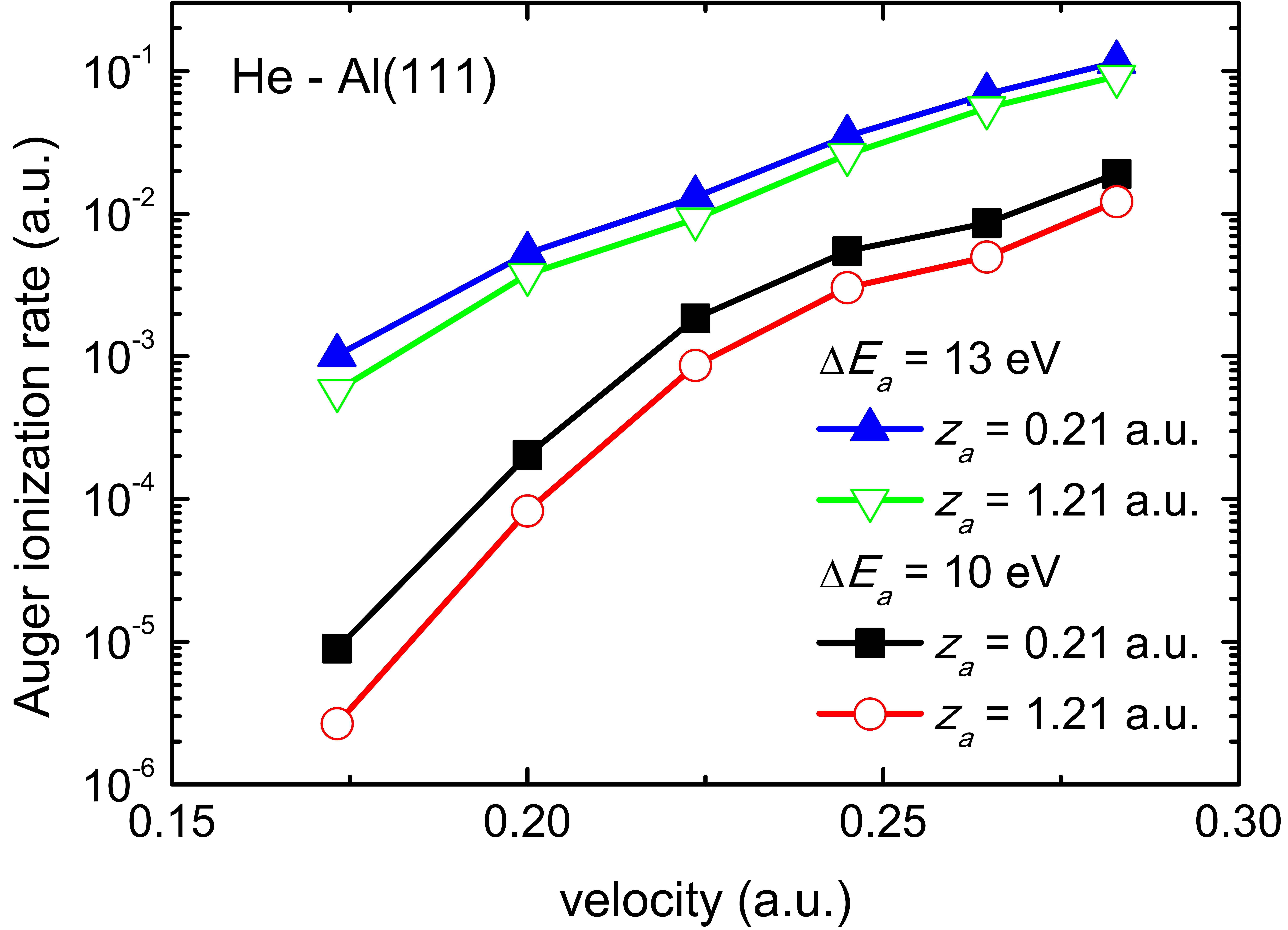}
\caption{} 
\label{fig_AI_rate_vs_v}
\end{figure}

\begin{figure}[htbp]
\centering
\includegraphics[width=9cm]{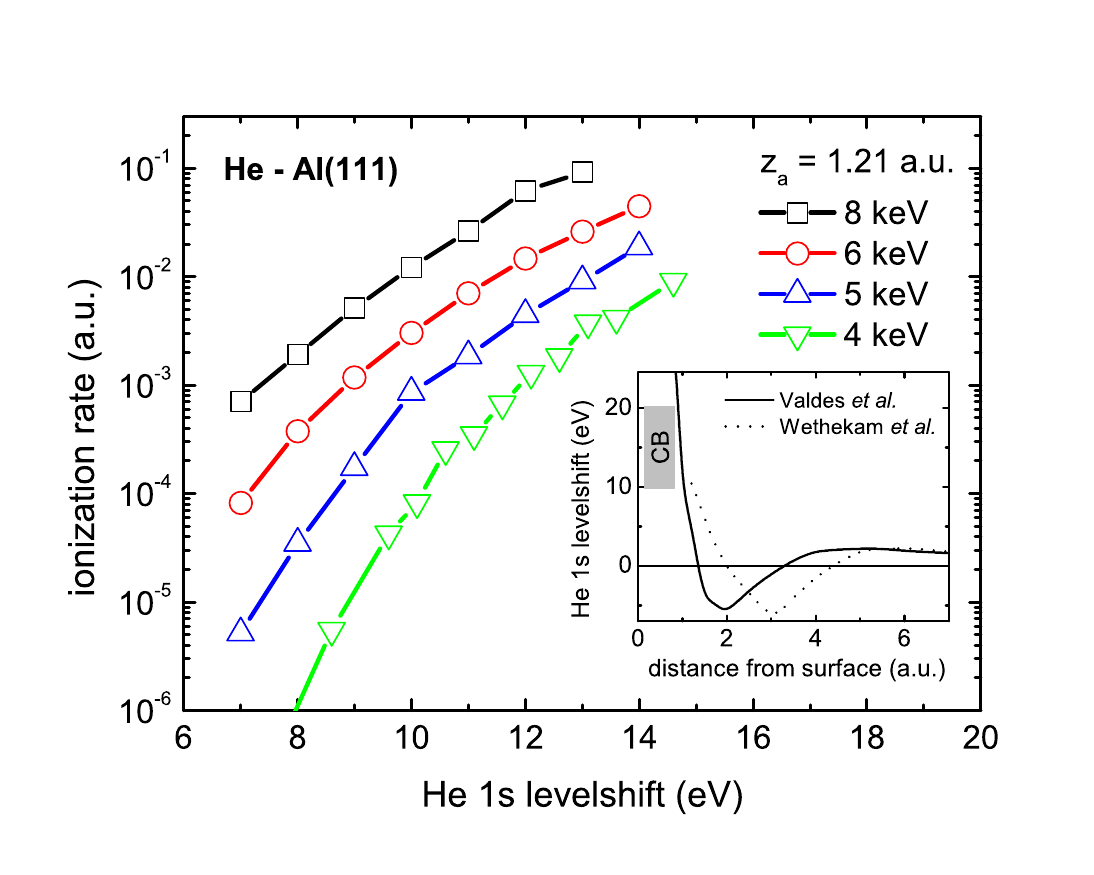}
\caption{} 
\label{fig_AI_rate_vs_E}
\end{figure}

\newpage

\begin{figure}[htbp]
\centering
\includegraphics[width=7cm]{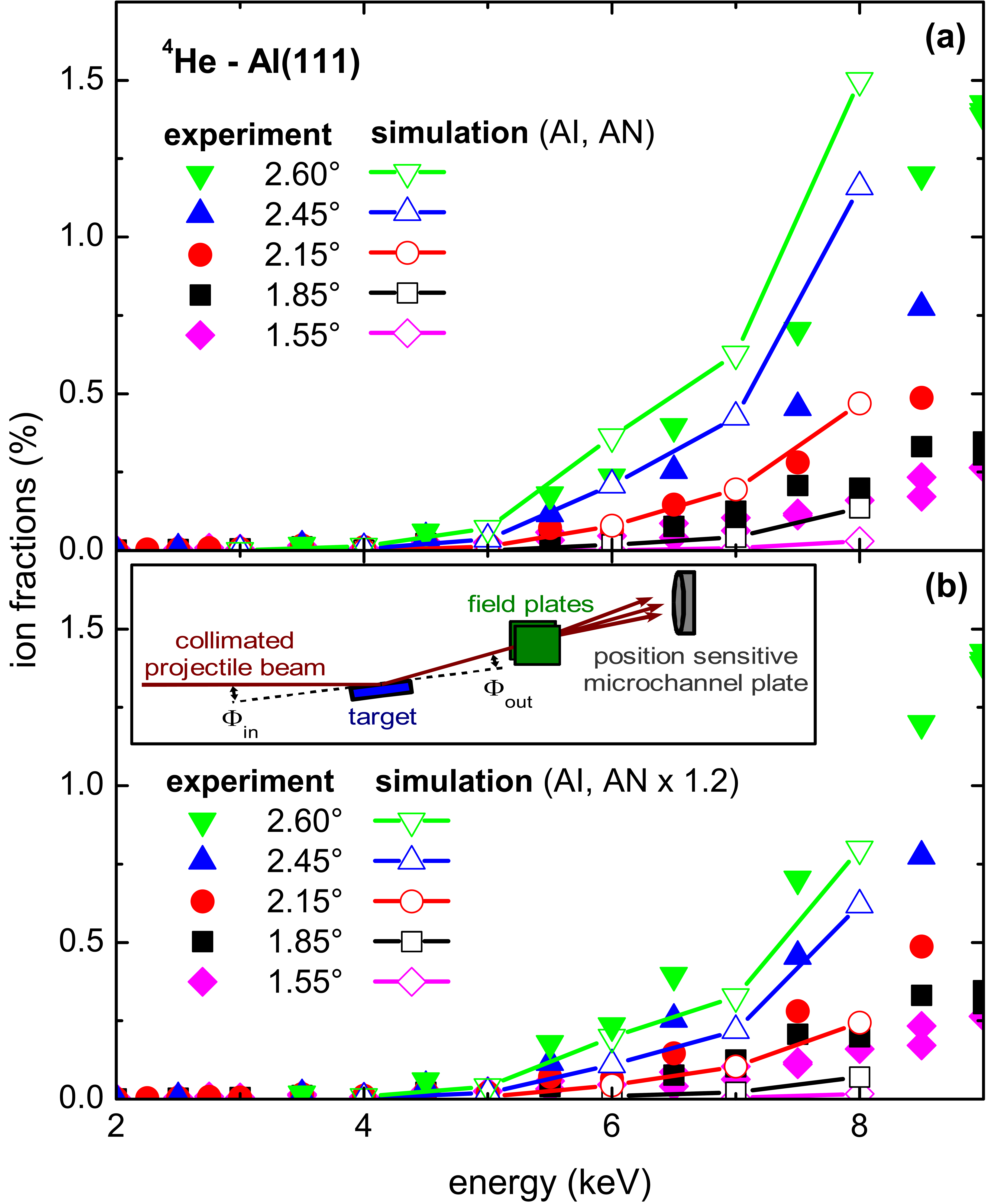}
\caption{} 
\label{fig_AI_ionfraction}
\end{figure}

\newpage

\begin{figure}[htbp]
\centering
\includegraphics[width=10cm]{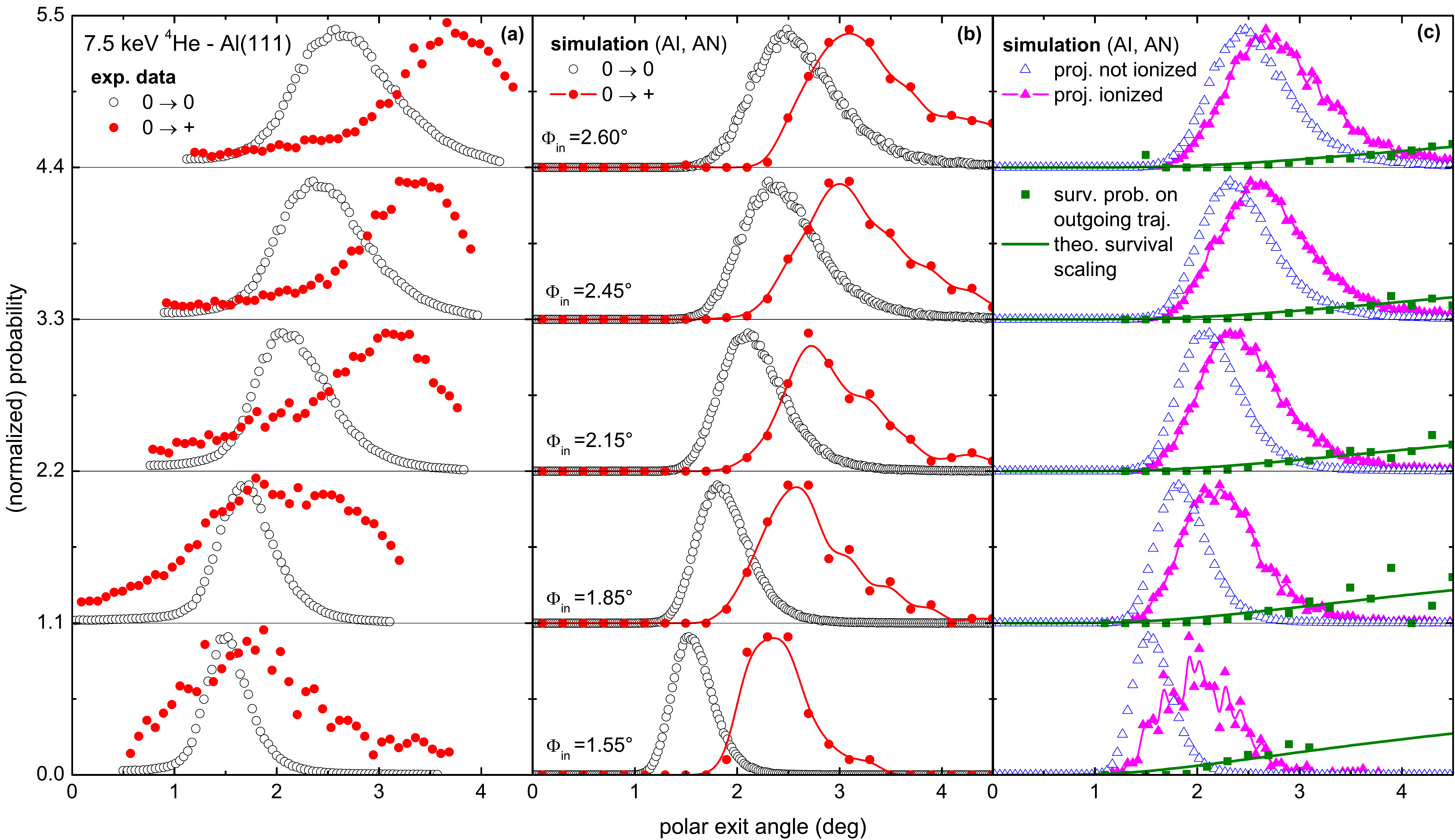}
\caption{} 
\label{fig_AI_exp_polar}
\end{figure}

\newpage

\begin{figure}[htbp]
\centering
\includegraphics[width=7cm]{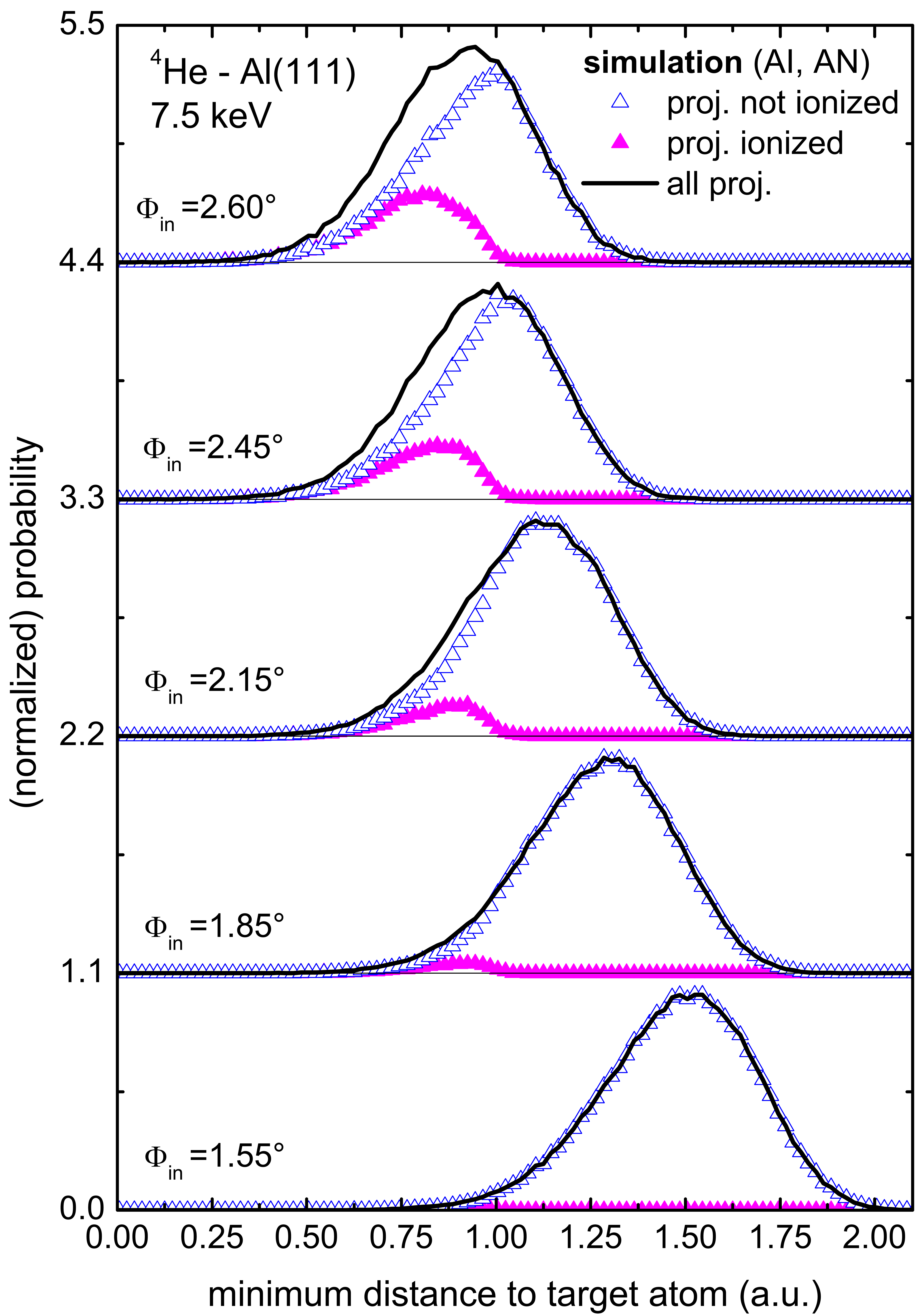}
\caption{} 
\label{fig_AI_proj_zmin}
\end{figure}

\newpage

\begin{figure}[htbp]
\centering
\includegraphics[width=7cm]{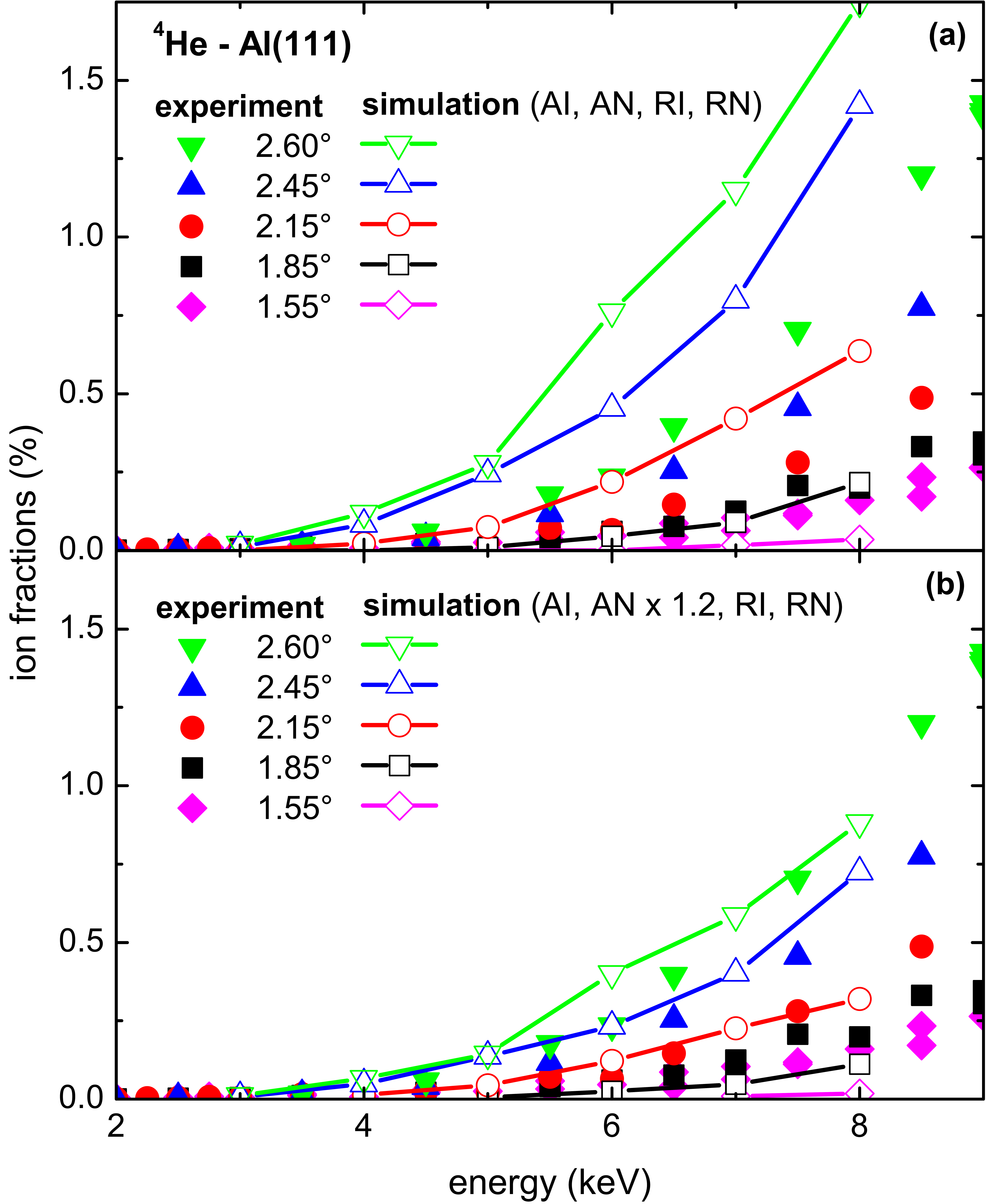}
\caption{} 
\label{fig_AI_ionfraction_AR}
\end{figure}

\end{document}